\documentclass[a4paper, 11pt]{article}
\usepackage{indentfirst}
\usepackage{graphicx}
\usepackage[dvips]{hyperref}
\usepackage[hang]{subfigure}
\usepackage{amsmath, amssymb, amsthm}
\usepackage{color}
\usepackage{psfrag}
\usepackage[mathscr]{euscript}

\newcommand\bbR{\mathbb{R}}
\newcommand\bbN{\mathbb{N}}
\newcommand\bxi{\boldsymbol{\xi}}
\newcommand\bx{\boldsymbol{x}}
\newcommand\bv{\boldsymbol{v}}
\newcommand\bu{\boldsymbol{u}}

\newcommand\bn{\boldsymbol{n}}
\newcommand\bF{\boldsymbol{F}}

\newcommand\bq{\boldsymbol{q}}
\newcommand\bC{\boldsymbol{C}}

\newcommand\dd{\,\mathrm{d}}
\newcommand\He{\mathit{He}}
\newcommand\Kn{\mathit{Kn}}

\newcommand\NRxx{NR$xx$ }
\newcommand\bNRxx{\texorpdfstring{{\bf NR}$\boldsymbol{xx}$ }{NR$xx$ }}

\numberwithin{equation}{section}
\graphicspath{{images/}}

{\theoremstyle{remark} \newtheorem{remark}{Remark}}
\newtheorem{proposition}{Proposition}

\title{\NRxx Simulation of Microflows with Shakhov Model}

\author{Zhenning Cai\thanks{School of Mathematical Sciences, Peking
    University, Beijing, China, email: {\tt caizn@pku.edu.cn}.},~~
  Ruo Li\thanks{CAPT, LMAM \& School of Mathematical Sciences, Peking
    University, Beijing, China, email: {\tt rli@math.pku.edu.cn}.},~~
  Zhonghua Qiao\thanks{Institute for Computational Mathematics \&
    Department of Mathematics, Hong Kong Baptist University, Kowloon
    Tong, Hong Kong, email: {\tt zqiao@hkbu.edu.hk}.}}

\begin{document}
\maketitle
% vim: tw=70:spell
\begin{abstract}
  In this paper, we propose a method to simulate the microflows with
  Shakhov model using the \NRxx method developed in \cite{NRxx, Cai,
    NRxx_new}. The equation under consideration is the Boltzmann
  equation with force terms and the Shakhov model is adopted to
  achieve the correct Prandtl number. As the focus of this paper, we
  derive a uniform framework for different order moment systems on the
  wall boundary conditions, which is a major difficulty in the moment
  methods. Numerical examples for both steady and unsteady problems
  are presented to show the convergence in the number of moments.

\vspace*{4mm}
\end{abstract}

\section{Introduction}
In the kinetic theory, the degree of rarefaction of a gas is often
characterized by the dimensionless Knudsen number $\Kn = \lambda / L$,
where $\lambda$ is the mean free path and $L$ is the relevant
characteristic length. The classic Navier-Stokes-Fourier (NSF)
equations are accurate only when $\Kn < 0.01$. However, the ongoing
miniaturization of technical devices requires modelling of gas in
microscopic channels, for which the characteristic length $L$ is so
small that even under normal density and temperature, the Knudsen
number is beyond the available region of NSF equations. Meanwhile, in
the transitional regime ($0.1 < \Kn < 10$), the traditional no-slip
wall boundary condition is no longer valid. In order to match the
physical experimentation, the interaction between wall and gas should
be carefully conducted. We refer the readers to \cite{Karniadakis} for
more details.

For microflows, it is known that the Boltzmann equation with Maxwell
boundary conditions \cite{Maxwell} is able to accurately describe the
flow state. However, on the computational perspective, the cost for
solving Boltzmann equation directly is unacceptable in the general
case. Grad \cite{Grad} did a pioneer work which extends Euler
equations to a thirteen-moment system, which opened a new way for
modelling rarefied gas flow called as moment method. However, it was
discovered by Grad himself in \cite{Grad1952} that this system fails
to give smooth shock profiles when the Mach number is larger than
$1.65$. To remedy this drawback, some authors tend to construct a
parabolic system similar to the NSF equations. In this field, some
methods such as Jin-Slemrod \cite{Jin}, COET \cite{Reitebuch} and R13
\cite{Grad1958, Struchtrup2003} were subsequently raised.
Concurrently, increasing attention is attracted to systems with more
than 13 moments (e.g.  \cite{Shock, Struchtrup2002}). As a combination
of these two directions, R20 and R26 equations were respectively
studied in \cite{Mizzi} and \cite{Emerson}. In \cite{NRxx}, a
general method for numerically solving the regularized moment
equations of arbitrary order was proposed, and it was improved in
\cite{NRxx_new, Cai} and abbreviated as \NRxx method in \cite{Cai} for
convenience. On the other hand, the boundary condition for the moment
methods is a major obstacle for applications of moment methods in the
field of microflows. In Grad's paper \cite{Grad}, the basic idea for
the modelling of Maxwell boundary conditions in the framework of
moment method is raised. The idea was also used in
\cite{Torrilhon2008, Emerson} for R13 and R26 equations. However, for
general moment equations, the numerical method to process the boundary
conditions for \NRxx method is unavailable yet.

The major concern of this paper is to supply suitable boundary
conditions for \NRxx method. Before that, the \NRxx method is first
improved such that it is able to predict stress and heat flux
correctly in the dense case. This is achieved by replacing the BGK
collision model \cite{BGK} used in \cite{NRxx, NRxx_new, Cai} with the
Shakhov model \cite{Shakhov}. Recall that for the BGK model, collision
term can be analytically solved when using the \NRxx method.
Similarly, analytical solution for each moment can also be obtained
when using the Shakhov model. At the same time, the force term is also
applied to the \NRxx method, and one can find that this term only
affects the momentum equation that it is turned to be trivial when
splitting method is employed.

As to the wall boundary conditions, we follow the idea of Grad
\cite{Grad} and try to approximate Maxwell boundary condition using
moment method. The Maxwell boundary condition is a linear combination
of specular reflection and diffusive reflection. According to Grad's
theory, only the moments of odd order in the normal microscopic
velocity are controlled by boundary conditions. These moments for the
specularly reflective part vanish. For the diffusive reflection, the
incidence part and the emergence part are considered separately. For
the incidence part, one need to calculate the moments of a
distribution cut off by a half space. Since the distribution is
expressed by a finite expansion of Hermite series, the cut-off turns
out to be quite intricate. We eventually derive a simple recursive
formula to obtain these moments with careful investigation into the
detailed expressions. The obtained formula brings only slight
increment of the computational cost. For the emergence part, which is
a half Maxwellian, the moments are obtained by direct integration, and
the result is also given in a recursive form. The overall boundary
condition is the summation of both the specular part and diffusive
part, which is rearranged into a simple formulation. It is numerically
implemented by first constructing a set of moments satisfying the
boundary conditions, and then approximating the flow state in the
ghost cell with a first order extrapolation of each moment. Thus,
boundary conditions for the \NRxx method of all orders are collected
into a uniform framework, which avoids separate and involved
implementation for different systems with sophisticated expressions
\cite{Thatcher, Torrilhon2008, Emerson}.

A number of numerical examples are presented to show the validity of
the boundary conditions. Both steady and unsteady problems are
studied. Numerical simulations up to $455$-moment system are carried
out. The classic symmetric planar Couette flow and force-driven
Poiseuille flow are investigated as examples for steady problems. All
the numerical results exhibit the convergence of the \NRxx method as
the number of moments increases.

The layout of this paper is as follows: in Section \ref{sec:NRxx}, we
give a brief introduction to the Boltzmann equation and the \NRxx
method. In Section \ref{sec:Shakhov}, the Shakhov collision model and
the force-induced acceleration terms are coupled with the \NRxx
method. In Section \ref{sec:BC}, the derivation of boundary conditions
are carried out. Numerical examples are shown in Section
\ref{sec:example}, and some discussions on the validity and accuracy of
the \NRxx method are given in Section \ref{sec:discussion}. Finally,
we make some conclusion in Section \ref{sec:conclusion}.

%%% Local Variables: 
%%% mode: latex
%%% TeX-master: "article"
%%% End: 

% vim: tw=70:spell
\section{The Boltzmann equation and the \bNRxx method}
\label{sec:NRxx}

The Boltzmann equation is the basic equation in the kinetic theory,
where a distribution function $f(t, \bx, \bxi)$ is introduced to
provide a statistical description for the motion of molecules. Here
$t\in \bbR^+$ is the time, and $\bx, \bxi \in \bbR^3$ are the position
and velocity of particles. The Boltzmann equation reads
\begin{equation}
\frac{\partial f}{\partial t} + \bxi \cdot \nabla_{\bx} f +
  \bF \cdot \nabla_{\bxi} f = Q(f,f),
\end{equation}
where $\bF$ is the acceleration of particles caused by external
forces. The detailed expression of the collision term $Q(f,f)$ is not
presented here due to its complexity, but we stress that $Q(f,f)$
contains a five-dimensional integration which causes great difficulty
in the numerical simulation. Instead, simplified collision models such
as the BGK model \cite{BGK} and the Shakhov model \cite{Shakhov} are
adopted in this paper. These models read:
\begin{enumerate}
\item \it BGK model:
  \begin{equation}
  \frac{\partial f}{\partial t} + \bxi \cdot \nabla_{\bx} f +
    \bF \cdot \nabla_{\bxi} f = \frac{1}{\tau}(f_M - f);
  \end{equation}
\item \it Shakhov model:
  \begin{equation} \label{eq:Shakhov}
  \frac{\partial f}{\partial t} + \bxi \cdot \nabla_{\bx} f +
    \bF \cdot \nabla_{\bxi} f =
    \frac{1}{\tau} \left\{ \left[ 1 + 
      \frac{(1 - \mathrm{Pr})(\bxi - \bu) \cdot \bq}{5 \rho \theta^2}
        \left( \frac{|\bxi - \bu|^2}{\theta} - 5 \right)
    \right] f_M - f \right\}.
  \end{equation}
\end{enumerate}
Here $\rho$, $\bu$, $\theta$ and $\bq$ denote the density, mean
velocity, temperature and heat flux respectively, and these
macroscopic variables are related with the distribution function $f$
by
\begin{equation}
\begin{gathered}
\rho = \int_{\bbR^3} f \dd\bxi,
\quad \bu = \frac{1}{\rho} \int_{\bbR^3} \bxi f \dd\bxi, \\
\theta = \frac{1}{3\rho} \int_{\bbR^3} |\bxi - \bu|^2 f \dd\bxi, \quad
\bq = \frac{1}{2} \int_{\bbR^3} |\bxi - \bu|^2 (\bxi - \bu) f \dd\bxi.
\end{gathered}
\end{equation}
Besides, $\tau$ is the relaxation time and $f_M$ is the local
Maxwellian which can be analytically formulated by
\begin{equation}
f_M = \frac{\rho}{(2\pi \theta)^{3/2}}
  \exp \left( -\frac{|\bxi - \bu|^2}{2\theta} \right).
\end{equation}
In \eqref{eq:Shakhov}, $\mathrm{Pr}$ stands for the Prandtl number
which is a constant. One can easily observe that if $\mathrm{Pr} = 1$,
then the Shakhov model reduces to the BGK model, which agrees with the
common knowledge that the BGK model predicts an incorrect Prandtl
number $1$.

The \NRxx method is a numerical tool for solving large moment
equations. It originated in \cite{NRxx} and was simplified in \cite
{NRxx_new}. The basic idea is to expand the distribution function $f$
into the Hermite series:
\begin{equation} \label{eq:expansion}
f(t, \bx, \bxi) = \sum_{\alpha \in \bbN^3} f_{\alpha}(t, \bx)
  \mathcal{H}_{\theta,\alpha} \left(
    \frac{\bxi - \bu(t, \bx)}{\sqrt{\theta(t, \bx)}}
  \right),
\end{equation}
where $\mathcal{H}_{\theta,\alpha}$ is the basis function defined as
\begin{equation} \label{eq:H}
\mathcal{H}_{\theta,\alpha}(\bv) =
  \prod_{d=1}^3 \frac{1}{\sqrt{2\pi}}
  \theta^{-\frac{\alpha_d + 1}{2}}
  \He_{\alpha_d}(v_d) \exp\left( -\frac{v_d^2}{2} \right),
  \quad \forall \alpha \in \bbN^3,
\end{equation}
and $\He_n$ is the Hermite polynomials
\begin{equation} \label{eq:He}
\He_n(x) = (-1)^n \exp \left( \frac{x^2}{2} \right)
  \frac{\mathrm{d}^n}{\mathrm{d} x^n}
  \exp \left( -\frac{x^2}{2} \right).
\end{equation}
For convenience, we let $\He_n(x) \equiv 0$ if $n < 0$. Thus $\mathcal
{H}_{\theta,\alpha}(\bv)$ is zero when any of the components of
$\alpha$ is negative.

With the expansion \eqref{eq:expansion}, the coefficients $f_{\alpha}$
can be considered as a set of infinite moments, and we have the
following relations:
\begin{equation} \label{eq:low_order_moments}
\begin{gathered}
f_0 = \rho, \quad f_{e_i} = 0, \quad \sum_{d=1}^3 f_{2e_d} = 0, \\
\sigma_{ij} = f_{e_i + e_j}, \quad \sigma_{ii} = 2 f_{2e_i}, \quad
q_i = 2 f_{3e_i} + \sum_{d=1}^3 f_{2e_d + e_i},
\end{gathered}
\end{equation}
where $i,j = 1,2,3$ and $i \neq j$, and $\sigma_{ij}$ is the stress
tensor or pressure deviators, which can be deduced from the
distribution function $f$ by
\begin{equation}
\sigma_{ij} = p_{ij} - \frac{1}{3} \delta_{ij} \sum_{d=1}^3 p_{dd},
  \quad \text{with} \quad
p_{ij} = \int_{\bbR^3} (\xi_i - u_i)(\xi_j - u_j) f \dd \bxi,
  \qquad i,j = 1,2,3.
\end{equation}

In order to implement \eqref{eq:expansion} numerically, a positive
integer $M \geqslant 3$ is chosen and only the coefficients $\{%
f_{\alpha}(t, \bx)\}_{|\alpha| \leqslant M}$ are stored. Due to the
absence of higher order moments, the resulting moment system is not
closed. According to \cite{NRxx_new}, the $(M+1)$-st order moments are
approximated by
\begin{equation} \label{eq:reg}
\begin{split}
f_{\alpha} &= \tau \Bigg\{
  \frac{1}{\rho} \sum_{j=1}^D
    \frac{\partial (\rho \theta)}{\partial x_j} f_{\alpha-e_j}
    + \frac{\theta}{D} \left(
      \sum_{j=1}^D \frac{\partial u_j}{\partial x_j}
    \right) \sum_{d=1}^D f_{\alpha-2e_d}
    - \sum_{j=1}^D \Bigg[ \theta
      \frac{\partial f_{\alpha-e_j}}{\partial x_j} \\
& \qquad {} + \sum_{d=1}^D \left(
  \frac{\partial u_d}{\partial x_j} \theta f_{\alpha-e_d-e_j}
  + \frac{1}{2} \frac{\partial \theta}{\partial x_j}
    (\theta f_{\alpha-2e_d-e_j} + (\alpha_j+1) f_{\alpha-2e_d+e_j})
  \right)
\Bigg] \Bigg\}.
\end{split}
\end{equation}
Here $f_{\alpha}$ is taken as zero when any of $\alpha$'s components
is negative. The numerical scheme for the force-free BGK model has
been constructed in \cite{Cai} based on the finite volume scheme with
linear reconstruction and the fractional step method. Suppose the
problem is in 1D and the grid is uniform with cell size $\Delta x$. We
denote the cell centers as $x_j$, and then a full time step of the
scheme can be sketched as follows:
\begin{enumerate}
\setlength\itemsep{0cm}
\item Determine the time step size $\Delta t$.
\item Reconstruct the first $M$-th order moments for the distribution
  functions on cell boundaries $x_{j \pm 1/2}$ with a conservative
  linear reconstruction.
\item Get the $(M+1)$-st order moments for the distribution functions
  on cell boundaries with a direct discretization of \eqref{eq:reg}.
\item \label{item:fvm} Apply the HLL scheme to solve the purely
  advective equation $\partial_t f + \bxi \cdot \nabla_{\bx} f = 0$
  over a time step of length $\Delta t$.
\item \label{item:BGK} Analytically solve the pure collision equation
  of the BGK model $\partial_t f = (f_M - f) / \tau$ over a time step
  of length $\Delta t$.
\end{enumerate}
We refer the readers to \cite{NRxx, NRxx_new, Cai} for details of the
algorithm. Here we only note that the Step \ref{item:fvm} is
nontrivial since two distributions cannot be added up directly, and in
Step \ref{item:BGK}, the reason why the collision-only equation can be
directly solved is that $f_M$ can be expressed in the Hermite series
$\{\mathcal{H}_{\theta,\alpha}\}$ trivially as $f_M = f_0 \mathcal{H}
_{\theta,0} \left( (\bxi - \bu) / \sqrt{\theta} \right)$.

%%% Local Variables: 
%%% mode: latex
%%% TeX-master: "article"
%%% End: 

% vim: tw=70:spell
\section{The \bNRxx method for Shakhov model with force terms}
\label{sec:Shakhov}
As is well known, the Prandtl number for monatomic gases is around
$2/3$, while the BGK model gives a Prandtl number $1$, which causes
incorrect prediction of the stress tensor $\sigma_{ij}$ or heat flux
$\bq$ for a dense gas. As a remedy, the Shakhov model was introduced
in \cite{Shakhov} as a generalization of the BGK model. The difference
between these two models has been investigated in \cite{Yang,
  Kudryavtsev}. In this section, we extend the \NRxx method in
\cite{NRxx_new} to the Shakhov model, and the force terms in
\eqref{eq:Shakhov} is added.

\subsection{The governing equations} \label{sec:gov_eq}
The moment system for the Shakhov model \eqref{eq:Shakhov} with moment
set $\{f_{\alpha}(t,\bx)\}_{|\alpha| \leqslant M}$ will be deduced
here. As in \cite{NRxx_new}, the strategy is to expand
\eqref{eq:Shakhov} into Hermite series, and then match the
coefficients for the same basis functions. In order to simplify the
notation, we define
\begin{equation}
\begin{split}
A &= \frac{\partial f}{\partial t} + \bxi \cdot \nabla_{\bx} f, \\
B &= \bF \cdot \nabla_{\bxi} f, \\
C &= \frac{1}{\tau} \left\{ \left[
  1 + \frac{(1 - \mathrm{Pr})(\bxi - \bu) \cdot \bq}{5 \rho \theta^2}
  \left( \frac{|\bxi - \bu|^2}{\theta} - 5 \right)
\right] f_M - f \right\}.
\end{split}
\end{equation}
It has been deduced in \cite{NRxx_new} that the Hermite expansion of
$A$ is
\begin{equation} \label{eq:A}
\begin{split}
A &= \sum_{\alpha \in \bbN^3} \Bigg\{ \left(
    \frac{\partial f_{\alpha}}{\partial t} +
    \sum_{d=1}^3 \frac{\partial u_d}{\partial t} f_{\alpha-e_d} +
    \frac{1}{2} \frac{\partial \theta}{\partial t}
      \sum_{d=1}^3 f_{\alpha-2e_d}
  \right) \\
& \qquad + \sum_{j=1}^3 \Bigg[ \left(
    \theta \frac{\partial f_{\alpha - e_j}}{\partial x_j} +
    u_j \frac{\partial f_{\alpha}}{\partial x_j} +
    (\alpha_j + 1) \frac{\partial f_{\alpha+e_j}}{\partial x_j}
  \right) \\
& \qquad \qquad + \sum_{d=1}^3 \frac{\partial u_d}{\partial x_j}
    \left( \theta f_{\alpha-e_d-e_j} + u_j f_{\alpha-e_d}
      + (\alpha_j + 1) f_{\alpha-e_d+e_j} \right) \\
& \qquad \qquad + \frac{1}{2} \frac{\partial \theta}{\partial x_j}
    \sum_{d=1}^3 \left(
      \theta f_{\alpha-2e_d-e_j} + u_j f_{\alpha-2e_d}
      + (\alpha_j + 1) f_{\alpha-2e_d+e_j}
    \right)
  \Bigg] \Bigg\} \mathcal{H}_{\theta,\alpha} \left(
    \frac{\bxi - \bu}{\sqrt{\theta}}
  \right).
\end{split}
\end{equation}
Using the differential relation of the Hermite polynomials, we have
\begin{equation}
\frac{\partial}{\partial \xi_d} \mathcal{H}_{\theta,\alpha} \left(
  \frac{\bxi - \bu}{\sqrt{\theta}}
\right) = -\mathcal{H}_{\theta,\alpha+e_d} \left(
  \frac{\bxi - \bu}{\sqrt{\theta}}
\right).
\end{equation}
Thus the Hermite expansion of the force term $B$ can be easily deduced
as
\begin{equation} \label{eq:B}
B = -\sum_{\alpha \in \bbN^3} \sum_{d=1}^3 F_d f_{\alpha-e_d}
  \mathcal{H}_{\theta,\alpha} \left(
    \frac{\bxi - \bu}{\sqrt{\theta}}
  \right).
\end{equation}
The expansion of the collision term $C$ can also be obtained by direct
calculation. The result is
\begin{equation} \label{eq:C}
C = \frac{1}{\tau} \left[
  \frac{1 - \mathrm{Pr}}{5} \sum_{i=1}^3 \sum_{j=1}^3
    q_i \mathcal{H}_{\theta, e_i + 2e_j} \left(
      \frac{\bxi - \bu}{\sqrt{\theta}}
    \right)
  - \sum_{|\alpha| \geqslant 2} f_{\alpha}
    \mathcal{H}_{\theta, \alpha} \left(
      \frac{\bxi - \bu}{\sqrt{\theta}}
    \right)
\right].
\end{equation}
Putting \eqref{eq:A}\eqref{eq:B} and \eqref{eq:C} into the
Boltzmann-Shakhov equation $A+B=C$ and extracting coefficients for all
basis functions, with a slight rearrangement, we get the following
general moment equations for Shakhov model:
\begin{equation} \label{eq:mnt_eq}
\begin{split}
& \frac{\partial f_{\alpha}}{\partial t} +
  \sum_{d=1}^3 \left(
    \frac{\partial u_d}{\partial t} +
    \sum_{j=1}^3 u_j \frac{\partial u_d}{\partial x_j} - F_d
  \right) f_{\alpha-e_d} + \frac{1}{2} \left(
    \frac{\partial \theta}{\partial t} +
    \sum_{j=1}^3 u_j \frac{\partial \theta}{\partial x_j}
  \right) \sum_{d=1}^3 f_{\alpha-2e_d} \\
& \quad + \sum_{j,d=1}^3 \left[
    \frac{\partial u_d}{\partial x_j} \left(
      \theta f_{\alpha-e_d-e_j} + (\alpha_j + 1) f_{\alpha-e_d+e_j}
    \right) + \frac{1}{2} \frac{\partial \theta}{\partial x_j} \left(
      \theta f_{\alpha-2e_d-e_j} + (\alpha_j + 1) f_{\alpha-2e_d+e_j}
    \right)
  \right] \\
& \quad + \sum_{j=1}^3 \left(
    \theta \frac{\partial f_{\alpha - e_j}}{\partial x_j} +
    u_j \frac{\partial f_{\alpha}}{\partial x_j} +
    (\alpha_j + 1) \frac{\partial f_{\alpha+e_j}}{\partial x_j}
  \right) = \frac{1}{\tau} \left(
    \frac{1 - \mathrm{Pr}}{5} \sum_{i,j=1}^3
      \delta_{ij}(\alpha) q_i - \delta(\alpha) f_{\alpha}
  \right),
\end{split}
\end{equation}
where $\delta_{ij}(\alpha)$ and $\delta(\alpha)$ are defined by
\begin{equation}
\delta_{ij}(\alpha) = \left\{ \begin{array}{ll}
  1, & \text{if } \alpha = e_i + 2e_j, \\
  0, & \text{otherwise,}
\end{array} \right. \qquad
\delta(\alpha) = \left\{ \begin{array}{ll}
  1, & \text{if } |\alpha| \geq 2, \\
  0, & \text{otherwise.}
\end{array} \right.
\end{equation}

Now we will explore something more from \eqref{eq:mnt_eq}. Noting
that $f_{e_j} = 0$, $\forall j = 1,2,3$, the following relation can be
obtained if we put $\alpha = 0$ into \eqref{eq:mnt_eq}:
\begin{equation}
\frac{\partial f_0}{\partial x_j} +
  \sum_{j=1}^3 \left(
    u_j \frac{\partial f_0}{\partial x_j} +
    f_0 \frac{\partial u_j}{\partial x_j}
  \right) = 0.
\end{equation}
This is the mass conservation law. If we set $\alpha = e_d$, $d =
1,2,3$, the equations are
\begin{equation}
f_0 \left(
  \frac{\partial u_d}{\partial t} +
  \sum_{j=1}^3 u_j \frac{\partial u_d}{\partial x_j} - F_d
\right) + f_0 \frac{\partial \theta}{\partial x_d} 
  + \theta \frac{\partial f_0}{\partial x_d}
  + \sum_{j=1}^3 (\delta_{jd} + 1)
    \frac{\partial f_{e_d + e_j}}{\partial x_j} = 0.
\end{equation}
This equation can be simplified as
\begin{equation} \label{eq:mtm}
f_0 \left(
  \frac{\partial u_d}{\partial t} +
  \sum_{j=1}^3 u_j \frac{\partial u_d}{\partial x_j} - F_d
\right) + \sum_{j=1}^3 \frac{\partial p_{jd}}{\partial x_j} = 0.
\end{equation}
Now we consider the case of $|\alpha| \geqslant 2$. Substituting
\eqref{eq:mtm} into \eqref{eq:mnt_eq}, the temporal differentiation of
$\bu$ can be eliminated. In order to eliminate the temporal
differentiation of $\theta$, we multiply \eqref{eq:Shakhov} by $|\bxi
- \bu|^2$ on both sides and then integrate on $\bbR^3$ with respect to
$\bxi$. The result is
\begin{equation} \label{eq:energy}
f_0 \left( \frac{\partial \theta}{\partial t}
  + \sum_{j=1}^3 u_j \frac{\partial \theta}{\partial x_j} \right)
  + \frac{2}{3} \sum_{j=1}^3 \left(
    \frac{\partial q_j}{\partial x_j} +
    \sum_{d=1}^3 p_{jd} \frac{\partial u_d}{\partial x_j}
  \right) = 0.
\end{equation}
Note that the force term does not appear in this equation, since
\begin{equation}
\int_{\bbR^3} |\bxi-\bu|^2 \frac{\partial f}{\partial \xi_j}
  \dd \bxi = -2 \int_{\bbR^3} (\xi_j - u_j) f \dd \bxi = 0.
\end{equation}
Thus, the final form of equations for $|\alpha| \geqslant 2$ reads
\begin{equation} \label{eq:mnt_system}
\begin{split}
& \frac{\partial f_{\alpha}}{\partial t} - \frac{1}{f_0}
  \sum_{d=1}^3 \sum_{j=1}^3
    \frac{\partial p_{jd}}{\partial x_j} f_{\alpha-e_d}
  - \frac{1}{3f_0} \sum_{j=1}^3 \left(
    \frac{\partial q_j}{\partial x_j} +
    \sum_{d=1}^3 p_{jd} \frac{\partial u_d}{\partial x_j}
  \right) \sum_{d=1}^3 f_{\alpha-2e_d} \\
& \quad + \sum_{j,d=1}^3 \left[
    \frac{\partial u_d}{\partial x_j} \left(
      \theta f_{\alpha-e_d-e_j} + (\alpha_j + 1) f_{\alpha-e_d+e_j}
    \right) + \frac{1}{2} \frac{\partial \theta}{\partial x_j} \left(
      \theta f_{\alpha-2e_d-e_j} + (\alpha_j + 1) f_{\alpha-2e_d+e_j}
    \right)
  \right] \\
& \quad + \sum_{j=1}^3 \left(
    \theta \frac{\partial f_{\alpha - e_j}}{\partial x_j} +
    u_j \frac{\partial f_{\alpha}}{\partial x_j} +
    (\alpha_j + 1) \frac{\partial f_{\alpha+e_j}}{\partial x_j}
  \right) = \frac{1}{\tau} \left(
    \frac{1 - \mathrm{Pr}}{5} \sum_{i,j=1}^3
      \delta_{ij}(\alpha) q_i - \delta(\alpha) f_{\alpha}
  \right).
\end{split}
\end{equation}
In order to get a closed system, we collect \eqref{eq:reg},
\eqref{eq:mtm}, \eqref{eq:energy} and \eqref{eq:mnt_system} with $2
\leqslant |\alpha| \leqslant M$ together. Then the governing system
for the \NRxx method with Shakhov model and force terms is formed.

\begin{remark} \label{rem:reg}
In the Shakhov model, the equation \eqref{eq:reg}, the prediction of
$f_{\alpha}$ with $|\alpha| = M + 1$ derived for the BGK model, is
still available. In \cite{NRxx_new}, \eqref{eq:reg} is deduced in the
following two steps:
\begin{enumerate}
\item Determine the orders of magnitude for all $f_{\alpha}$ using
  Maxwellian iteration.
\item \label{item:remove_hot} For $|\alpha| = M + 1$, remove all the
  high order terms in the equations containing only $-f_{\alpha} /
  \tau$ in their right hand sides.
\end{enumerate}
The Maxwellian iteration can also be applied to \eqref{eq:mnt_system},
and after the first iteration step, we immediately get
\begin{equation}
f_{\alpha} = O(\tau), \qquad |\alpha| = 2, \quad \text{or} \quad
  \alpha = e_i + 2e_j, \quad i,j=1, 2, 3,
\end{equation}
and other moments with $|\alpha| \geqslant 2$ remain to be zero.
This result is the same as that we have derived in the BGK model.
We note that for $|\alpha| > 3$ and $\alpha = (1,1,1)$,
\eqref{eq:mnt_system} is just the corresponding equation for the BGK
model. Thus, in the view of order of magnitude, the subsequent
iterations are identical to the BGK case. Moreover, when $M \geqslant
3$, which we have assumed in the last section, step
\ref{item:remove_hot} is also identical for both models. Hence
\eqref{eq:reg} still applies for the Shakhov model.
\end{remark}

\subsection{The numerical approach}
The acceleration $\bF$ only appears in \eqref{eq:mtm} in the governing
system, thus a splitting method can be applied as follows:
\begin{enumerate}
\setlength\itemsep{0cm}
\item \emph{Transportation:} solve the force-free Shakhov equation
  over a time step of length $\Delta t$.
\item \emph{Acceleration:} solve $\partial_t \bu = \bF$ over a time
  step of length $\Delta t$.
\end{enumerate}
In order to solve the force-free Shakhov equation, another splitting
of the convection and collision part is needed. For the convection
part, the method is identical to that used in the BGK model. We refer
the readers to \cite{NRxx,Cai} for details. For the collision part,
since a new collision model is adopted, the procedure is slightly
different.

Now we consider the pure collision model, where $\rho, \bu$ and
$\theta$ are not changed while time evolves. Therefore, the collision
terms only exist in \eqref{eq:mnt_system} with $|\alpha| \geqslant 2$.
Two cases are considered below:

(1) $\alpha = e_i + 2e_j$, $i,j = 1,2,3$. In these cases, the pure
collision equations are written as
\begin{equation} \label{eq:coll}
\begin{split}
\frac{\partial f_{e_i + 2e_j}}{\partial t} &=
  \frac{(1-\mathrm{Pr}) q_i - 5 f_{e_i + 2e_j}}{5 \tau} \\
&= \frac{1}{\tau} \left[
    \frac{1-\mathrm{Pr}}{5} \left(
      2f_{3e_i} + \displaystyle\sum_{j=1}^3 f_{e_i + 2e_j}
    \right) - f_{e_i + 2e_j}
  \right], \qquad i,j = 1,2,3.
\end{split}
\end{equation}
In the general case, $\tau$ only depends on $\rho$ and $\theta$. Thus
it is invariant in the collision-only system. This turns
\eqref{eq:coll} into a linear \emph{ordinary} differential system with
9 equations, which can be analytically integrated as
\begin{equation} \label{eq:sol1}
f_{e_i + 2e_j}(t) = \frac{1}{5} q_i(t_0) \exp \left(
  -\frac{\mathrm{Pr} (t - t_0)}{\tau}
\right) - \left( \frac{1}{5} q_i(t_0) - f_{e_i+2e_j}(t_0) \right)
  \exp \left( -\frac{t - t_0}{\tau} \right),
\end{equation}
where $t_0$ denotes the initial time.

(2) Other cases. For other $\alpha$'s, the collision-only equation is
the same as the BGK model:
\begin{equation}
\frac{\partial f_{\alpha}}{\partial t} = -\frac{1}{\tau} f_{\alpha}.
\end{equation}
The solution is
\begin{equation} \label{eq:sol2}
f_{\alpha}(t) = f_{\alpha}(t_0)
  \exp \left( -\frac{t - t_0}{\tau} \right).
\end{equation}

When \eqref{eq:sol1} and \eqref{eq:sol2} are used in the numerical
scheme, we replace $t$ and $t_0$ with $t_{n+1}$ and $t_n$
respectively. Note that when $\tau$ is independent of $\bu$, the
acceleration and collision do not coupled with each other, thus the
splitting is applied only once rather than twice. This makes it more
efficient when the Strang splitting is employed.

%%% Local Variables: 
%%% mode: latex
%%% TeX-master: "article"
%%% End: 

% vim: tw=70:spell
\section{Boundary conditions} \label{sec:BC}
In the moment methods, the boundary condition is always a complicated
issue when simulating microflows. As discussed in \cite{Grad,
Struchtrup2000,Thatcher,Torrilhon2008,Emerson} and the references
therein, delicate derivations and careful numerical techniques are
needed for a solid wall. In this section, a numerical way for dealing
with boundary conditions in the \NRxx method is introduced, which
appears to be uniform for all orders of moment systems.

\subsection{The kinetic boundary condition}
In the kinetic theory, the most extensively used boundary condition is
the one proposed by Maxwell in \cite{Maxwell}. According to the common
hyperbolic theory, for \eqref{eq:Shakhov}, the boundary condition is
only needed when $\bxi \cdot \bn < 0$, where $\bn$ is the outer normal
vector of the boundary. For a point $\bx$ on the wall, supposing the
velocity and temperature of the wall to be $\bu^W(t, \bx)$ and
$\theta^W(t, \bx)$ at time $t$, Maxwell proposed the following
boundary condition:
\begin{equation} \label{eq:bc}
f(t, \bx, \bxi) = \left\{ \begin{array}{ll}
  \chi f_M^W(t,\bx,\bxi) + (1 - \chi) f(t, \bx, \bxi^*),
    & \bC^W \cdot \bn < 0, \\[5pt]
  f(t, \bx, \bxi), & \bC^W \cdot \bn \geqslant 0,
\end{array} \right.
\end{equation}
where $\chi \in [0,1]$ is a parameter for different gases and walls,
and
\begin{gather}
\bxi^* = \bxi - 2(\bC^W \cdot \bn) \bn,
  \quad \bC^W = \bxi - \bu^W(t, \bx), \\
\label{eq:f_M}
f_M^W(t, \bx, \bxi) =
  \frac{\rho^W(t, \bx)}{(2\pi \theta^W(t, \bx))^{3/2}}
  \exp \left(
    -\frac{|\bxi - \bu^W(t, \bx)|^2}{2\theta^W(t, \bx)}
  \right).
\end{gather}
The functions $\bu^W(t, \bx)$ and $\theta^W(t, \bx)$ are prescribed
and stand for the wall velocity and temperature at time $t$ and
position $\bx$, and $\rho^W(t, \bx)$ ensures the conservation of the
mass at the wall, that is,
\begin{equation} \label{eq:mass_csv}
\begin{split}
& \int_{\bbR^3} (\bC^W \cdot \bn) f(t, \bx, \bxi) \dd \bxi \\
={} & \chi \left(
  \int_{\bC^W \cdot \bn < 0}
    (\bC^W \cdot \bn) f_M^W(t, \bx, \bxi) \dd \bxi +
  \int_{\bC^W \cdot \bn \geqslant 0}
    (\bC^W \cdot \bn) f(t, \bx, \bxi) \dd \bxi
\right) = 0.
\end{split}
\end{equation}

For this boundary condition, the normal velocity of gas on the
boundary is the same as the normal velocity of the wall. However, in
the case of shear flow, velocity slip and temperature jump will appear
on the boundary.

\subsection{The boundary conditions for the \bNRxx method}
The boundary condition can be derived by taking moments on both sides
on \eqref{eq:bc}. Before that, we define
\begin{equation}
C_{\theta,\alpha} = \frac{(2\pi)^{3/2} \theta^{|\alpha| + 3}}
  {\alpha_1! \alpha_2!  \alpha_3!},
  \qquad \forall \theta > 0, \quad \alpha \in \bbN^3.
\end{equation}
This definition leads to
\begin{equation}
g_{\alpha} = C_{\theta,\alpha} \int_{\bbR^3}
  g(\bxi) \mathcal{H}_{\theta,\alpha}(\bv) \exp(|\bv|^2 / 2)
\dd \bv,
\end{equation}
where $\bv = (\bxi - \bu) / \sqrt{\theta}$ and $g(\bxi)$ is a
distribution function expanded into Hermite series as $g(\bxi) =
\sum_{\alpha \in \bbN^3} g_{\alpha} \mathcal{H}_{\theta,\alpha}(\bv)$.
In order to simplify the calculation, we suppose $\bn = (0, 1, 0)^T$.
Thus, taking moments for \eqref{eq:bc} requires half-space integration
\begin{equation} \label{eq:hs_int}
C_{\theta,\alpha} \int_{\xi_2 \geqslant u_2^W}
  g(\bxi) \mathcal{H}_{\theta,\alpha}(\bv) \exp(|\bv|^2/2) \dd \bv.
\end{equation}

Suppose an $M$-th order system is used in the \NRxx method; that is,
an $(M+1)$-st order approximation of the distribution can be obtained
through \eqref{eq:reg}. This approximation is directly used in \eqref
{eq:hs_int} so that the integral can be actually worked out.
Concretely speaking, \eqref{eq:hs_int} is approximated as
\begin{equation} \label{eq:sum}
\sum_{|\beta| \leqslant M+1} g_{\beta} C_{\theta,\alpha}
  \int_{\xi_2 \geqslant u_2^W}
    \mathcal{H}_{\theta,\alpha}(\bv)
    \mathcal{H}_{\theta,\beta}(\bv) \exp(|\bv|^2 / 2)
  \dd \bv.
\end{equation}
Since $u_2 = u_2^W$ on the boundary, the region of integration can be
written as $\{v_2 \geqslant 0\}$. Thus, we only need to calculate 
\begin{equation} \label{eq:I_def}
I_{\alpha,\beta}(\theta) =  C_{\theta,\alpha}
  \int_{v_2 \geqslant 0}
    \mathcal{H}_{\theta,\alpha}(\bv)
    \mathcal{H}_{\theta,\beta}(\bv) \exp(|\bv|^2 / 2)
  \dd \bv.
\end{equation}
The details can be found in Appendix \ref{sec:hs_int}, and the result
is
\begin{equation} \label{eq:I}
I_{\alpha,\beta}(\theta) =
  S(\alpha_2,\beta_2) \theta^{\frac{\alpha_2 - \beta_2}{2}}
  \cdot \delta_{\alpha_1 \beta_1} \delta_{\alpha_3 \beta_3}
\end{equation}
and
\begin{equation} \label{eq:S}
S(m,n) = \left\{ \begin{array}{ll}
  1/2, & m = n = 0, \\
  K(1,n-1), & m = 0 \text{ and } n \neq 0, \\
  K(m,0), & m \neq 0 \text{ and } n = 0, \\
  K(m,n) + S(m-1,n-1) \cdot n / m, & \text{otherwise},
\end{array} \right.
\end{equation}
where
\begin{equation} \label{eq:K}
K(m,n) = \frac{(2\pi)^{-1/2}}{m!} \He_{m-1}(0) \He_n(0).
\end{equation}
The above deduction leads to the following proposition:
\begin{proposition} \label{prop:hs}
Suppose $g(\bv)$ is a function defined on $\bbR^3$ which can be
denoted by a finite expansion of Hermite basis functions
\begin{equation}
g(\bv) = \sum_{|\alpha| \leqslant M + 1}
  g_{\alpha} \mathcal{H}_{\theta,\alpha}(\bv).
\end{equation}
for some $\theta > 0$. Let $\tilde{g}(\bv)$ be a half-space cut-off of
$g(\bv)$ as
\begin{equation}
\tilde{g}(\bv) = \left\{ \begin{array}{ll}
  g(\bv), & v_2 \geqslant 0, \\
  0, & v_2 < 0. \\
\end{array} \right.
\end{equation}
Then $\tilde{g}$ can also be expanded into Hermite series as
\begin{equation} \label{eq:g_exp}
\tilde{g}(\bv) = \sum_{\alpha \in \bbN^3}
  \sum_{|\beta| \leqslant M + 1} g_{\beta} I_{\alpha,\beta}(\theta)
    \mathcal{H}_{\theta,\alpha}(\bv),
\end{equation}
where $I_{\alpha,\beta}(\theta)$ is defined in \eqref{eq:I}---\eqref
{eq:K}.
\end{proposition}
\begin{proof}
It is already known in \cite{NRxx} that $\{\mathcal{H}_{\theta,\alpha}
(\bv)\}_{\alpha \in \bbN^3}$ is an orthogonal basis of the weighted
$L^2$ space $L^2(\bbR^3; \exp(|\bv|^2 / 2) \dd \bv)$. Since
\begin{equation}
\begin{split}
& \int_{\bbR^3} |\tilde{g}(\bv)|^2 \exp(|\bv|^2/2) \dd \bv =
  \int_{v_2 \geqslant 0} |g(\bv)|^2 \exp(|\bv|^2/2) \dd \bv \\
\leqslant {} & \int_{\bbR^3} |g(\bv)|^2 \exp(|\bv|^2/2) \dd \bv =
  \sum_{|\alpha| \leqslant M + 1}
    C_{\theta,\alpha}^{-1} |g_{\alpha}|^2 < +\infty,
\end{split}
\end{equation}
$\tilde{g}(\bv)$ also lies in $L^2(\bbR^3; \exp(|\bv|^2 / 2) \dd
\bv)$. Thus the validity of \eqref{eq:g_exp} can be naturally
obtained.
\end{proof}

The following proposition depicts the sparsity of $I_{\alpha,\beta}$.
\begin{proposition} \label{prop:sparsity}
If $I_{\alpha,\beta}(\theta)$ is nonzero, then (1) $\alpha_1 =
\beta_1$; (2) $\alpha_3 = \beta_3$; (3) $\alpha_2 - \beta_2$ is zero
or odd. When $\alpha = \beta$, $I_{\alpha,\beta}(\theta)$ is equal to
$1/2$.
\end{proposition}
\begin{proof}
If $I_{\alpha,\beta}(\theta)$ is nonzero, \eqref{eq:I} directly gives
$\alpha_1 = \beta_1$ and $\alpha_3 = \beta_3$. If $\alpha_2 - \beta_2$
is a nonzero even integer, $K(\alpha_2, \beta_2)$ is zero since $\He%
_n(0)$ is zero when $n$ is odd. In order to prove $I_{\alpha,\beta}%
(\theta) = 0$ in this case, according to \eqref{eq:I}, we only need to
prove $S(\alpha_2, \beta_2) = 0$. This can be done by induction:

(1) If $\alpha_2 = 0$ or $\beta_2 = 0$, $\alpha_2$ and $\beta_2$ must
be both even but one of them must be positive. Equation \eqref{eq:S}
shows $S(\alpha_2, \beta_2) = 0$ directly.

(2) Suppose $S(\alpha_2 - 1, \beta_2 - 1) = 0$. Then, according to the
last case in \eqref{eq:S}, $S(\alpha_2, \beta_2)$ is also zero.

Finally, when $\alpha = \beta$, $\eqref{eq:I}$ gives $I_{\alpha,\beta}
(\theta) = S(\alpha_2, \beta_2)$. The subsequent proof can also be
done by induction, since $S(0,0) = 1/2$ and $K(n,n) = 0$ for $n > 0$.
\end{proof}

According to Proposition \ref{prop:sparsity}, we find that only
$(\lceil \alpha_2 / 2 \rceil + 1)$ terms are nonzero in the summation
\eqref{eq:sum}. This greatly reduces the computational cost.

Now let us return to the boundary conditions. According to Grad's
theory \cite{Grad, Grad1958}, in order to ensure the continuity of
boundary conditions when $\chi \rightarrow 0$, only a subset of
moments $\{f_{\alpha} \mid |\alpha| \leqslant M + 1 \text{ and }
\alpha_2 \text{ is odd}\}$ should be used to formulate boundary
conditions. This will be completed in the following three subsections.
Later in this section, for conciseness, the variables $t$ and $\bx$
are omitted in our statement if not specified, and all spatially
dependent functions are considered to be on the boundary.

\subsubsection{Determination of $\rho^W$}
For simplicity, we factorize the right hand side of \eqref{eq:bc}
into three parts and consider each part independently. Define
\begin{equation}
\begin{gathered}
p(\bxi) = \left\{ \begin{array}{ll}
  f_M^W(\bxi), & \xi_2 < u_2^W, \\
  0, & \xi_2 \geqslant u_2^W,
\end{array} \right. \quad
q(\bxi) = \left\{ \begin{array}{ll}
  f(\bxi), & \xi_2 \geqslant u_2^W, \\
  0, & \xi_2 < u_2^W,
\end{array} \right. \quad
r(\bxi) = q(\bxi) + q(\bxi^*).
\end{gathered}
\end{equation}
Then \eqref{eq:bc} can be rewritten as
\begin{equation} \label{eq:factorization}
f(\bxi) = \chi p(\bxi) + \chi q(\bxi) + (1 - \chi) r(\bxi).
\end{equation}
Suppose the Hermite expansion of $f$ is
\begin{equation} \label{eq:f}
f(\bxi) = \sum_{|\alpha| \leqslant M + 1}
  f_{\alpha} \mathcal{H}_{\theta,\alpha}
  \left( \frac{\bxi - \bu}{\sqrt{\theta}} \right).
\end{equation}
Then $q(\bxi)$ can also be expanded into Hermite series according to
Proposition \ref{prop:hs} and \ref{prop:sparsity} as
\begin{equation} \label{eq:q}
q(\bxi) = \sum_{\alpha \in \bbN^3}
  q_{\alpha} \mathcal{H}_{\theta,\alpha}
  \left( \frac{\bxi - \bu}{\sqrt{\theta}} \right).
\end{equation}
Substituting \eqref{eq:f_M} and \eqref{eq:q} into \eqref{eq:mass_csv},
$\rho^W$ can be worked out as
\begin{equation} \label{eq:rho_W}
\rho^W = \sqrt{\frac{2\pi}{\theta^W}} q_{e_2}
  = \sqrt{\frac{2\pi}{\theta^W}}
    \sum_{k=0}^{\lceil M/2 \rceil} S(1,2k) \theta^{1/2-k} f_{2k e_2},
\end{equation}
where the expression of $q_{e_2}$ is derived from \eqref{eq:I}, \eqref
{eq:g_exp} and Proposition \ref{prop:sparsity}.

\subsubsection{The moments of $p$ and $r$}
Now the moments for $q(\bxi)$ have been calculated in \eqref{eq:q}, we
still need to get Hermite expansions of $p(\bxi)$ and $r(\bxi)$.
We suppose that $p(\bxi)$ can be expanded under the basis
$\big\{\mathcal{H}_{\theta, \alpha} \big( (\bxi - \bu) / \sqrt{\theta}
\big)\big\}_{\alpha \in \bbN^3}$ as
\begin{equation}
p(\bxi) = \sum_{\alpha \in \bbN^3}
  p_{\alpha} \mathcal{H}_{\theta, \alpha} \left(
    \frac{\bxi - \bu}{\sqrt{\theta}}
  \right).
\end{equation}
Then, according to \eqref{eq:f_M}, the coefficients can be formulated
by
\begin{equation} \label{eq:p}
p_{\alpha} = C_{\theta,\alpha} \int_{v_2 < 0}
  \frac{\rho^W}{(2\pi\theta^W)^{3/2}} \exp \left(
    - \frac{|\bxi - \bu^W|^2}{2\theta^W}
  \right) \mathcal{H}_{\theta,\alpha}(\bv) \exp \left(
    \frac{|\bv|^2}{2}
  \right) \dd \bv,
\end{equation}
where $\bxi = \sqrt{\theta} \bv + \bu$. Define
\begin{align}
\label{eq:J}
J_s(x) &= \frac{1}{s!} \theta^{\frac{s+1}{2}}
  \int_{-\infty}^{+\infty} \frac{1}{\sqrt{2\pi \theta^W}}
    \exp \left( -\frac{|\sqrt{\theta} y - x|^2}{2 \theta^W} \right)
    \He_s(y) \dd y, \\
\label{eq:tilde_J}
\tilde{J}_s(x) &= \frac{1}{s!} \theta^{\frac{s+1}{2}}
  \int_{-\infty}^0 \frac{1}{\sqrt{2\pi \theta^W}}
    \exp \left( -\frac{|\sqrt{\theta} y - x|^2}{2 \theta^W} \right)
    \He_s(y) \dd y.
\end{align}
Then $p_{\alpha}$ can be expressed by
\begin{equation} \label{eq:p_alpha}
p_{\alpha} = \rho^W
  J_{\alpha_1}(u_1^W - u_1)
  \tilde{J}_{\alpha_2}(u_2^W - u_2)
  J_{\alpha_3}(u_3^W - u_3).
\end{equation}
$J_s(x)$ and $\tilde{J}_s(x)$ can be calculated recursively as
\begin{align}
\label{eq:J_s}
J_s(x) &= \frac{1}{s} \left[
  (\theta^W - \theta) J_{s-2}(x) + x J_{s-1}(x)
\right], \quad s \geqslant 1; \\
\label{eq:tilde_J_s}
\tilde{J}_s(x) &= \frac{1}{s} \left[
  (\theta^W-\theta) \tilde{J}_{s-2}(x) + x \tilde{J}_{s-1}(x)
\right] - H_s(x), \quad s \geqslant 1; \\
H_s(x) &= -\frac{s-2}{s(s-1)} \theta H_{s-2}(x),
  \quad s \geqslant 2.
\end{align}
The starting values are
\begin{align}
\label{eq:J_start}
J_{-1}(x) = 0, & \qquad J_0(x) = 1, \\
\label{eq:tilde_J_start}
\tilde{J}_{-1}(x) = 0, & \qquad
  \tilde{J}_0(x) = \frac{1}{2} \mathrm{erfc} \left(
    \frac{x}{\sqrt{2 \theta^W}}
  \right), \\
\label{eq:H_start}
H_0(x) = 0, & \qquad 
  H_1(x) = \sqrt{\frac{\theta^W}{2\pi}} \exp \left(
    -\frac{x^2}{2 \theta^W}
  \right),
\end{align}
The detailed derivation of \eqref{eq:J_s}-\eqref{eq:H_start} can be
found in the Appendix \ref{sec:half_Max}. Noting that $u_2 = u_2^W$,
\eqref{eq:p_alpha} can be further simplified as
\begin{equation} \label{eq:p_alpha_simplified}
p_{\alpha} = \rho^W J_{\alpha_1}(u_1^W - u_1)
  \hat{J}_{\alpha_2} J_{\alpha_3}(u_3^W - u_3),
\end{equation}
where
\begin{equation} \label{eq:J_H}
\begin{gathered}
\hat{J}_s = \frac{1}{s}(\theta^W - \theta) \hat{J}_{s-2} - \hat{H}_s,
  \quad s \geqslant 1, \qquad
\hat{H}_s = -\frac{s-2}{s(s-1)} \theta \hat{H}_{s-2},
  \quad s \geqslant 2, \\
\hat{J}_{-1} = \hat{H}_0 = 0, \quad
  \hat{J}_0 = 1/2, \quad \hat{H}_1 = \sqrt{\frac{\theta^W}{2\pi}}.
\end{gathered}
\end{equation}
Here we emphasize that due to equation \eqref{eq:rho_W}, all
$p_{\alpha}$'s are only related with $\{f_{2ke_2}\}_{0 \leqslant k
\leqslant \lceil M/2 \rceil}$ besides $\bu$, $\bu^W$, $\theta$ and
$\theta^W$.

Now we turn to the moments of $r(\bxi)$. Note that only the moments
with odd $\alpha_2$ are needed. However, $r(\bxi)$ is an even function
with respect to $C_2^W$, which causes all its moments with odd
$\alpha_2$ vanished. This indicates that $r(\bxi)$ can be simply
neglected when discussing the boundary conditions.

\subsubsection{Construction of boundary conditions}
Now we take moments with odd $\alpha_2$ on both sides of \eqref
{eq:factorization}. Making use of Proposition \ref{prop:sparsity},
we have
\begin{equation}
f_{\alpha} = \chi p_{\alpha} + \chi q_{\alpha}
  = \chi p_{\alpha} + \frac{1}{2} \chi f_{\alpha} +
    \chi \sum_{k=0}^{K_2(\alpha)}
    S(\alpha_2, 2k) \theta^{\alpha_2/2 - k}
    f_{\alpha+(2k-\alpha_2)e_2},
\end{equation}
where $K_2(\alpha) = \lceil (M - \alpha_1 - \alpha_3) / 2 \rceil$. A
simple rearrangement gives
\begin{equation} \label{eq:mnt_bc}
f_{\alpha} = \frac{2\chi}{2 - \chi} \left[
  p_{\alpha} + \sum_{k=0}^{K_2(\alpha)}
    S(\alpha_2, 2k) \theta^{\alpha_2/2 - k}
    f_{\alpha+(2k-\alpha_2)e_2}
\right].
\end{equation}
Equations \eqref{eq:mnt_bc} with $|\alpha| \leqslant M+1$ and odd
$\alpha_2$, together with $u_2 = u_2^W$ form the boundary conditions
of the dynamic moment equations. Recalling
\begin{equation}
p_{\alpha} = p_{\alpha}(\bu, \bu^W, \theta, \theta^W,
  f_0, f_{2e_2}, \cdots, f_{2 \lceil M/2 \rceil e_2}),
\end{equation}
one can find that the terms which appear on the left hand side of
\eqref{eq:mnt_bc} never appear on its right hand side. Thus, if an
arbitrary distribution function denoted as \eqref{eq:f} is given, we
can define a functional $F^b$ which maps \eqref{eq:f} to another
distribution $f^b(\bxi)$:
\begin{equation} \label{eq:f_b}
f^b(\bxi) = \sum_{|\alpha| \leqslant M + 1}
  f_{\alpha}^b \mathcal{H}_{\theta^b, \alpha} \left(
    \frac{\bxi - \bu^b}{\sqrt{\theta^b}}
  \right),
\end{equation}
where $\bu^b = (u_1, u_2^W, u_3)$, $\theta^b = \theta$, and
\begin{equation} \label{eq:f_b_alpha}
f_{\alpha}^b = \left\{ \begin{array}{ll}
  f_{\alpha}, & \text{if $\alpha_2$ is even}, \\
  \text{the right hand side of \eqref{eq:mnt_bc}}, &
    \text{if $\alpha_2$ is odd}.
\end{array} \right.
\end{equation}
Thus $f_{\alpha}^b$ satisfies the boundary condition. The mapping
$F^b$ will be used in the numerical implementation of boundary
conditions.

At the end of this section, we prove that $\bu^b$ and $\theta^b$ are
the corresponding velocity and temperature of the distribution
function $f^b(\bxi)$. This is equivalent to the following proposition:
\begin{proposition} \label{prop:conservation}
If a distribution $f(\bxi)$ with expression \eqref{eq:f} satisfies
\begin{equation}
f_{e_1} = f_{e_2} = f_{e_3} = \sum_{d=1}^3 f_{2e_d} = 0,
\end{equation}
then $f^b = F^b(f)$ with expression \eqref{eq:f_b} also satisfies
\begin{equation}
f_{e_1}^b = f_{e_2}^b = f_{e_3}^b = \sum_{d=1}^3 f_{2e_d}^b = 0.
\end{equation}
\end{proposition}
\begin{proof}
Equation \eqref{eq:f_b_alpha} gives
\begin{equation}
f_{e_1}^b = f_{e_1}, \qquad
f_{e_3}^b = f_{e_3}, \qquad
f_{2e_d}^b = f_{2e_d}, \quad d=1,2,3.
\end{equation}
Thus it only remains to prove $f_{e_2}^b = 0$. According to \eqref
{eq:J_start}, \eqref{eq:p_alpha_simplified} and \eqref{eq:J_H},
$p_{e_2}$ can actually be expressed by
\begin{equation}
p_{e_2} = \rho^W J_0(u_1^W - u_1) \hat{J}_1 J_0(u_3^W - u_3)
  = \rho^W [(\theta^W - \theta) \hat{J}_{-1} - \hat{H}_1]
  = -\rho^W \sqrt{\frac{\theta^W}{2\pi}}.
\end{equation}
Since $K_2(e_2) = \lceil M/2 \rceil$, the above equation together with
\eqref{eq:rho_W} and \eqref{eq:mnt_bc} immediately gives $f_{e_2}^b =
0$.
\end{proof}

\subsection{Numerical implementation of boundary conditions}
In a finite volume scheme, the boundary conditions is often applied by
ghost cell techniques. Suppose the distribution function of the cell
on the boundary is denoted as \eqref{eq:f}. The distribution function
of the ghost cell can be constructed as follows:
\begin{enumerate}
\setlength\itemsep{0cm}
\item Apply $F^b$ on $f(\bxi)$ and suppose the result is
  \eqref{eq:f_b};
\item Construct the ghost cell distribution as
  \begin{equation} \label{eq:ghost}
    f^{\mathrm{ghost}}(\bxi) = \sum_{|\alpha| \leqslant M + 1}
      (2f_{\alpha}^b - f_{\alpha}) \mathcal{H}_{\theta, \alpha} \left(
        \frac{\bxi - (2\bu^b - \bu)}{\sqrt{\theta}}
      \right).
  \end{equation}
\end{enumerate}

Now we consider the time complexity of this operation. Suppose $N_M =
(M+2)(M+3)(M+4)/6$ is the number of moments involved in the boundary
condition. Obviously, \eqref{eq:ghost} requires $O(N_M)$ operations.
For the calculation of $F^b(f)$, we list the cost as follows:
\begin{enumerate}
\setlength\itemsep{0cm}
\item Half-space cut-off of $f$ \eqref{eq:q}: $O(M N_M)$ operations;
\item Calculation of $\rho^W$ \eqref{eq:rho_W}: $O(1)$ operations;
\item Calculation of $p_{\alpha}$ \eqref{eq:p_alpha_simplified}:
  $O(N_M)$ operations;
\item Evaluation of \eqref{eq:f_b_alpha}: $O(N_M)$ operations.
\end{enumerate}
Thus, the total computational cost is $O(M N_M)$, while the time
complexity is $O(N_M)$ if no boundary condition is considered.
However, since this procedure only takes place on the boundary, it
produces little increment of the computational time in real
computation.
\begin{remark}
  Proposition \ref{prop:conservation} indicates the conservation of
  mass on the boundary when using the HLL numerical flux as in
  \cite{Cai}.  One can find that when $u_2^W = 0$, saying a special
  reference coordinate system is used, the minimum and maximum signal
  speeds in need of the HLL flux are opposite numbers. Together with
  $\rho^{\mathrm{ghost}} = \rho$, $u_2^{\mathrm{ghost}} = -u_2$, the
  mass conservation of the HLL scheme follows naturally.
\end{remark}

%%% Local Variables: 
%%% mode: latex
%%% TeX-master: "article"
%%% End: 

% vim: tw=70:spell
\section{Numerical examples} \label{sec:example} In this section,
three numerical examples are presented to validate our algorithm. In
all these examples, a hard sphere gas is assumed, for which the
relaxation time is defined as
\begin{equation} \label{eq:tau}
\tau = \frac{5}{16} \sqrt{\frac{2\pi}{\theta}} \frac{\Kn}{\rho}
\end{equation}
following \cite{Bird}, where $\Kn$ is the Knudsen number. The CFL
number is always $0.95$. And for all the tests, the wall is set to be
a fully diffusive one ($\chi = 1$) with $\theta^W = 1$. The POSIX
multithreading technique is utilized in our simulation, and at most 8
CPU cores are used.

\subsection{The beginning of a shock wave's formation} \label{sec:reflection}
The first example is a simulation of the interaction of a coming flow
with a diffusive wall. The computational domain is $[-5,0]$ and the
global Knudsen number $\Kn$ used in \eqref{eq:tau} is set to be $0.5$.
The left boundary is a free boundary, and the right is a stationary
diffusive wall parallel to the $xz$-plane. The initial condition is
given by
\begin{equation}
\rho_0(y) = 1.0, \quad \bu_0(y) = (0, 0.5, 0)^T,
  \quad \theta_0(y) = 1.0, \qquad \forall y \in [-5, 0],
\end{equation}
and the gas is in equilibrium everywhere. A left-going shock wave will
form after a sufficiently long time. Here we stop the computation at
$t=1.0$ in order to check the validity of the boundary condition.  For
a reference solution, we solve the Shakhov equation \eqref{eq:Shakhov}
directly using a Conservative Discrete Velocity Method (CDVM)
introduced in \cite{Titarev}. For the computation of both \NRxx method
and CDVM, a uniform mesh with $500$ grids are used to discretize the
domain. For CDVM, the computational velocity domain is $[-10, 10]
\times [-10, 10] \times [-10, 10]$ and discretized by $50 \times 100
\times 50$ grids.

\begin{figure}[!ht]
\centering
\psfrag{dashed, NRxx}{\footnotesize{$\rho$, \NRxx}}
\psfrag{dashed, CDVM}{\footnotesize{$\rho$, CDVM}}
\psfrag{solid, NRxx}{\footnotesize{$\theta$, \NRxx}}
\psfrag{solid, CDVM}{\footnotesize{$\theta$, CDVM}}
\subfigure[$M=3$]{
  \includegraphics[width=.45\textwidth]{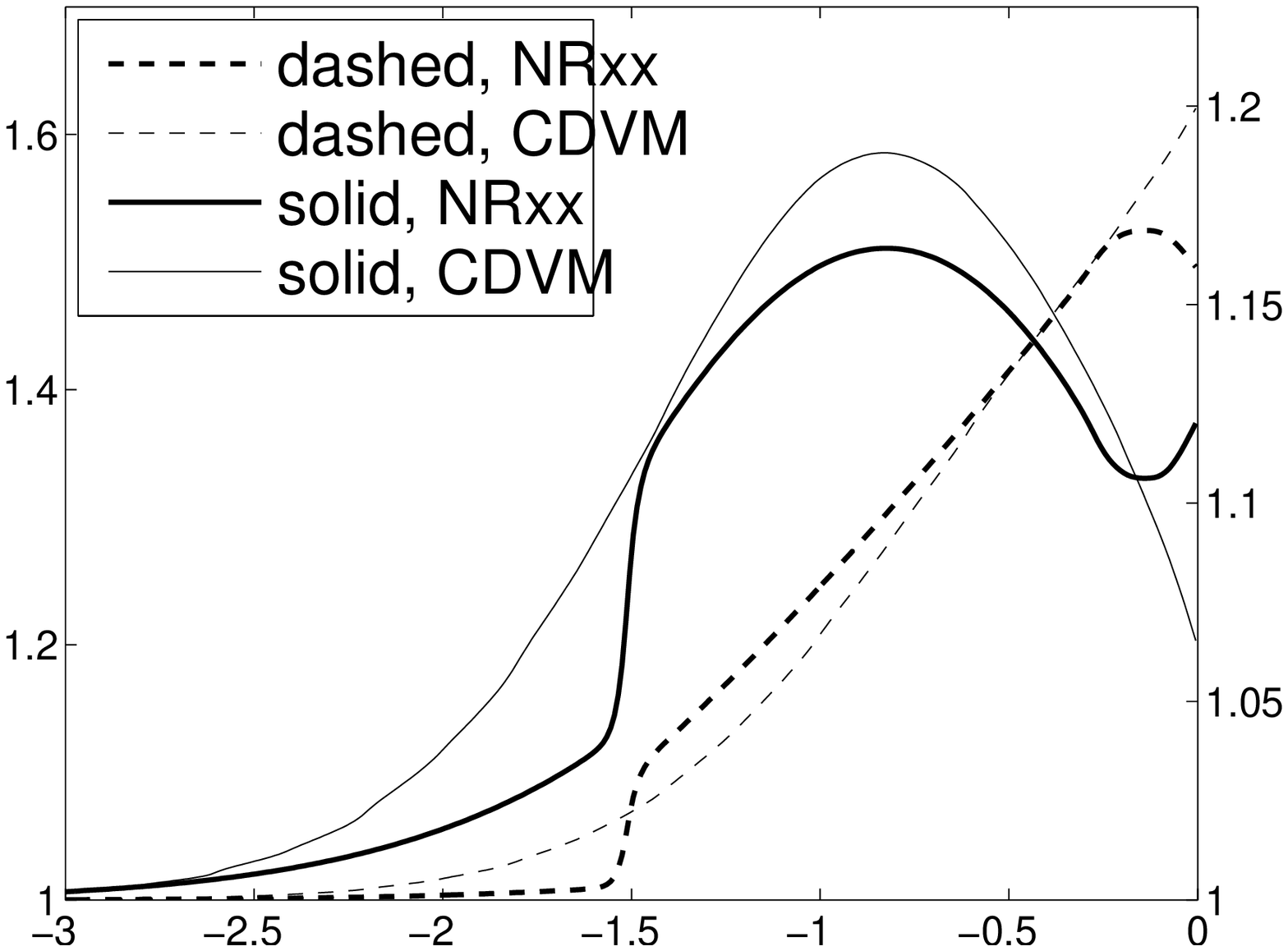}
}
\subfigure[$M=4$]{
  \includegraphics[width=.45\textwidth]{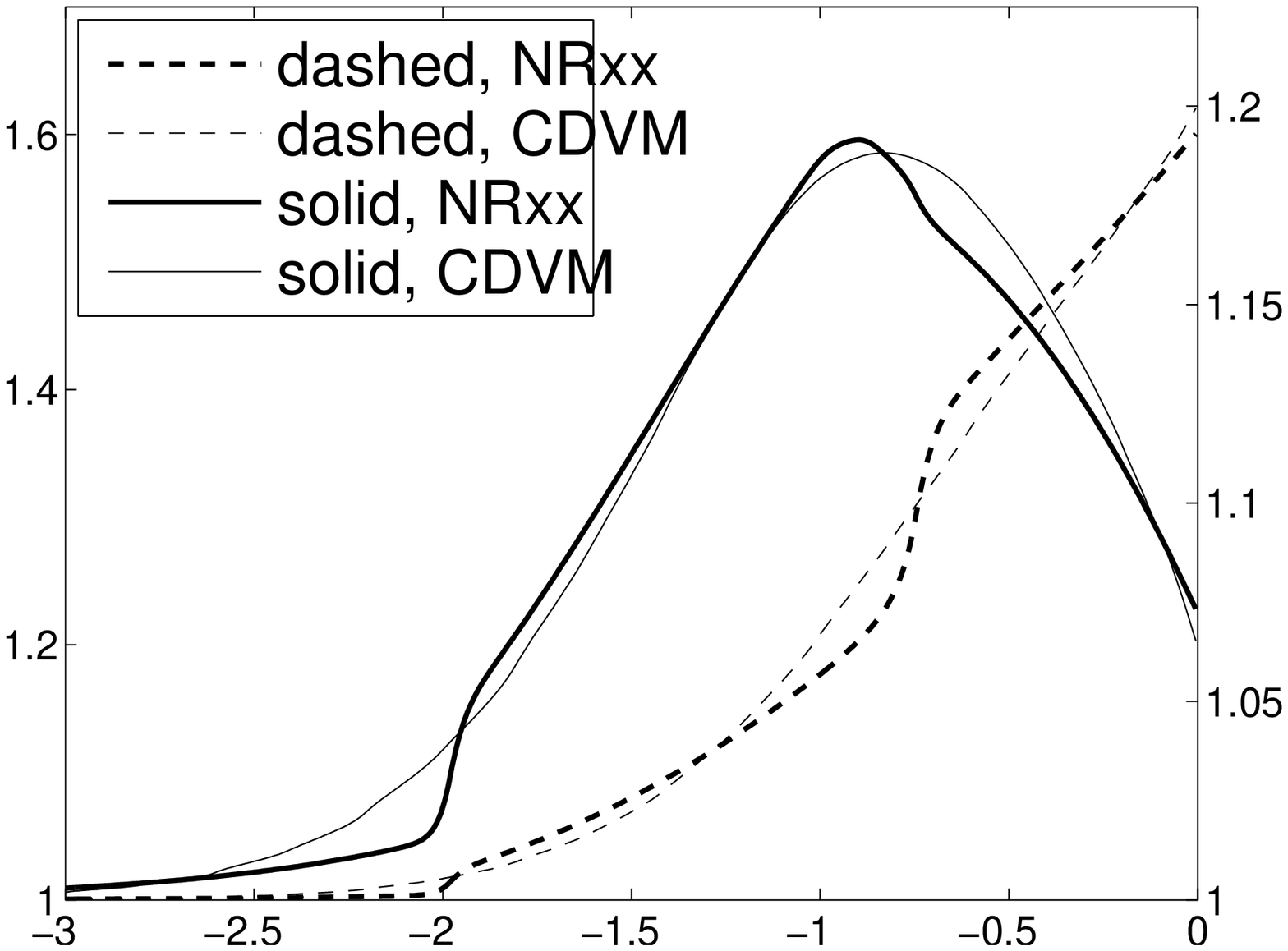}
}
\subfigure[$M=5$]{
  \includegraphics[width=.45\textwidth]{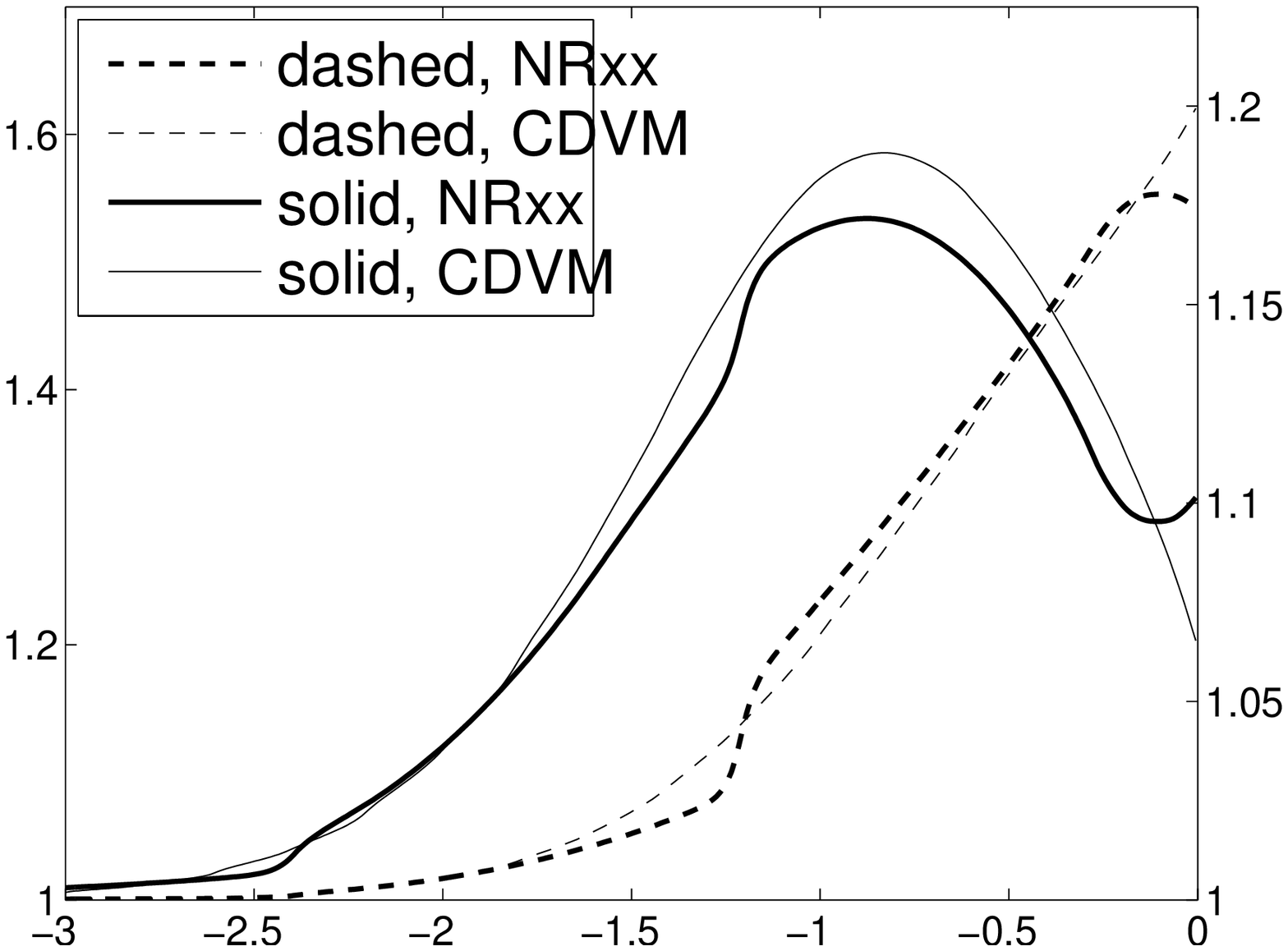}
}
\subfigure[$M=6$]{
  \includegraphics[width=.45\textwidth]{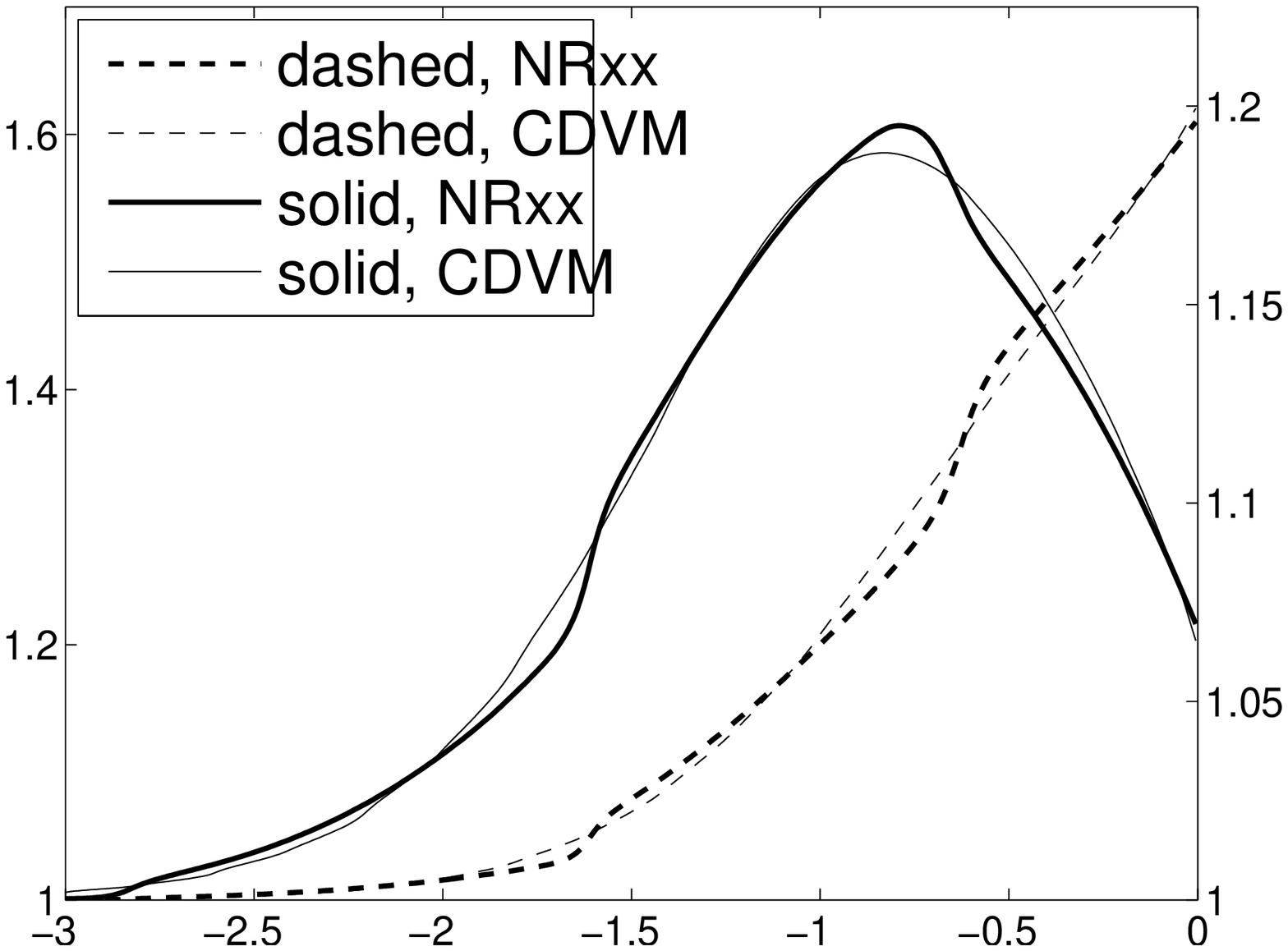}
}
\subfigure[$M=7$]{
  \includegraphics[width=.45\textwidth]{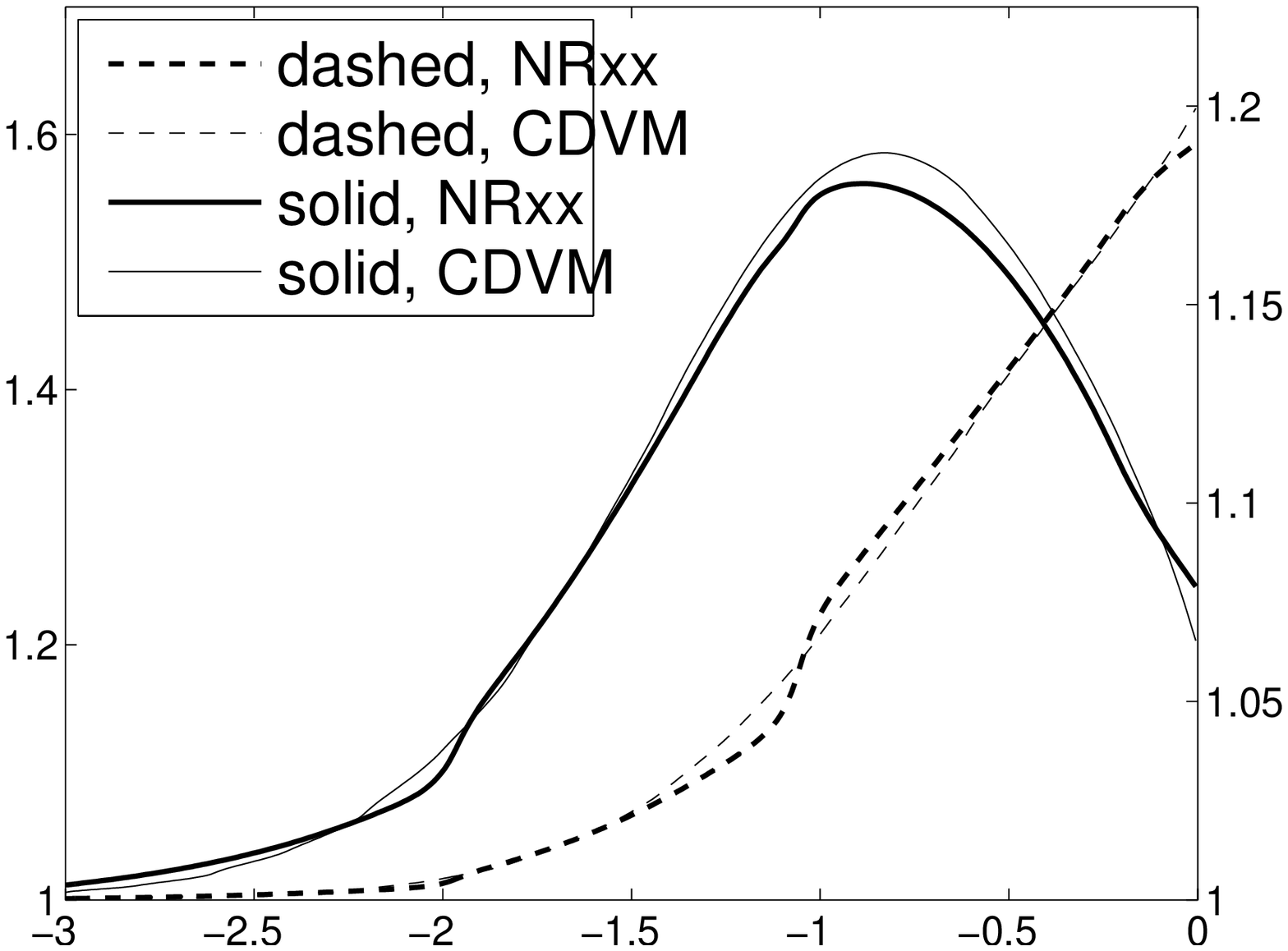}
}
\subfigure[$M=8$]{
  \includegraphics[width=.45\textwidth]{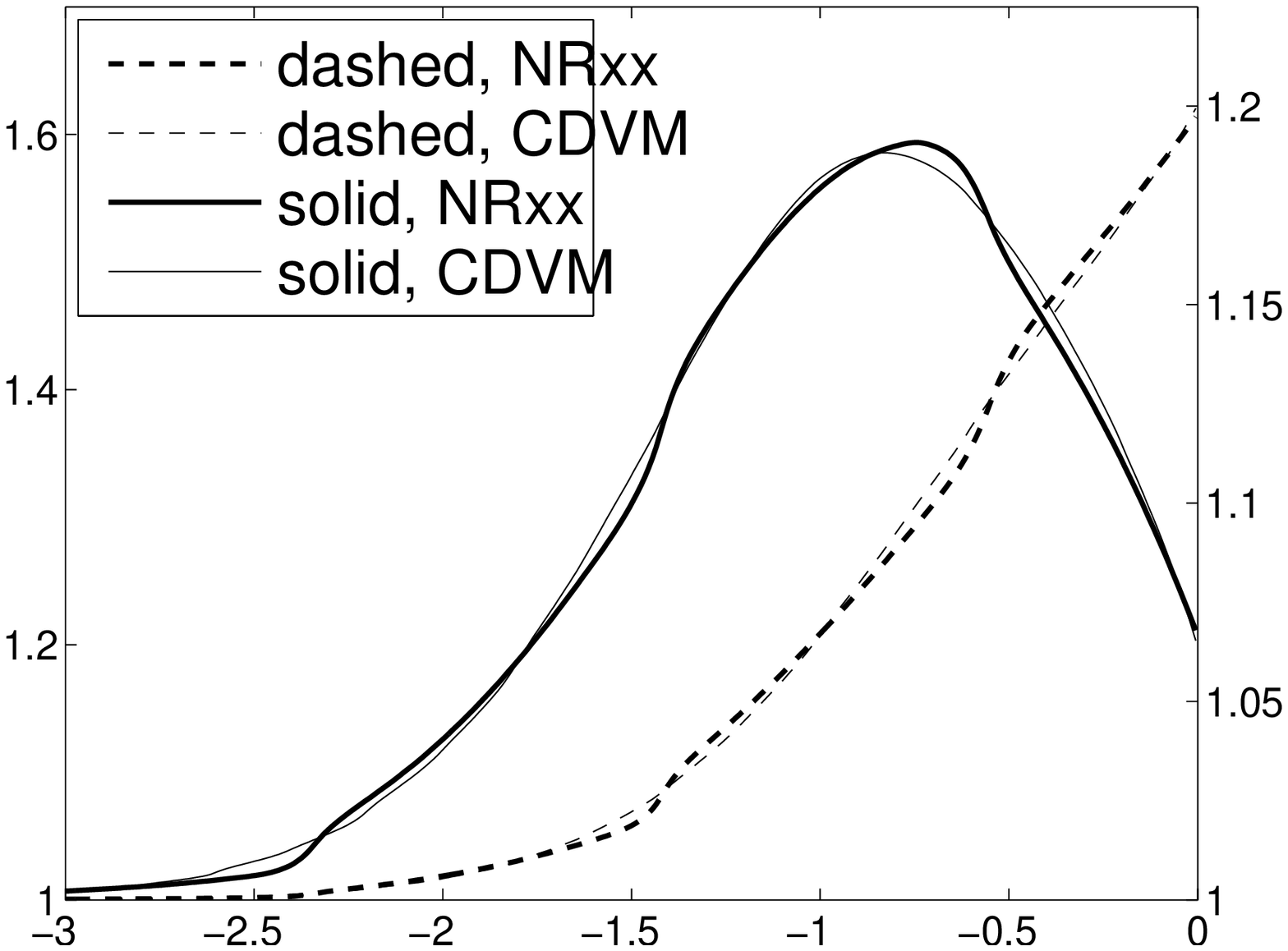}
}
\caption{Density and temperature plots for the problem in section
\ref{sec:reflection}. The left axis is for the dashed lines, and the
right axis is for the solid lines (to be continued).}
\label{fig:Ref_rho_theta}
\end{figure}

Figure \ref{fig:Ref_rho_theta} and \ref{fig:Ref_sigma_q} are the
results for CDVM and \NRxx method for $M=3$ to $12$. Only the part $y
\in [-3, 0]$ is shown since all variables for the remaining part are
almost constant. Since a large Knudsen number is considered,
predictions from lower order moment equations give very large
deviations, so the necessity of high order moment theory is obvious.
As the number of moments increases, all profiles get closer and closer
to the results of CDVM. When $M$ reaches $11$, the density and
temperature plots agree with the CDVM results very well, and the errors
in $\sigma_{22}$ and $q_2$ are much smaller than the low order cases,
though it is still observable. It is reasonable that higher order
moments converge more slowly than lower order moments, which is also
observed in \cite{Emerson}.

\addtocounter{figure}{-1}
\begin{figure}[!ht]
\centering
\psfrag{dashed, NRxx}{\footnotesize{$\rho$, \NRxx}}
\psfrag{dashed, CDVM}{\footnotesize{$\rho$, CDVM}}
\psfrag{solid, NRxx}{\footnotesize{$\theta$, \NRxx}}
\psfrag{solid, CDVM}{\footnotesize{$\theta$, CDVM}}
\setcounter{subfigure}{6}
\subfigure[$M=9$]{
  \includegraphics[width=.45\textwidth]{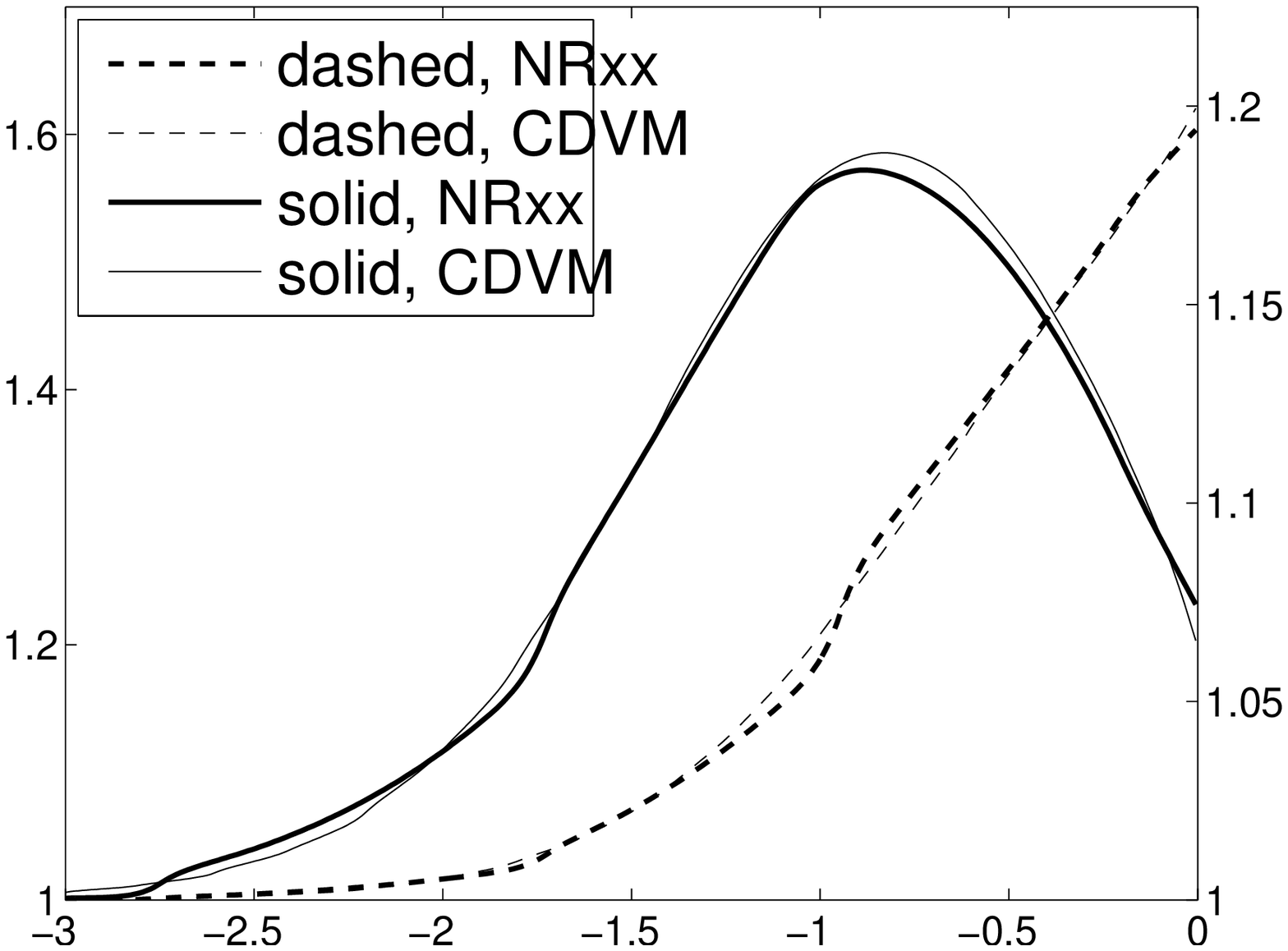}
}
\subfigure[$M=10$]{
  \includegraphics[width=.45\textwidth]{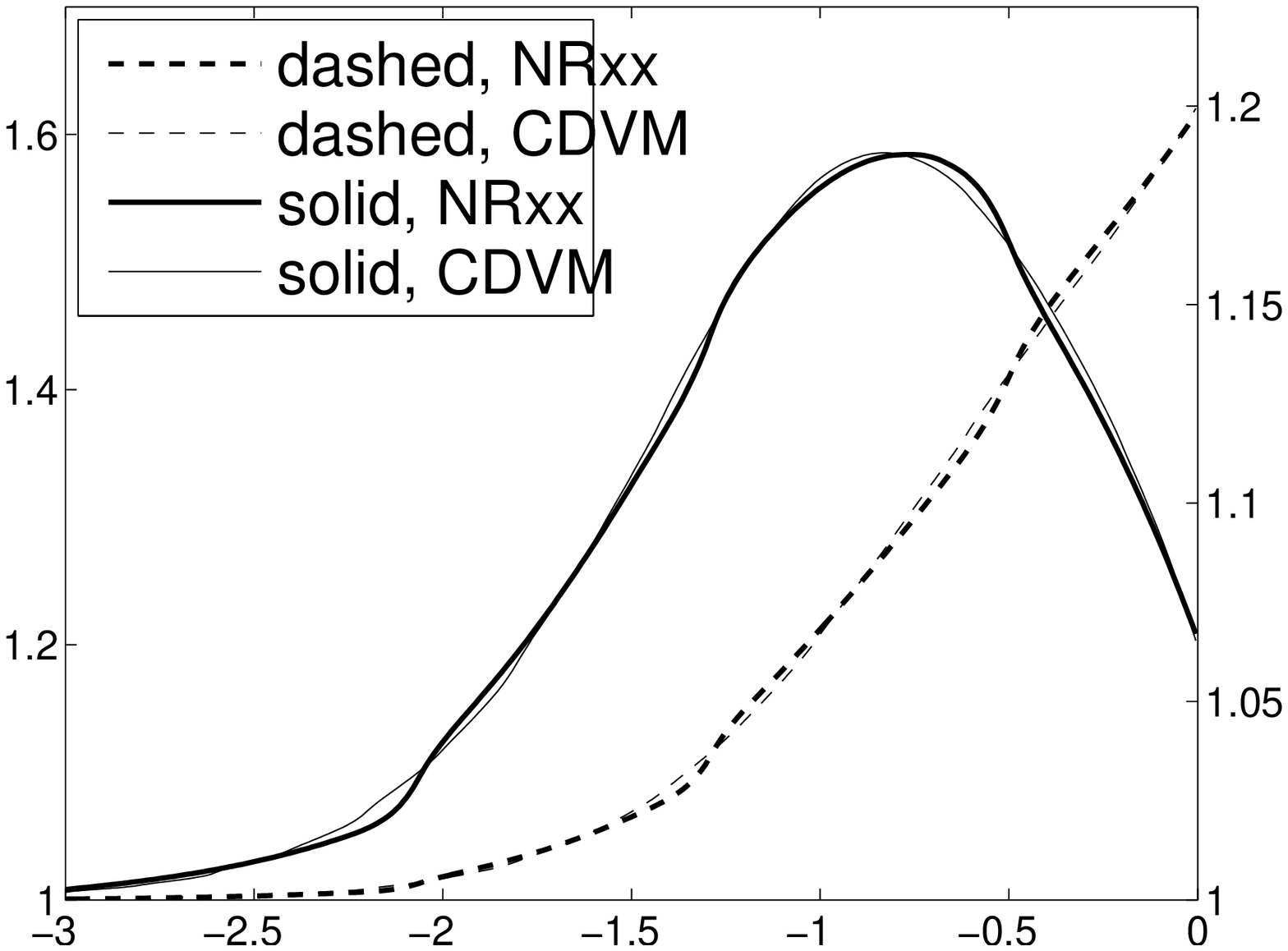}
}
\subfigure[$M=11$]{
  \includegraphics[width=.45\textwidth]{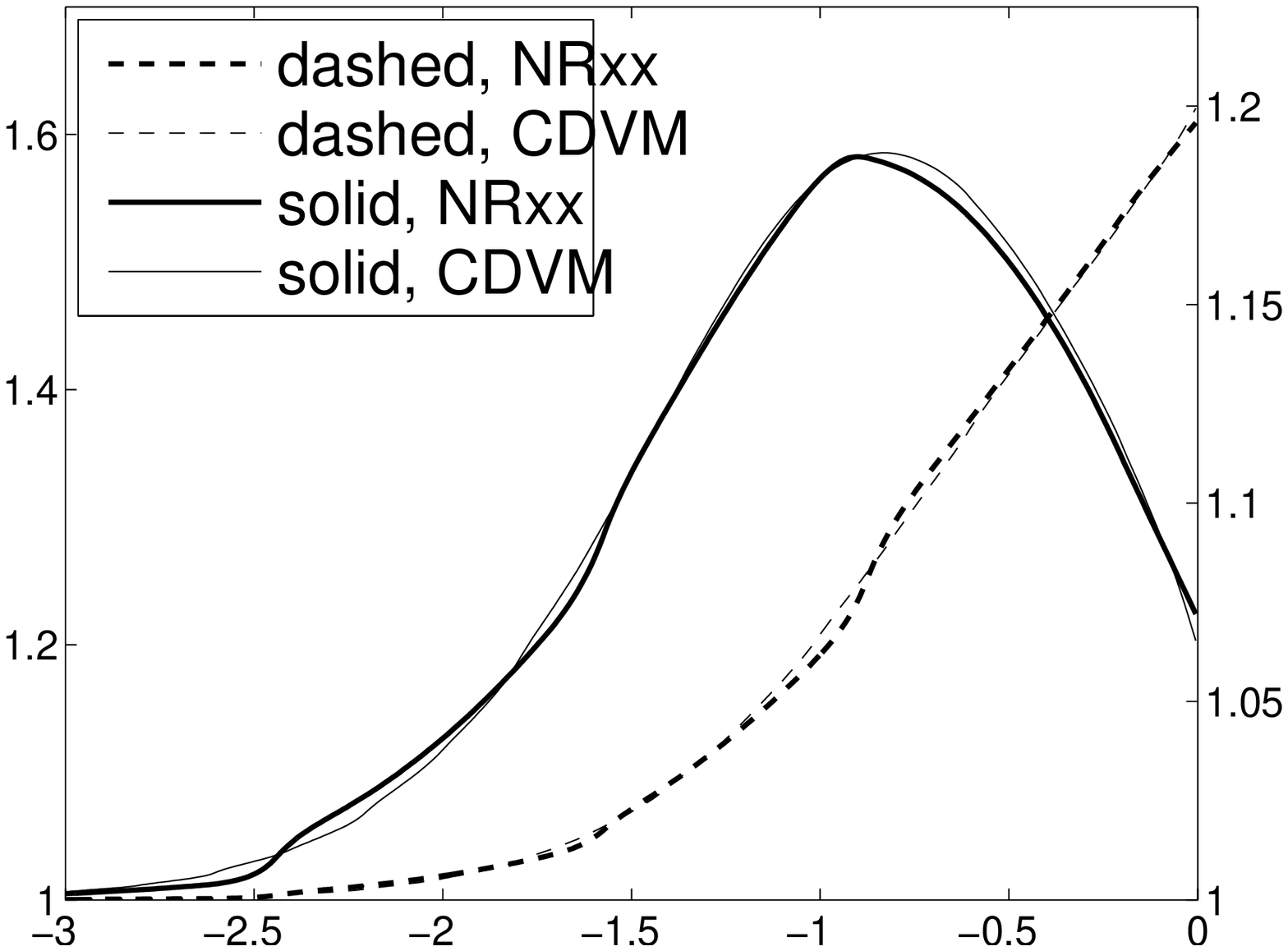}
}
\subfigure[$M=12$]{
  \includegraphics[width=.45\textwidth]{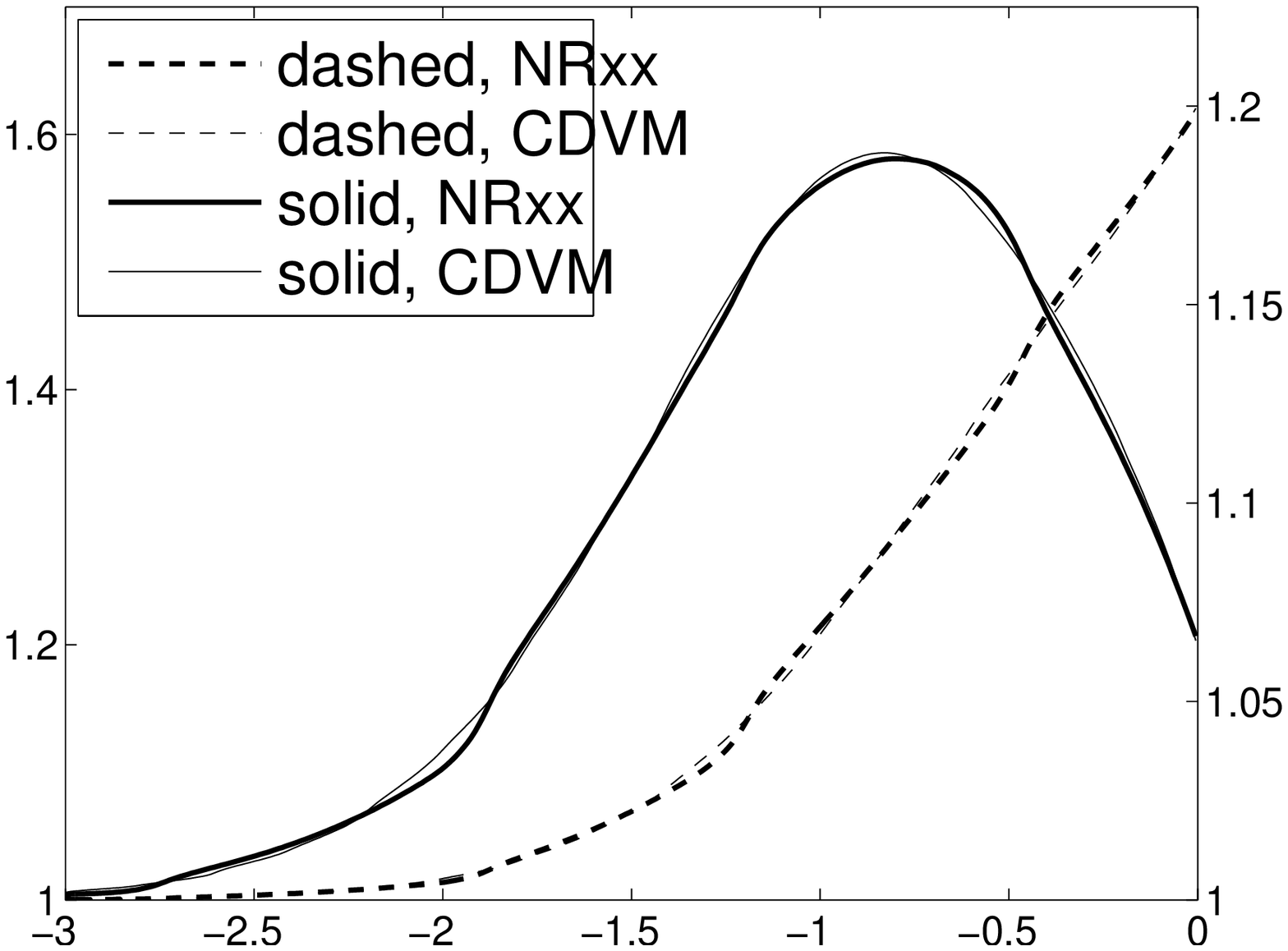}
}
\caption{Density and temperature plots for the problem in section
\ref{sec:reflection}. The left axis is for the dashed lines, and the
right axis is for the solid lines.}
\end{figure}

\begin{figure}[!ht]
\centering
\psfrag{dashed, NRxx}{\scalebox{0.8}{$\sigma_{22}$, \NRxx}}
\psfrag{dashed, CDVM}{\scalebox{0.8}{$\sigma_{22}$, CDVM}}
\psfrag{solid, NRxx}{\scalebox{0.8}{$q_2$, \NRxx}}
\psfrag{solid, CDVM}{\scalebox{0.8}{$q_2$, CDVM}}
\subfigure[$M=3$]{
  \includegraphics[width=.45\textwidth]{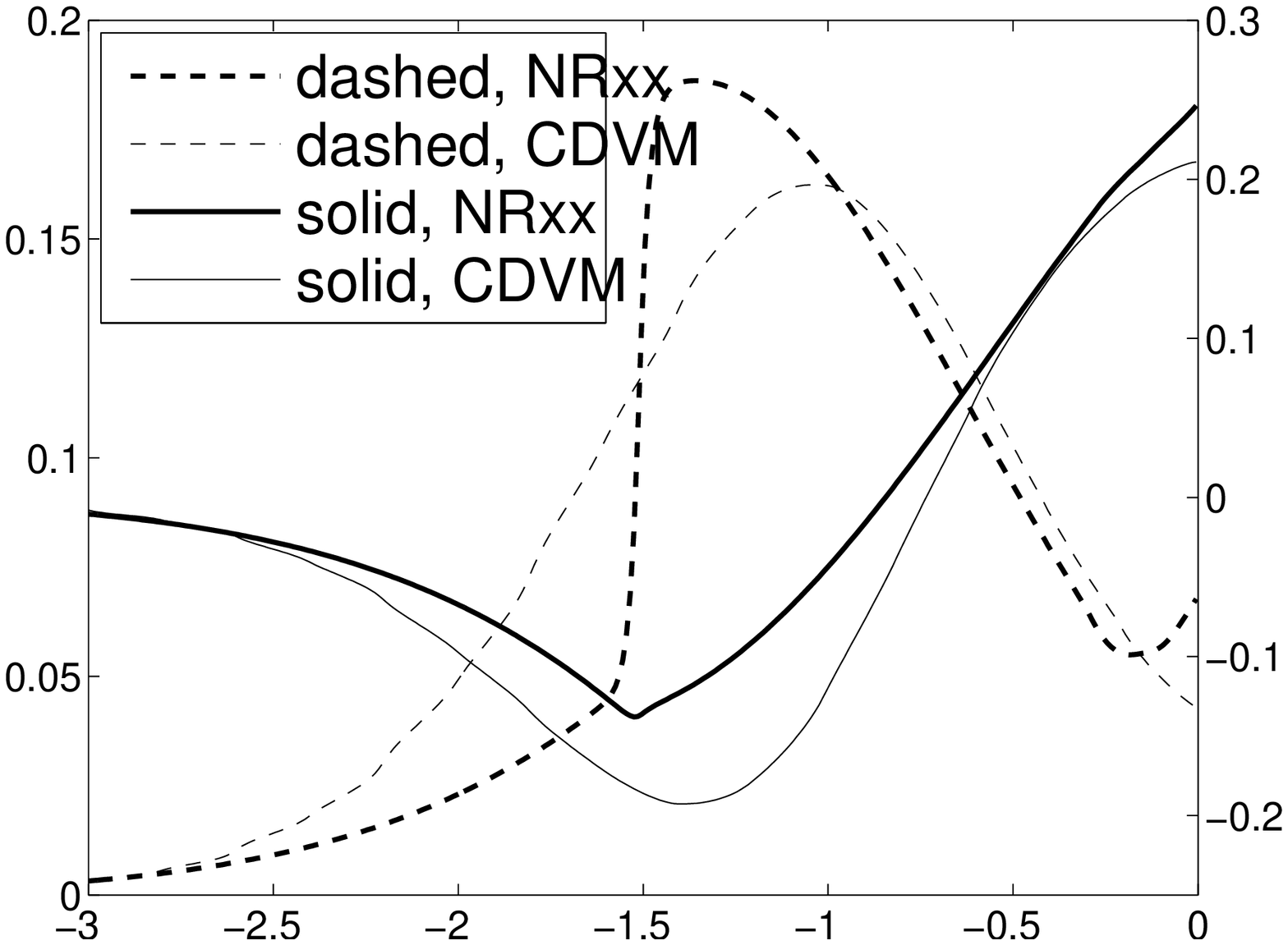}
}
\subfigure[$M=4$]{
  \includegraphics[width=.45\textwidth]{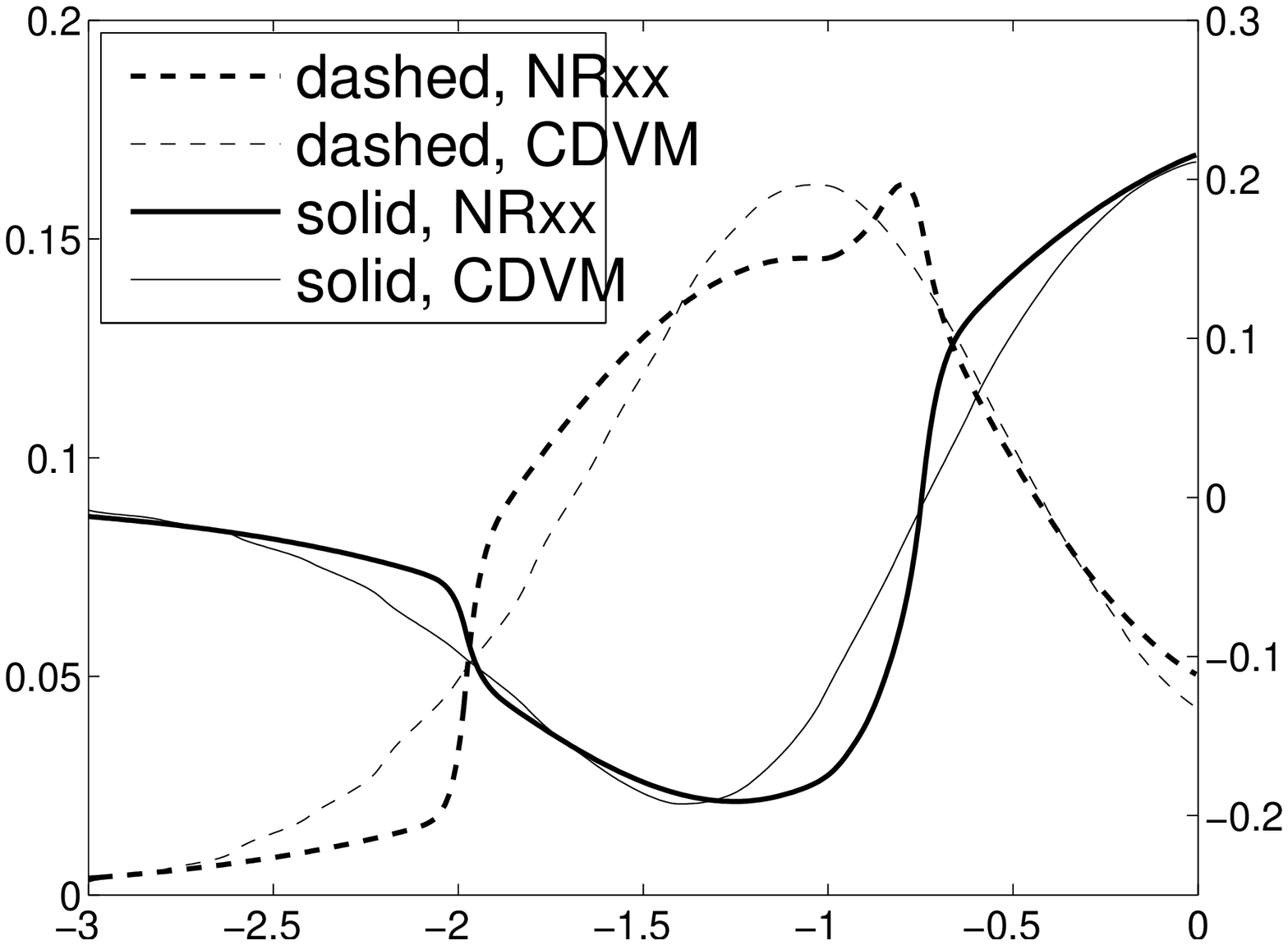}
}
\subfigure[$M=5$]{
  \includegraphics[width=.45\textwidth]{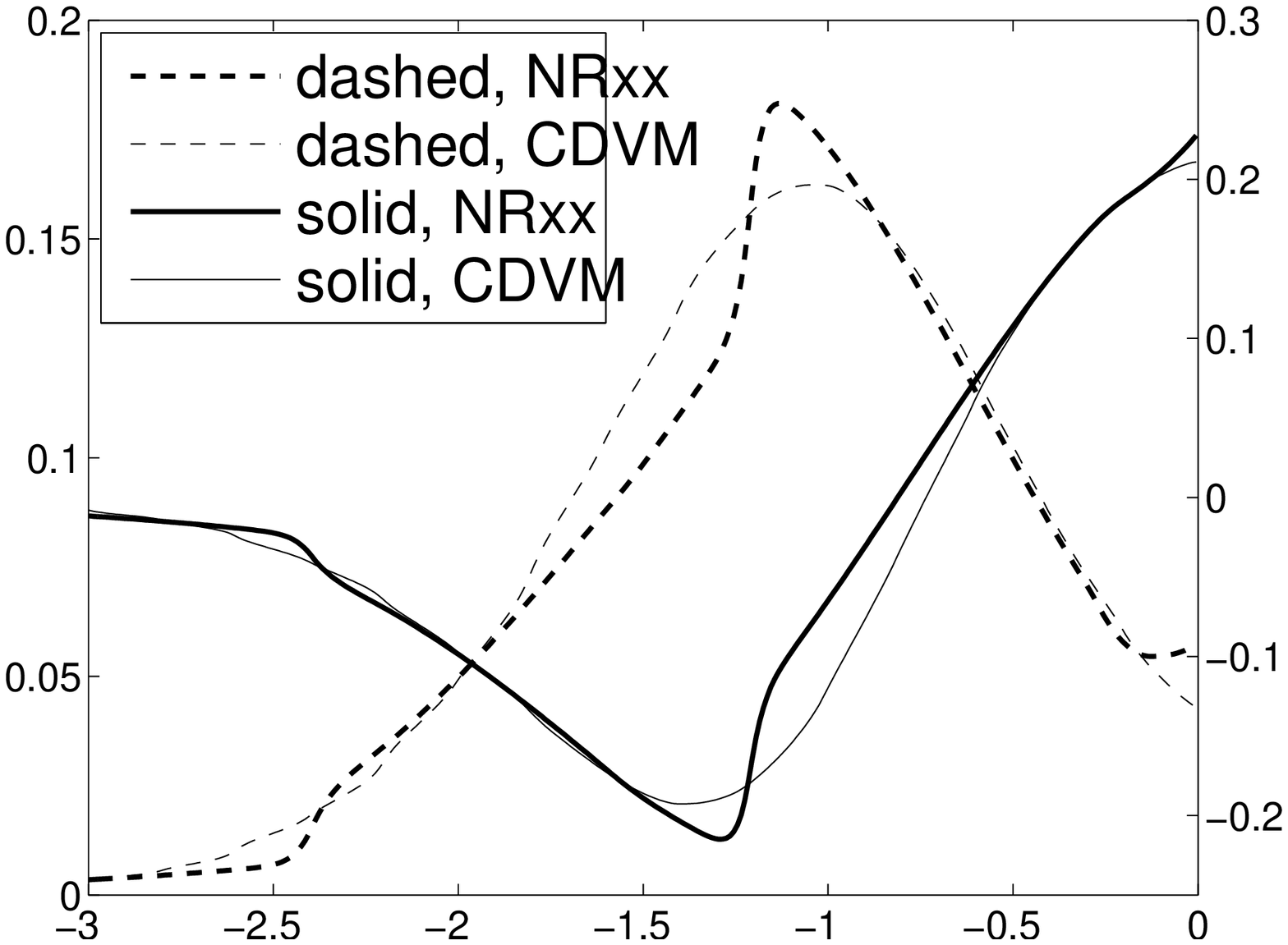}
}
\subfigure[$M=6$]{
  \includegraphics[width=.45\textwidth]{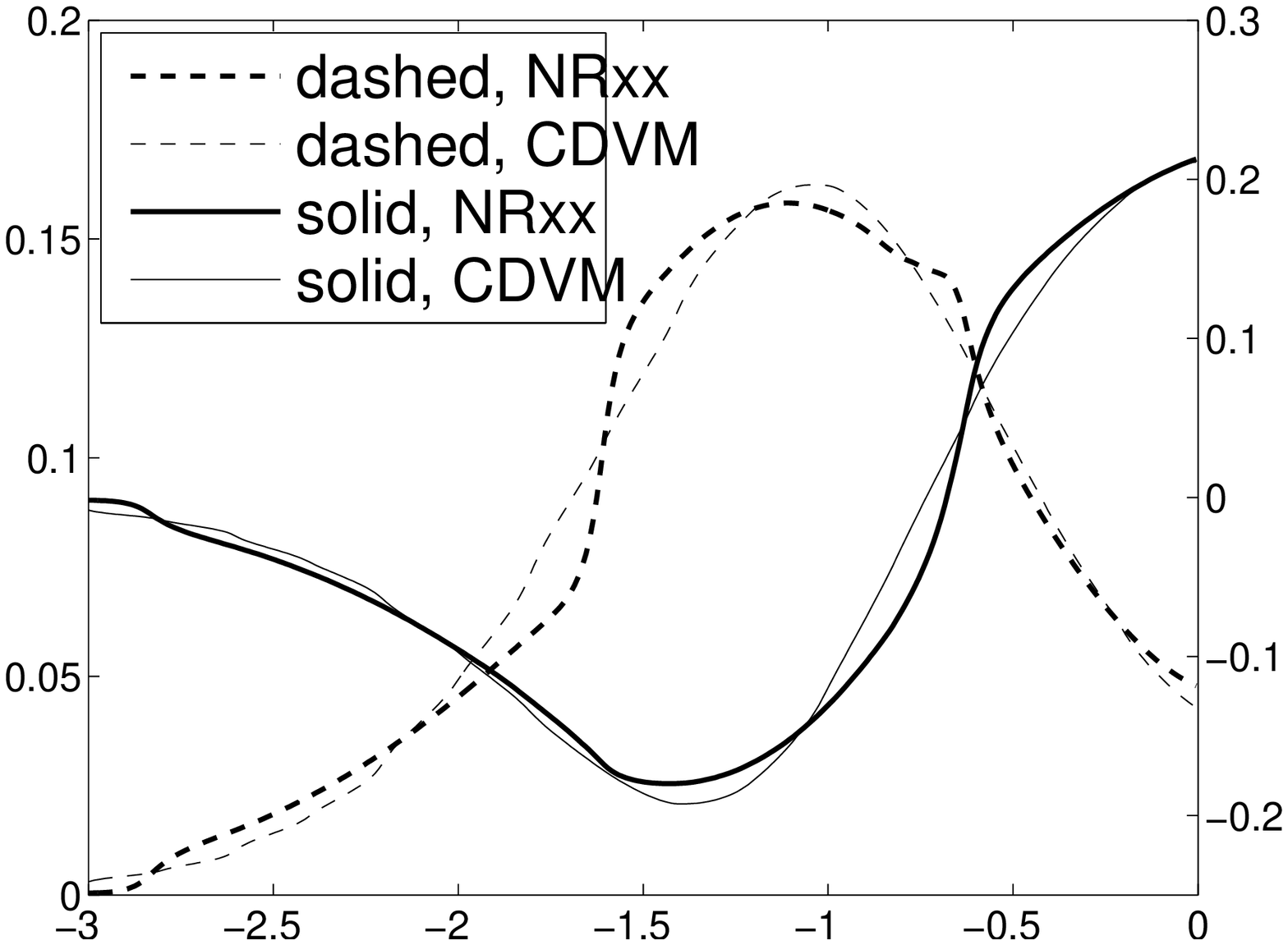}
}
\subfigure[$M=7$]{
  \includegraphics[width=.45\textwidth]{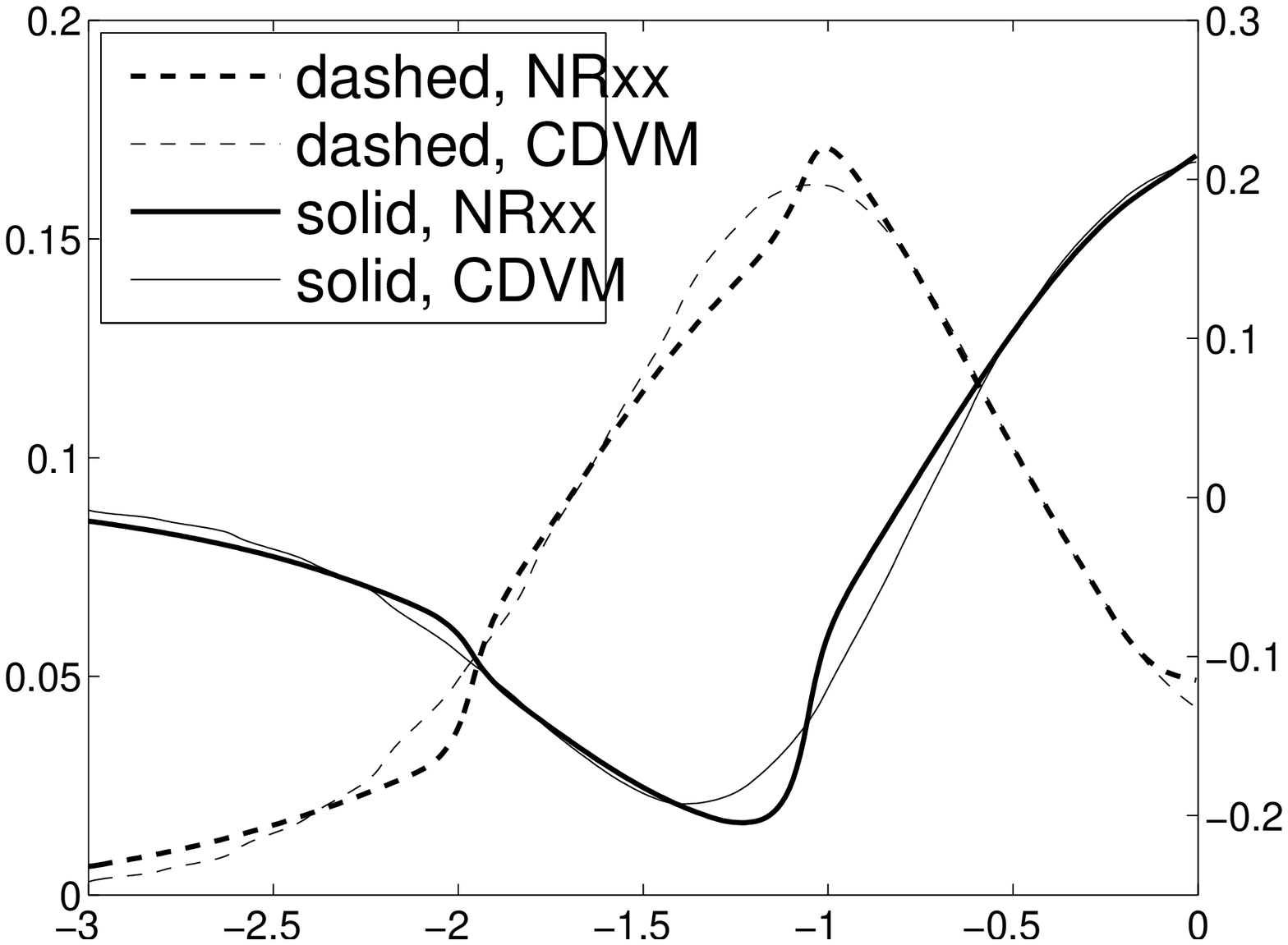}
}
\subfigure[$M=8$]{
  \includegraphics[width=.45\textwidth]{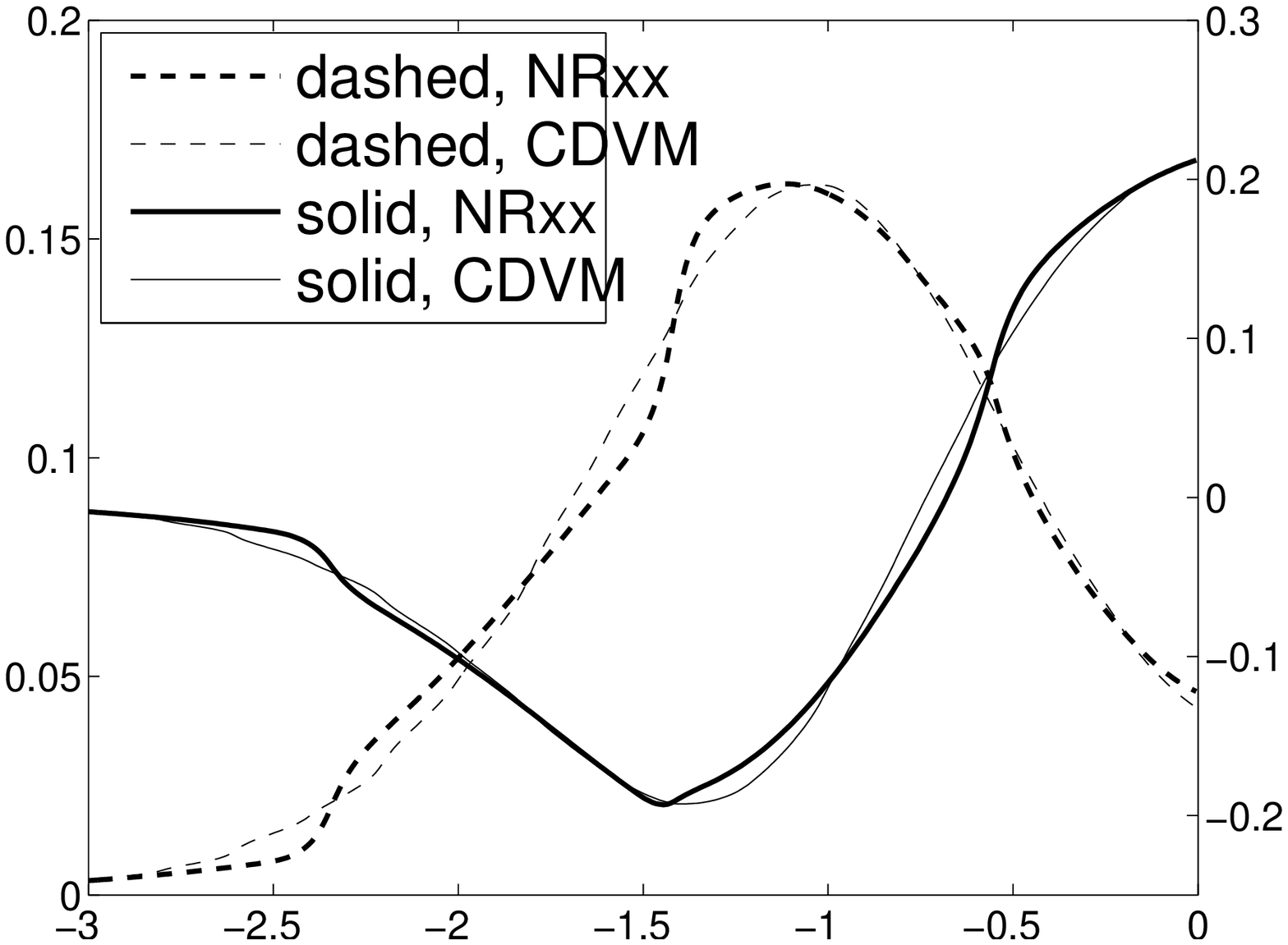}
}
\caption{Stress and heat flux plots for the problem in section
\ref{sec:reflection}. The left axis is for the dashed lines, and the
right axis is for the solid lines (to be continued).}
\label{fig:Ref_sigma_q}
\end{figure}
\addtocounter{figure}{-1}
\begin{figure}[!ht]
\centering
\psfrag{dashed, NRxx}{\scalebox{0.8}{$\sigma_{22}$, \NRxx}}
\psfrag{dashed, CDVM}{\scalebox{0.8}{$\sigma_{22}$, CDVM}}
\psfrag{solid, NRxx}{\scalebox{0.8}{$q_2$, \NRxx}}
\psfrag{solid, CDVM}{\scalebox{0.8}{$q_2$, CDVM}}
\setcounter{subfigure}{6}
\subfigure[$M=9$]{
  \includegraphics[width=.45\textwidth]{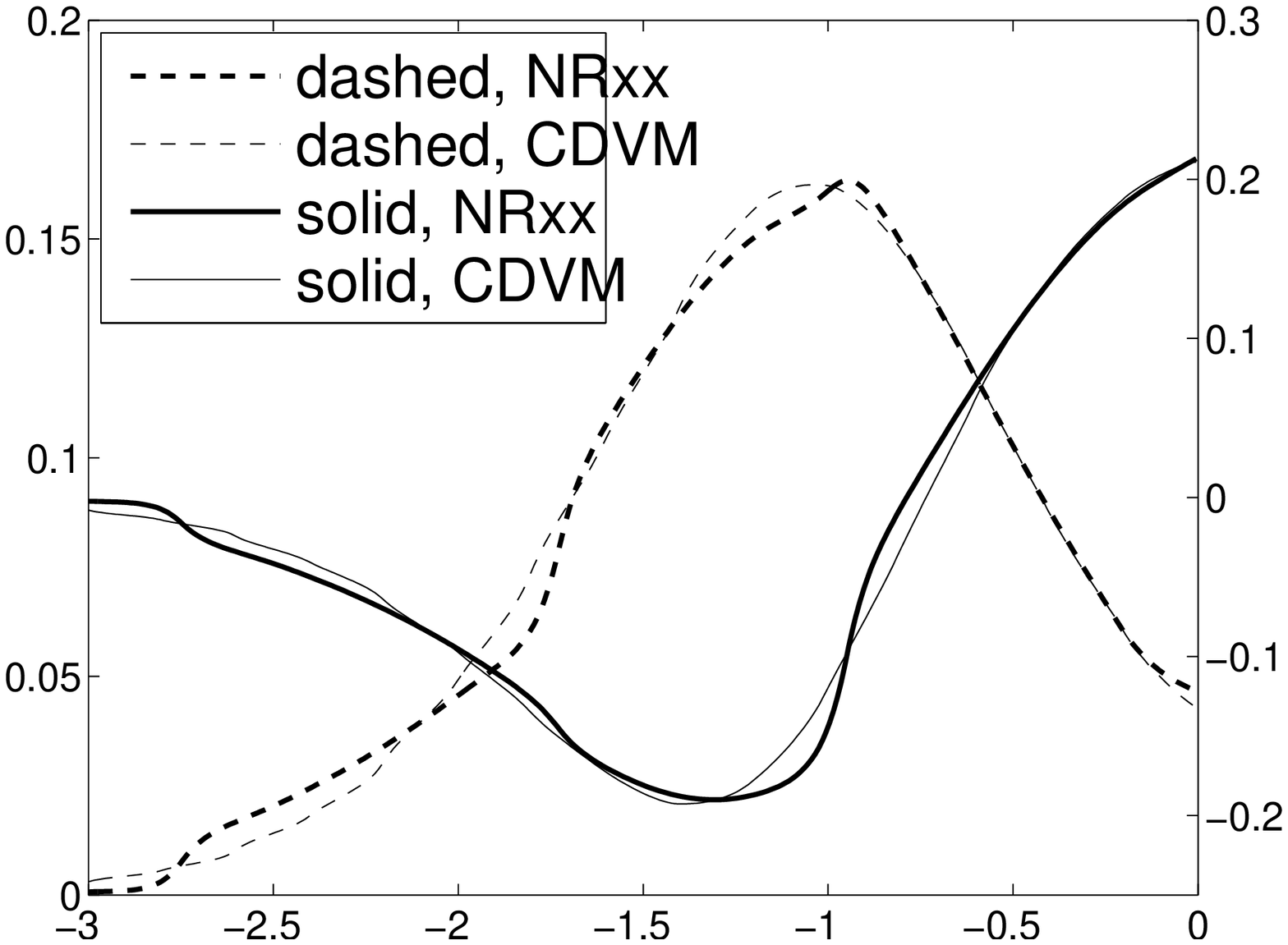}
}
\subfigure[$M=10$]{
  \includegraphics[width=.45\textwidth]{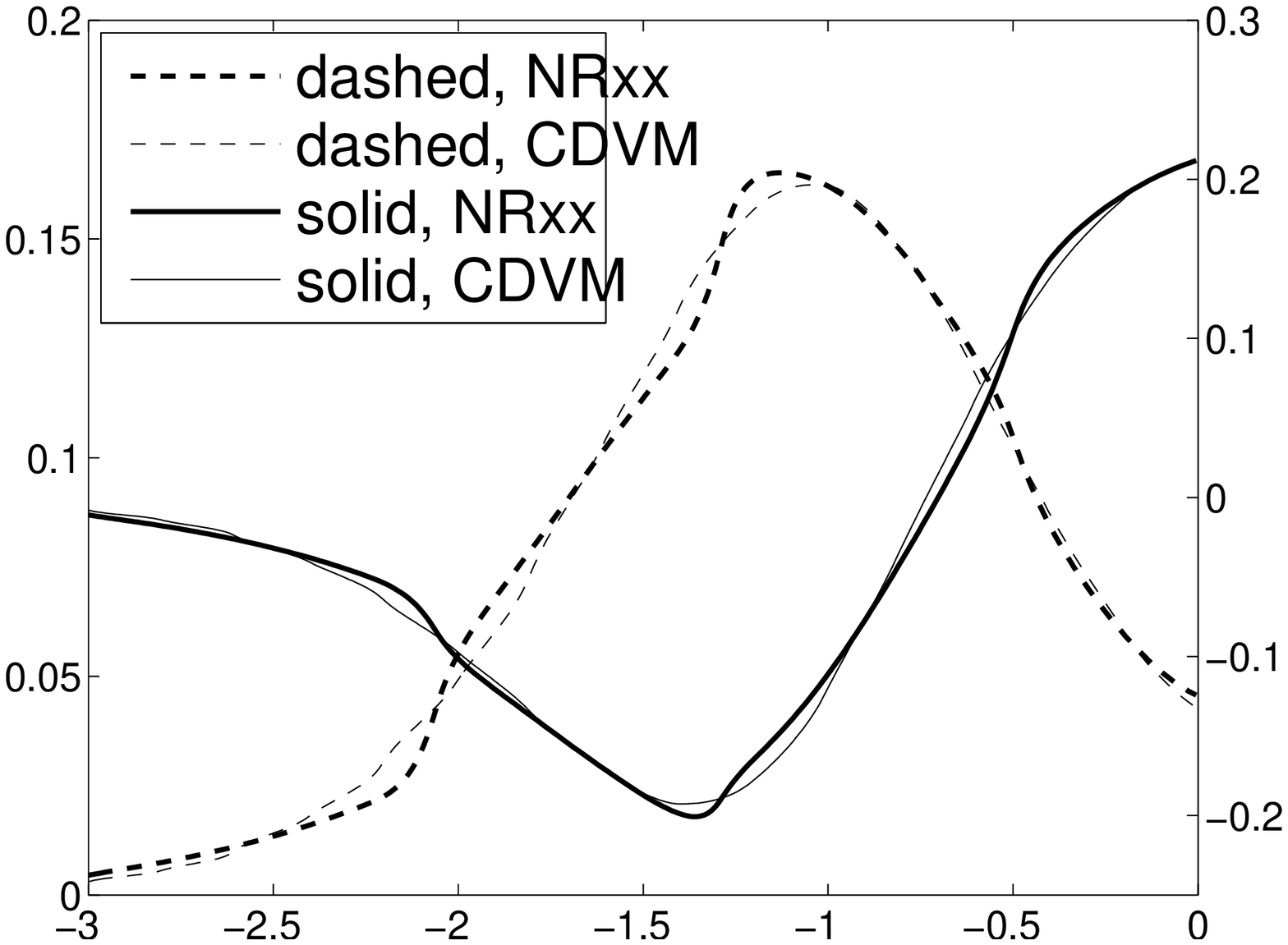}
}
\subfigure[$M=11$]{
  \includegraphics[width=.45\textwidth]{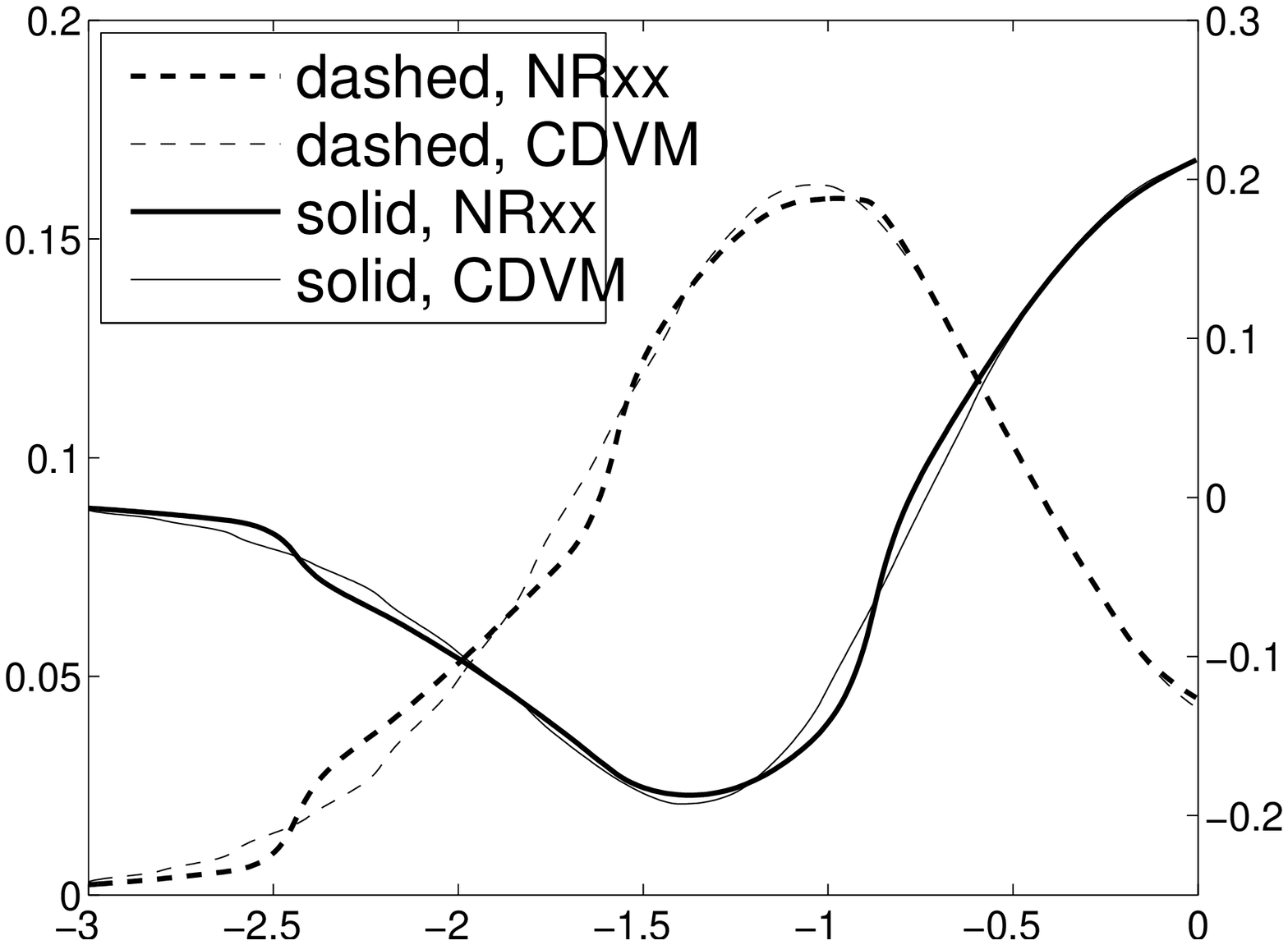}
}
\subfigure[$M=12$]{
  \includegraphics[width=.45\textwidth]{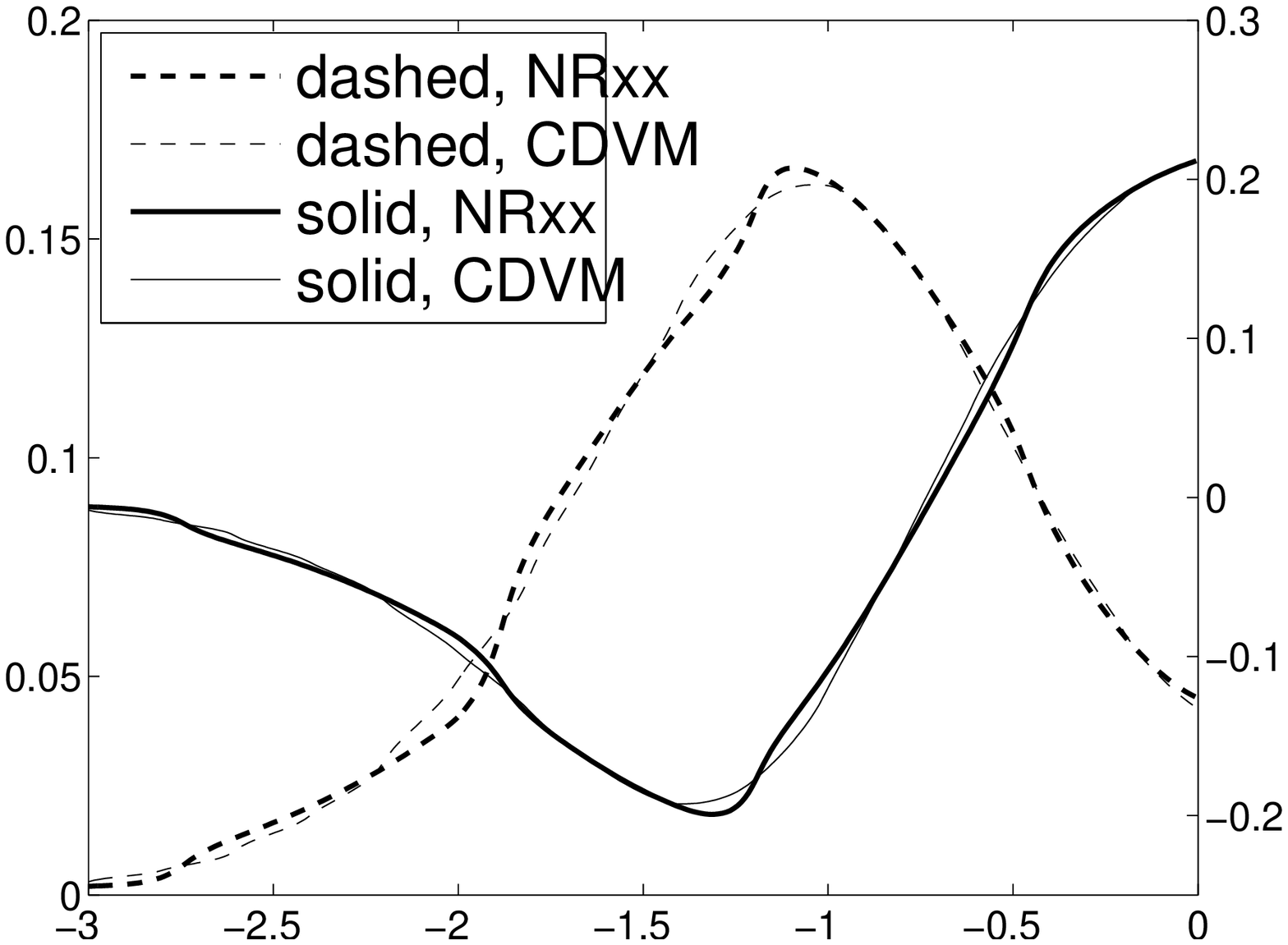}
}
\caption{Stress and heat flux plots for the problem in section
\ref{sec:reflection}. The left axis is for the dashed lines, and the
right axis is for the solid lines.}
\end{figure}

\subsection{Planar Couette flow} \label{sec:Couette}
The planar Couette flow is a classic benchmark test in the field of
microflows. The moment method for this problem has been investigated
in a lot of papers such as \cite{Tallec, Reitebuch1999, Thatcher, Gu,
Torrilhon2008, Emerson}. Here we consider the symmetric Couette flow.
The gas lies between two plates parallel to the $xz$-plane.  Two
plates move in the opposite direction with constant velocities within
their own planes. A steady state can be obtained for a fully developed
flow.

In this example, the computational domain is $[-0.5, 0.5]$. The
velocities of the left and right plates are
\begin{equation}
\bu_L^W = (-0.6296, 0, 0)^T, \quad \bu_R^W = (0.6296, 0, 0)^T.
\end{equation}
The initial state is a global equilibrium with
\begin{equation}
\rho_0(y) = 1, \quad \bu_0(y) = 0, \quad \theta_0(y) = 1,
  \qquad \forall y \in [-0.5, 0.5].
\end{equation}
The steady state can be achieved if the computational time is
sufficiently long. Also, both the \NRxx method and CDVM are applied to
this problem. Three different Knudsen numbers, $Kn = 0.1, 0.5, 1.0$,
are investigated. For CDVM, the computational velocity domain is
chosen as $[-10, 10] \times [-10, 10] \times [-10, 10]$, and $50
\times 50 \times 50$ grids are used. Here we note that such
discretization may not produce numerical results accurate enough as
the reference solution, but the computation is already extremely
slow.

\begin{figure}[!ht]
\centering
\setlength\subfigcapskip{10pt}
\begin{tabular}{r}
\subfigure[Density, $\rho$]{
  \includegraphics[scale=.75]{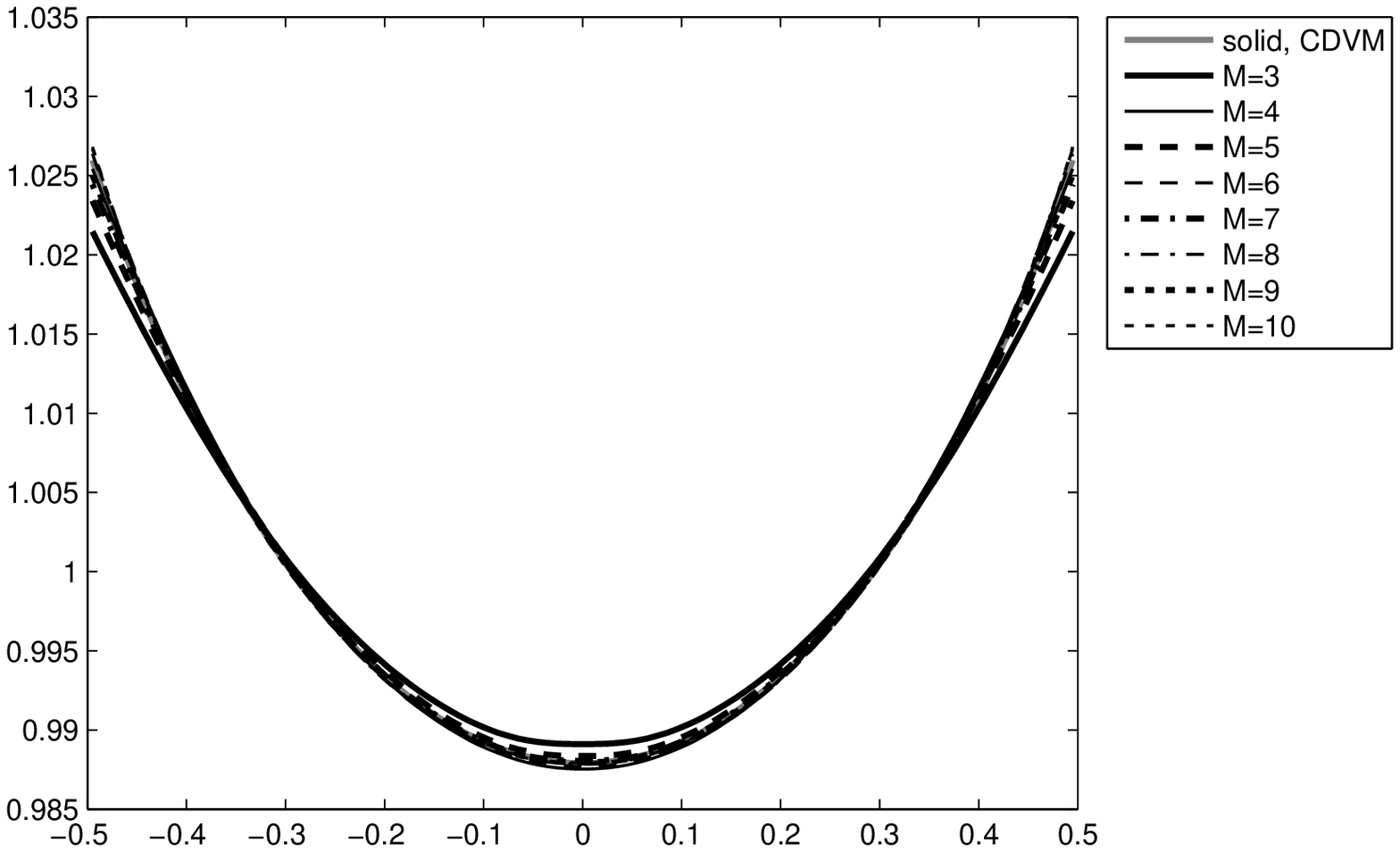}
} \\
\subfigure[Temperature, $\theta$]{
  \includegraphics[scale=.75]{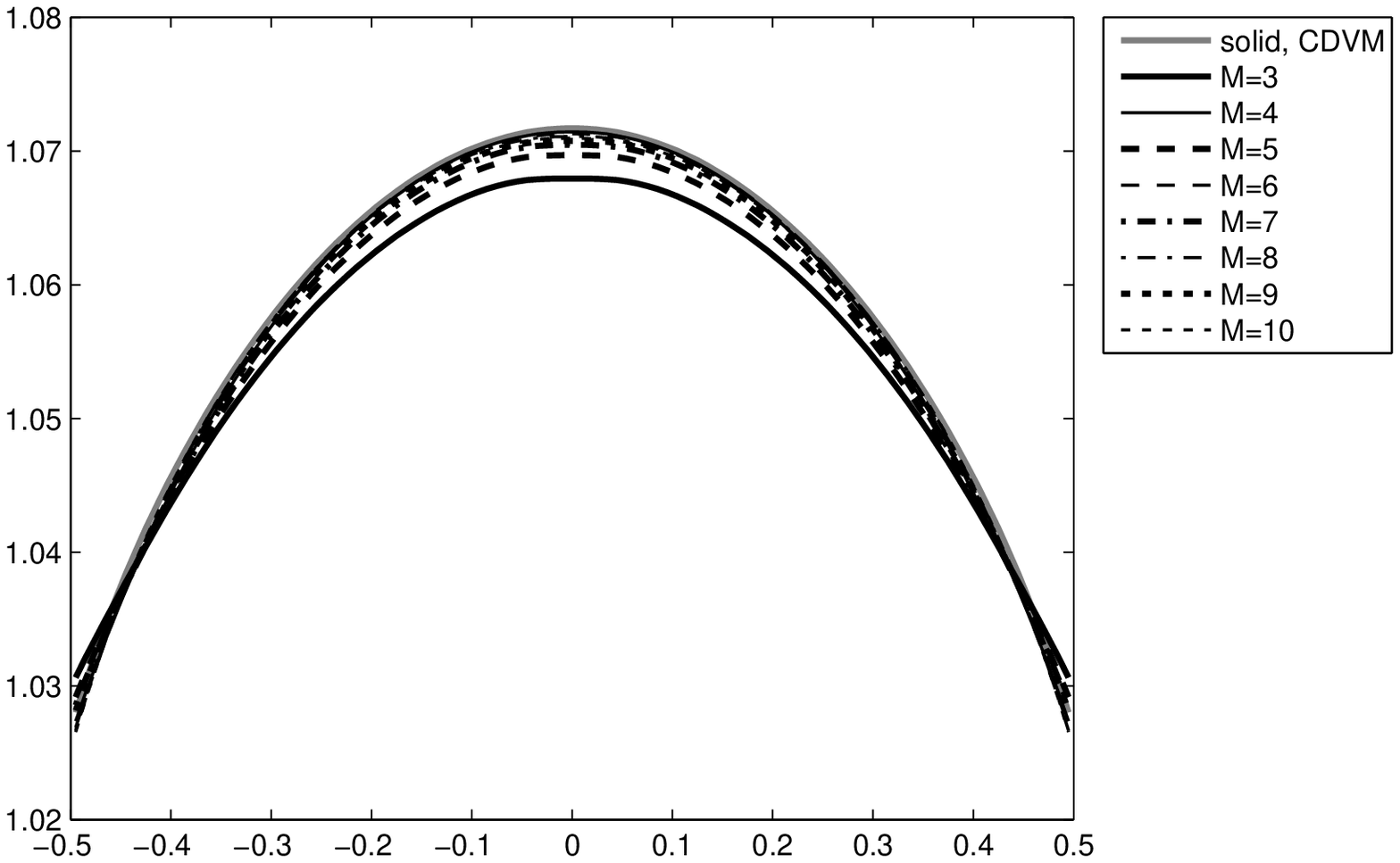}
}
\end{tabular}
\caption{Density and temperature plots for the planar Couette flow
with $\Kn=0.1$ (to be continued)}
\label{fig:rho_theta_Kn=0.1}
\end{figure}

\begin{figure}[!ht]
\centering
\setlength\subfigcapskip{10pt}
\begin{tabular}{r}
\subfigure[Shear stress, $\sigma_{12}$]{
  \includegraphics[scale=.75]{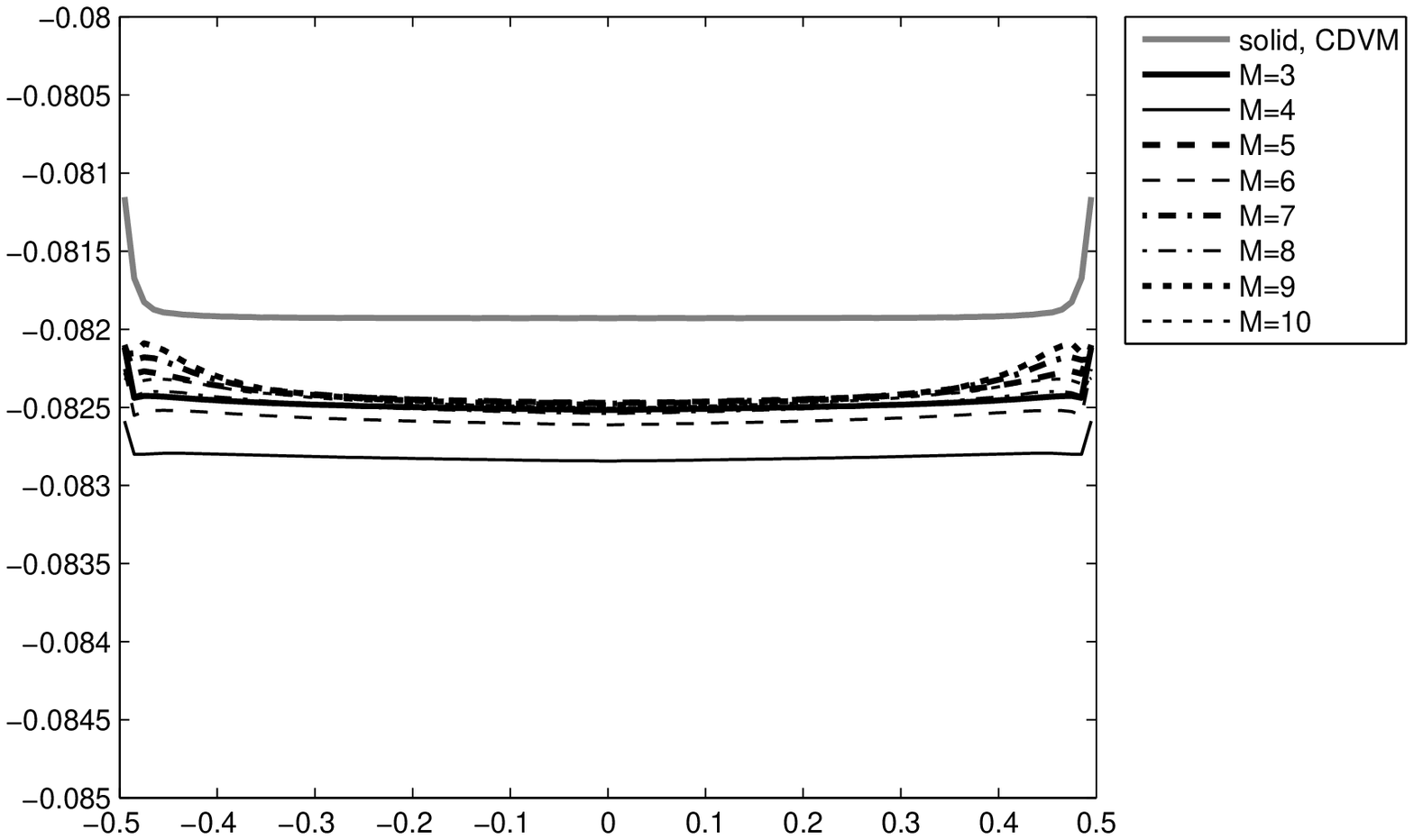}
} \\
\subfigure[Normal stress, $\sigma_{22}$]{
  \includegraphics[scale=.75]{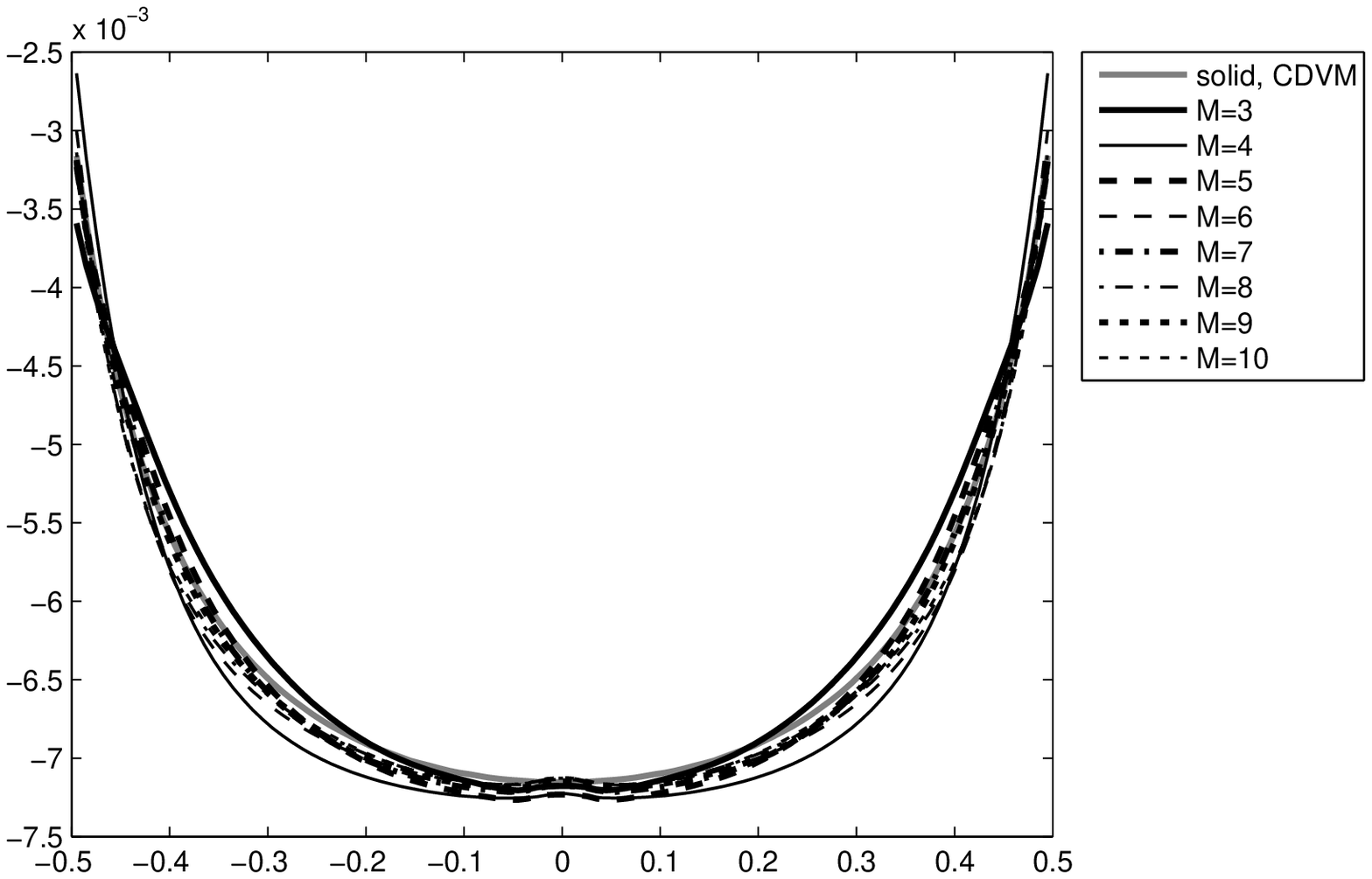}
}
\end{tabular}
\caption{Shear and normal stress plots for the planar Couette flow
with $\Kn=0.1$}
\label{fig:sigma_sigma_yy_Kn=0.1}
\end{figure}

Numerical results for $\Kn = 0.1$ are shown in Figure
\ref{fig:rho_theta_Kn=0.1} and \ref{fig:sigma_sigma_yy_Kn=0.1}. In
this case, most lines agree with each other. The convergence in the
number of moments can be observed, however, due to the numerical error
from both \NRxx method and CDVM, small deviations between the CDVM
results and the possible limit of the \NRxx method can be found. One
can disclose that lower order \NRxx results deviate from the CDVM
results more than higher order ones. This correctly reflects the
behavior of \NRxx method under low Knudsen numbers, as is also found
by \cite{NRxx}.

\begin{figure}[!ht]
\centering
\setlength\subfigcapskip{10pt}
\begin{tabular}{r}
\subfigure[Density, $\rho$]{
  \includegraphics[scale=.75]{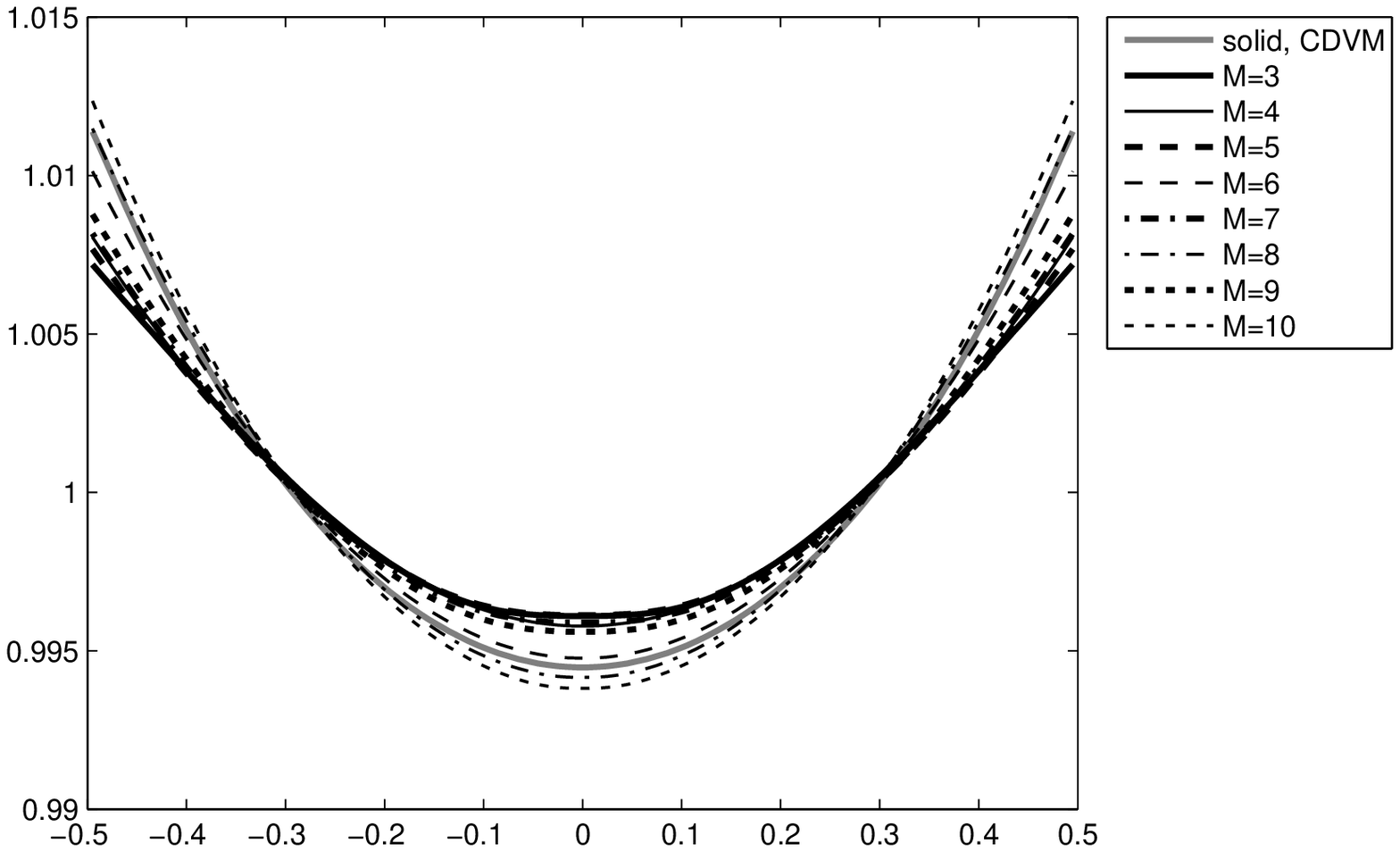}
} \\
\subfigure[Temperature, $\theta$]{
  \includegraphics[scale=.75]{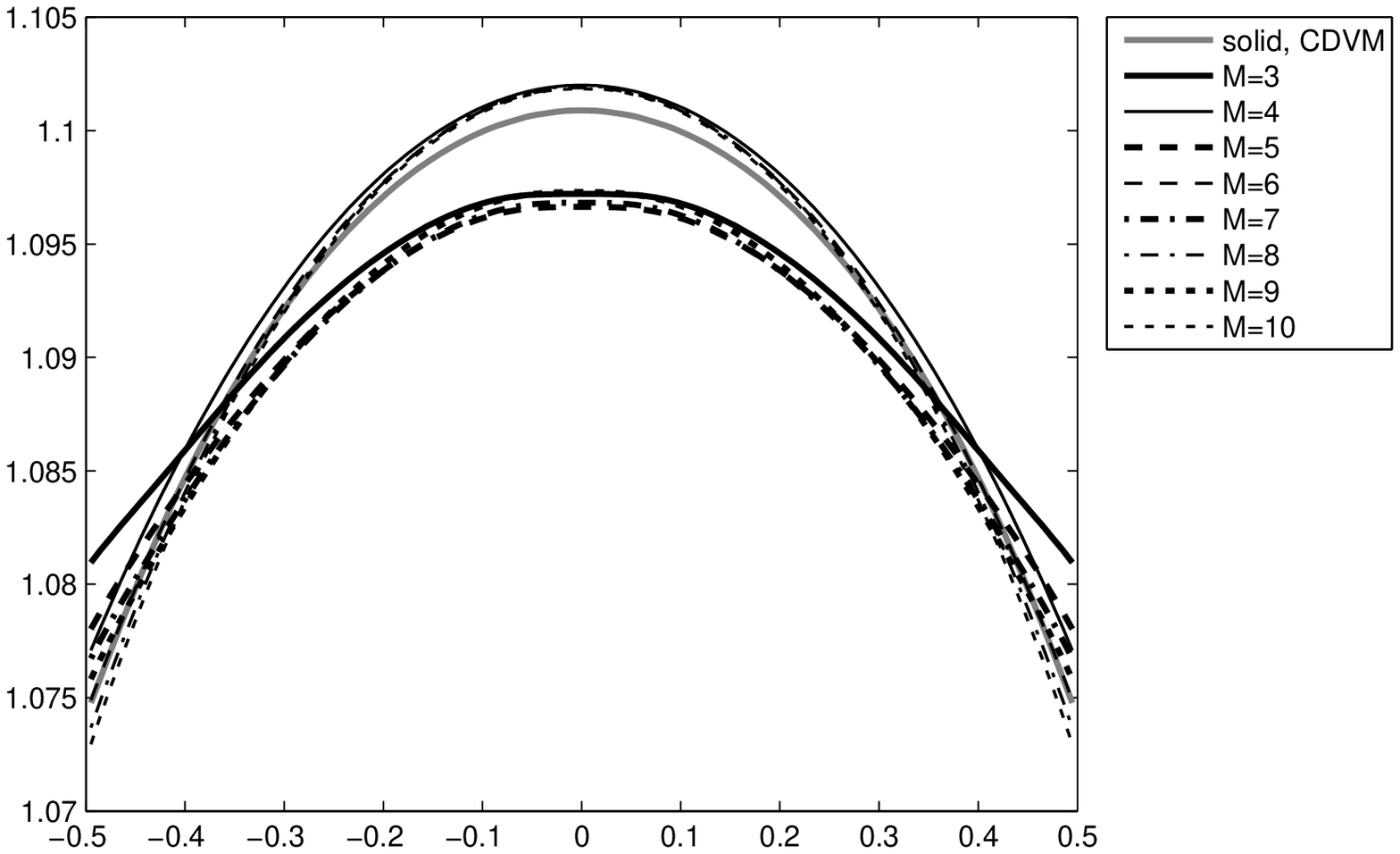}
}
\end{tabular}
\caption{Density and temperature plots for the planar Couette flow
with $\Kn=0.5$}
\label{fig:rho_theta_Kn=0.5}
\end{figure}

\begin{figure}[!ht]
\centering
\setlength\subfigcapskip{10pt}
\begin{tabular}{r}
\subfigure[Shear stress, $\sigma_{12}$]{
  \includegraphics[scale=.75]{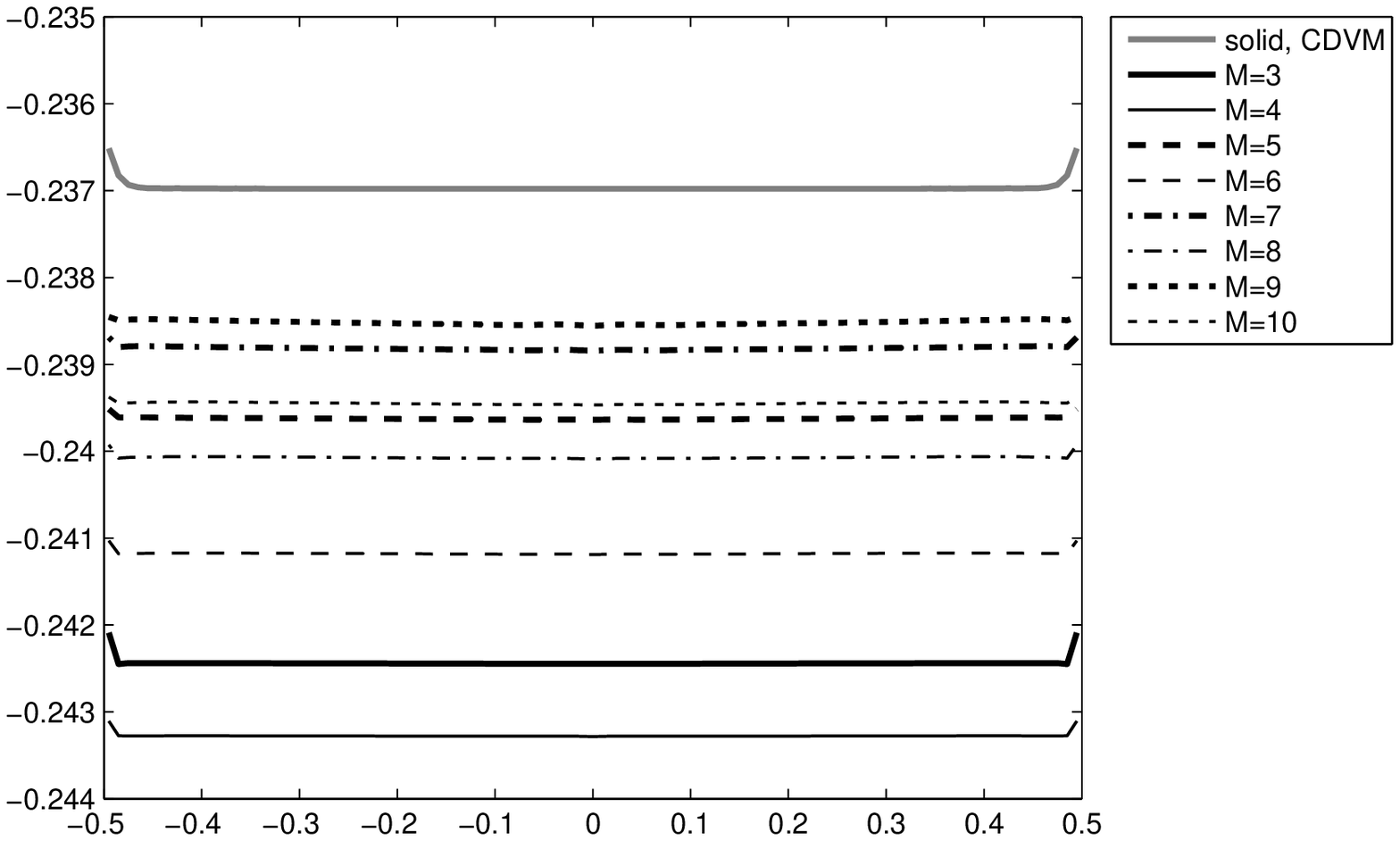}
} \\
\subfigure[Normal stress, $\sigma_{22}$]{
  \includegraphics[scale=.75]{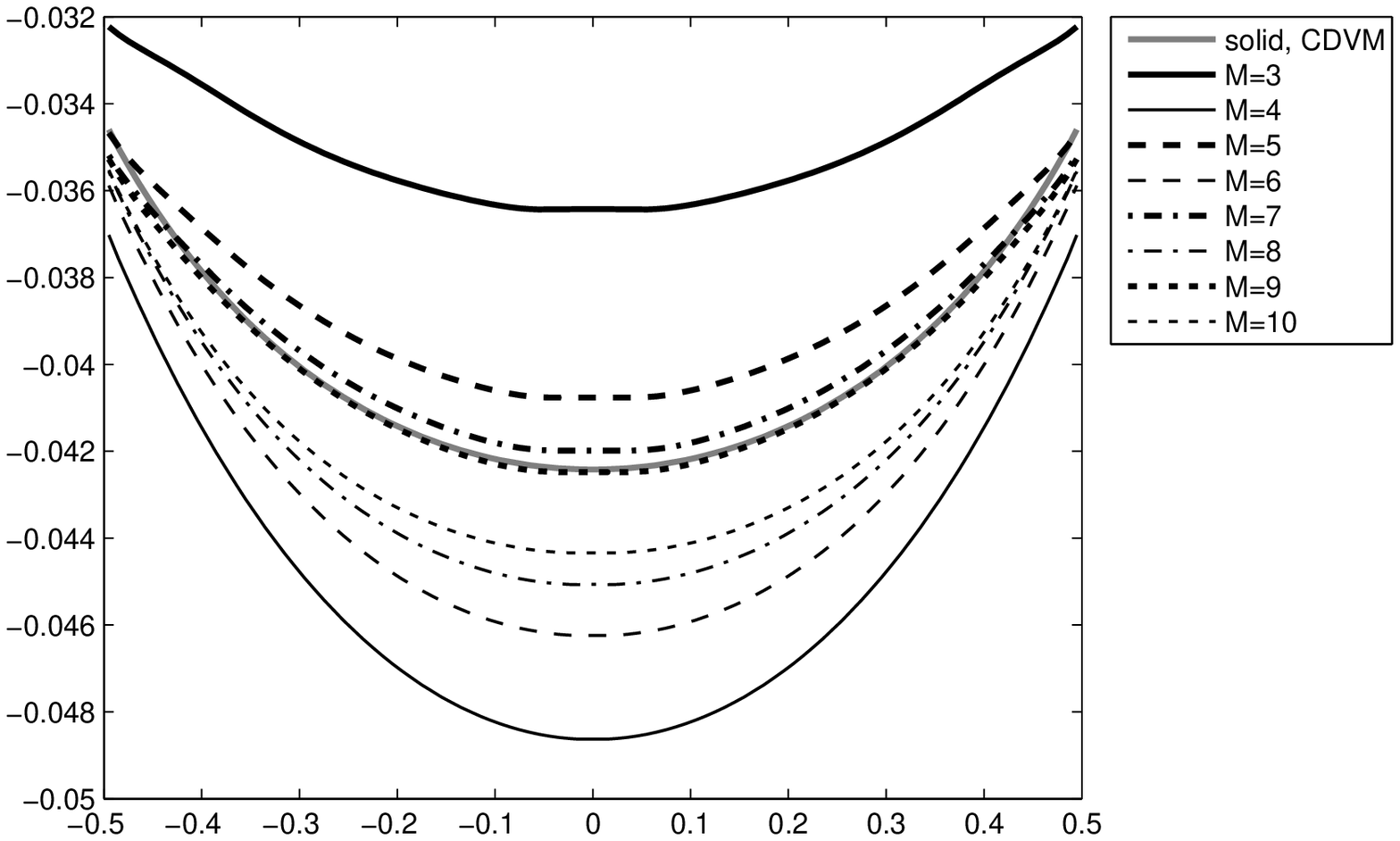}
}
\end{tabular}
\caption{Shear and normal stress plots for the planar Couette flow
with $\Kn=0.5$}
\label{fig:sigma_sigma_yy_Kn=0.5}
\end{figure}

Now a larger Knudsen number $\Kn = 0.5$ is considered, and the results
are given in Figure \ref{fig:rho_theta_Kn=0.5} and
\ref{fig:sigma_sigma_yy_Kn=0.5}. In this case, the results for odd and
even orders evidently break into two groups, and they approach closer
to the CDVM results separately. This can be also find in Figure \ref
{fig:Ref_rho_theta} and \ref{fig:Ref_sigma_q}. For $\rho$ and
$\theta$, the even group gives better results, while for $\sigma_{12}$
and $\sigma_{22}$, the odd group is more accurate. The reason remains
to be further explored. The two subfigures in Figure
\ref{fig:sigma_sigma_yy_Kn=0.5} clearly exhibit the convergence. In
\cite{Torrilhon2008, Emerson}, it was discovered that the normal
stress $\sigma_{22}$ is difficult to match by R13 and R26 equations.
Here one may find that when the number of moments is increasing, the
quality of the approximation to this quantity is improved
continuously. In case of $M=9$, the profile agrees with the CDVM
result quite well, and when $M=10$, the relative difference is below
$5\%$.

\begin{figure}[!ht]
\centering
\setlength\subfigcapskip{10pt}
\begin{tabular}{r}
\subfigure[Density, $\rho$]{
  \includegraphics[scale=.75]{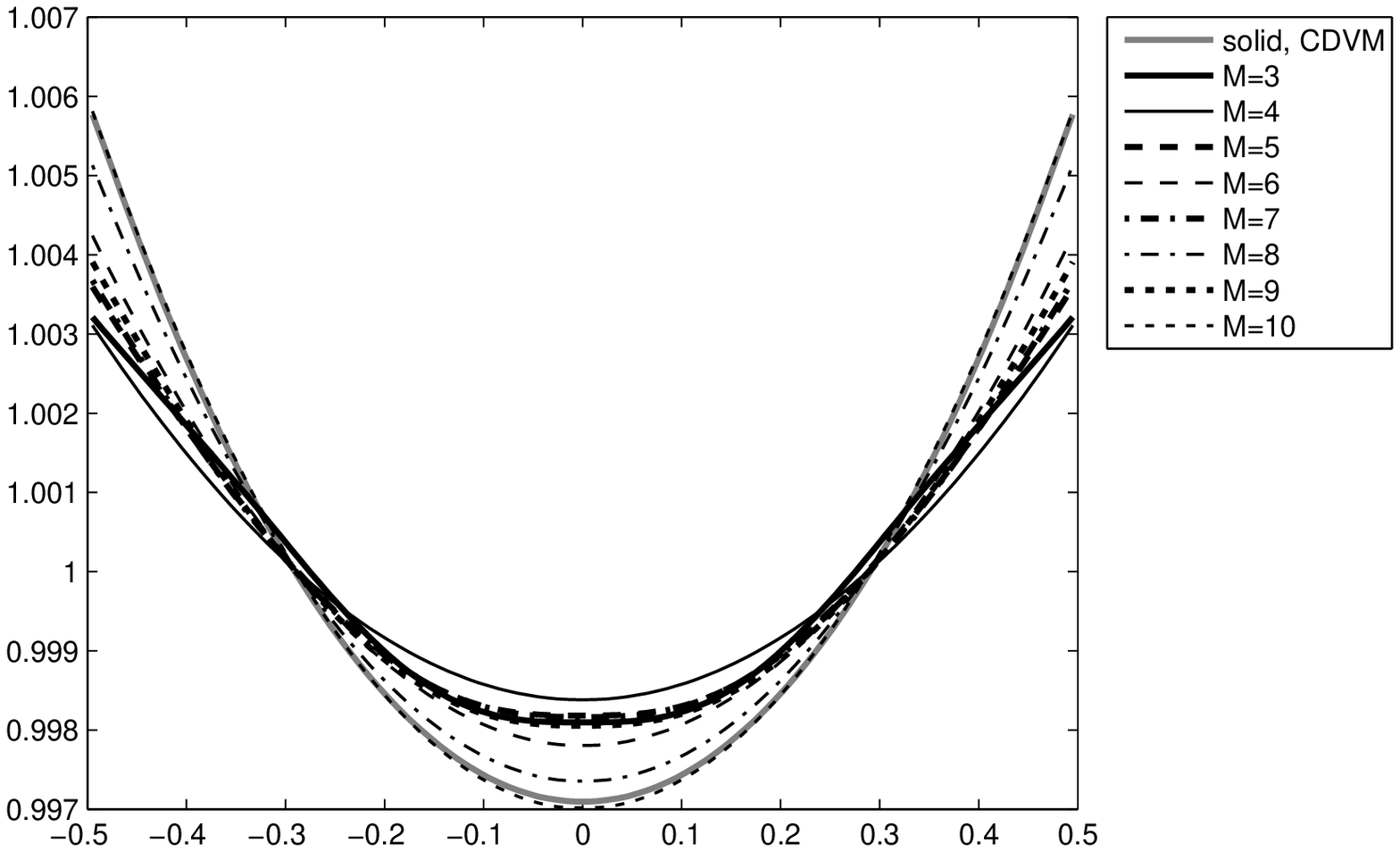}
} \\
\subfigure[Temperature, $\theta$]{
  \includegraphics[scale=.75]{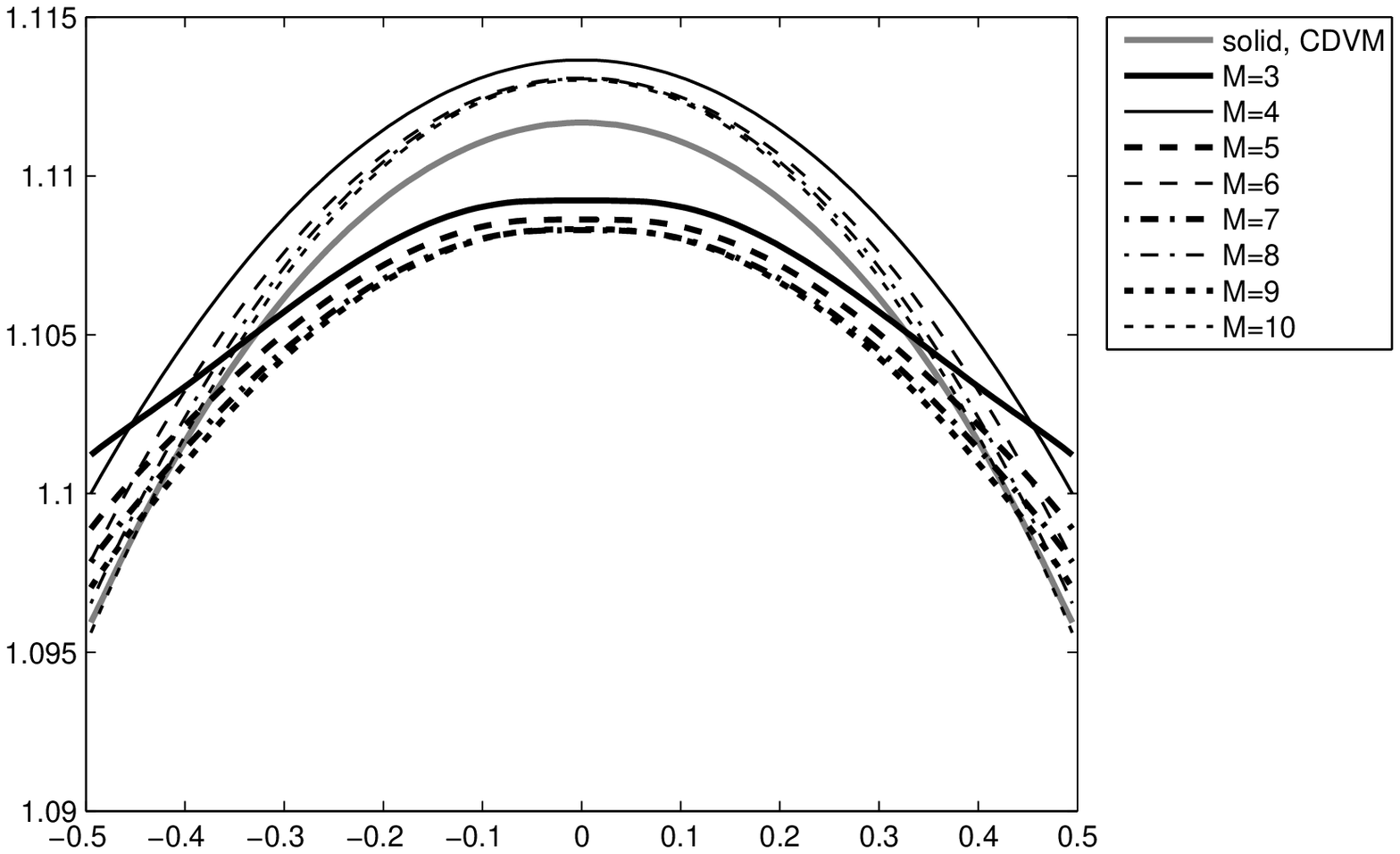}
}
\end{tabular}
\caption{Density and temperature plots for the planar Couette flow
with $\Kn=1.0$}
\label{fig:rho_theta_Kn=1.0}
\end{figure}

\begin{figure}[!ht]
\centering
\setlength\subfigcapskip{10pt}
\begin{tabular}{r}
\subfigure[Shear stress, $\sigma_{12}$]{
  \includegraphics[scale=.75]{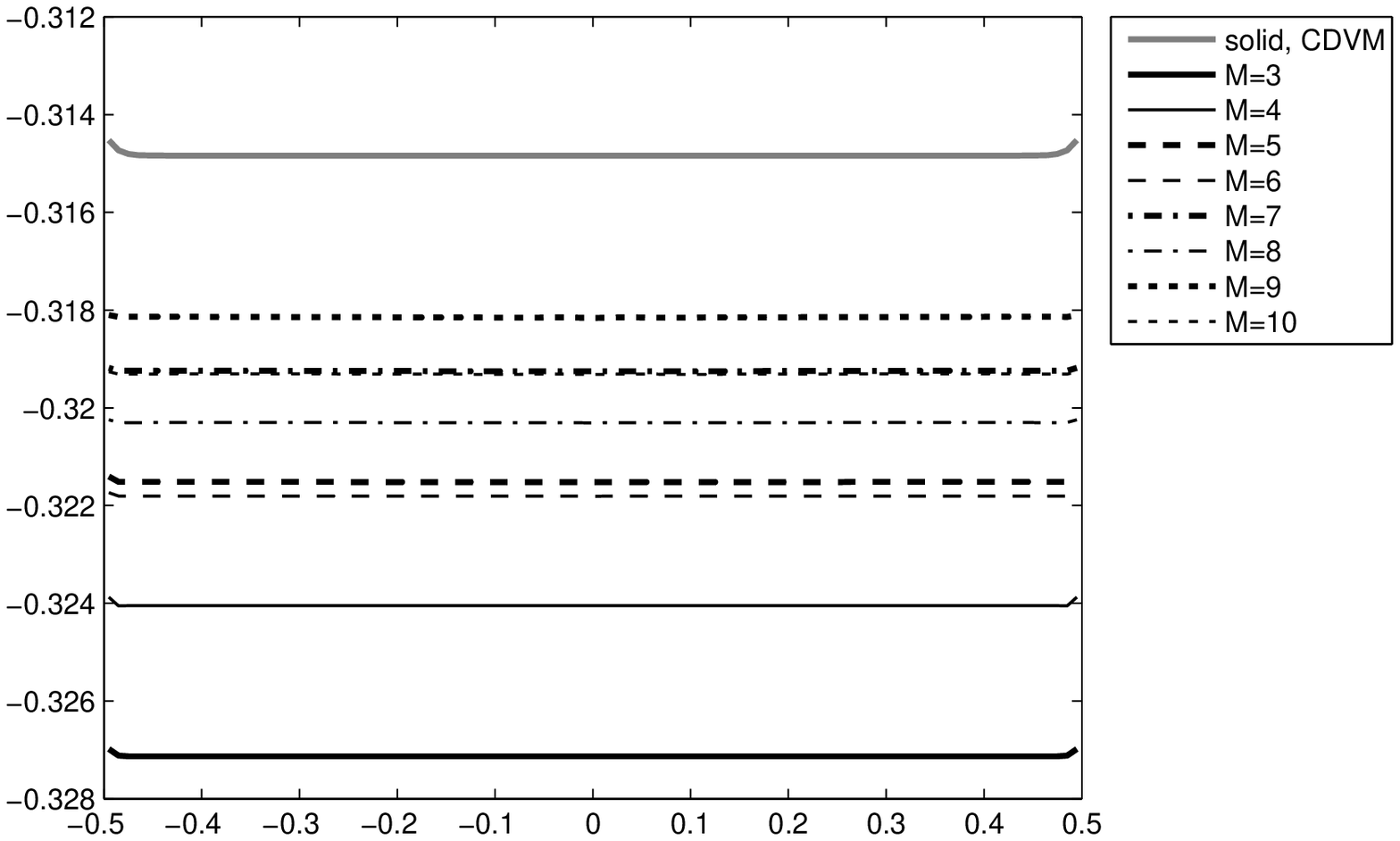}
} \\
\subfigure[Normal stress, $\sigma_{22}$]{
  \includegraphics[scale=.75]{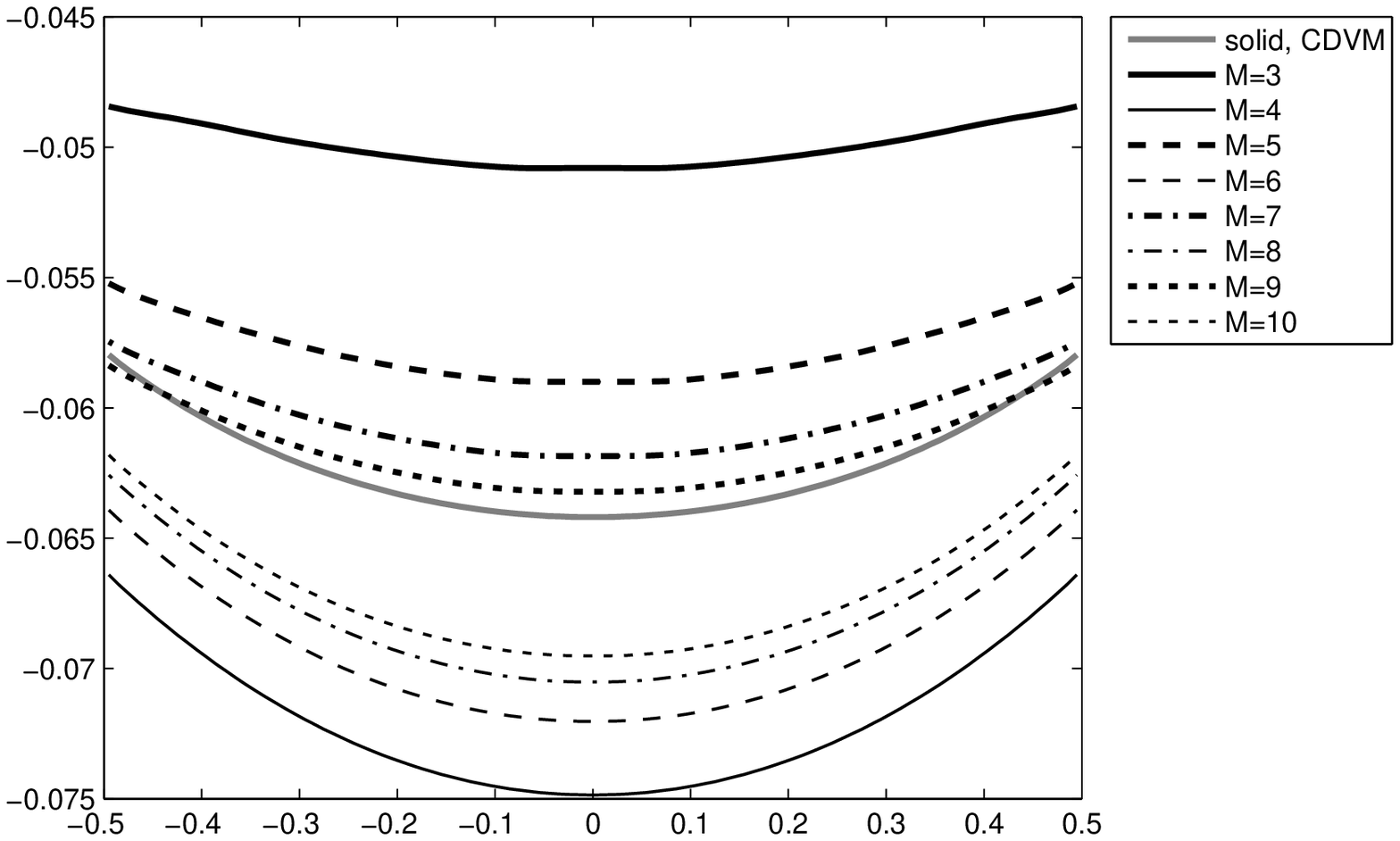}
}
\end{tabular}
\caption{Shear and normal stress plots for the planar Couette flow
with $\Kn=1.0$}
\label{fig:sigma_sigma_yy_Kn=1.0}
\end{figure}

The severe case $\Kn=1.0$ is also studied. Similar results with the
case $\Kn=0.5$ are obtained in Figure \ref{fig:rho_theta_Kn=1.0} and
\ref{fig:sigma_sigma_yy_Kn=1.0}, while the magnitude of the difference
is much larger. For $\sigma_{22}$, now the relative difference for
$M=9$ is about $10\%$. But the rate of convergence is still
encouraging --- compared with the result with $M=4$, the error is
halved.

\psfrag{M=3}{\footnotesize{$M=3$}}
\psfrag{M=4}{\footnotesize{$M=4$}}
\psfrag{M=5}{\footnotesize{$M=5$}}
\psfrag{M=6}{\footnotesize{$M=6$}}
\psfrag{M=7}{\footnotesize{$M=7$}}
\psfrag{M=8}{\footnotesize{$M=8$}}
\psfrag{M=9}{\footnotesize{$M=9$}}
\psfrag{M=10}{\footnotesize{$M=10$}}
\psfrag{solid, CDVM}{\footnotesize{CDVM}}
\psfrag{circle, DSMC}{\footnotesize{DSMC}}




\subsection{Force-driven Poiseuille flow}
This is another example which is frequently used to verify the
boundary conditions of moment methods \cite{Torrilhon2008, Emerson}.
Similar with the Couette flow, the gas also lies between two parallel
plates, but the plates are stationary and an external constant force
parallel to the plates causes the flow to reach a non-stationary
steady state. In our settings, the computational domain is again
$[-0.5, 0.5]$, and the Knudsen number is set to be $0.1$. The force
introduces an acceleration
\begin{equation}
\bF = (0.2555, 0, 0)^T.
\end{equation}
The initial condition is the same as the Couette flow. These settings
are the non-dimensional form of the test in \cite{Garcia}, where the
DSMC result is carried out, and this example is also considered in
\cite{Xu2007, Emerson}. Since it is quite difficult for us to exert
the force term in CDVM, we have to use the DSMC result in
\cite{Garcia} for comparison in spite of the difference in the
collision model.

\begin{figure}[p]
\centering
\setlength\subfigcapskip{10pt}
\begin{tabular}{r}
\subfigure[Density, $\rho$]{
  \includegraphics[scale=.6]{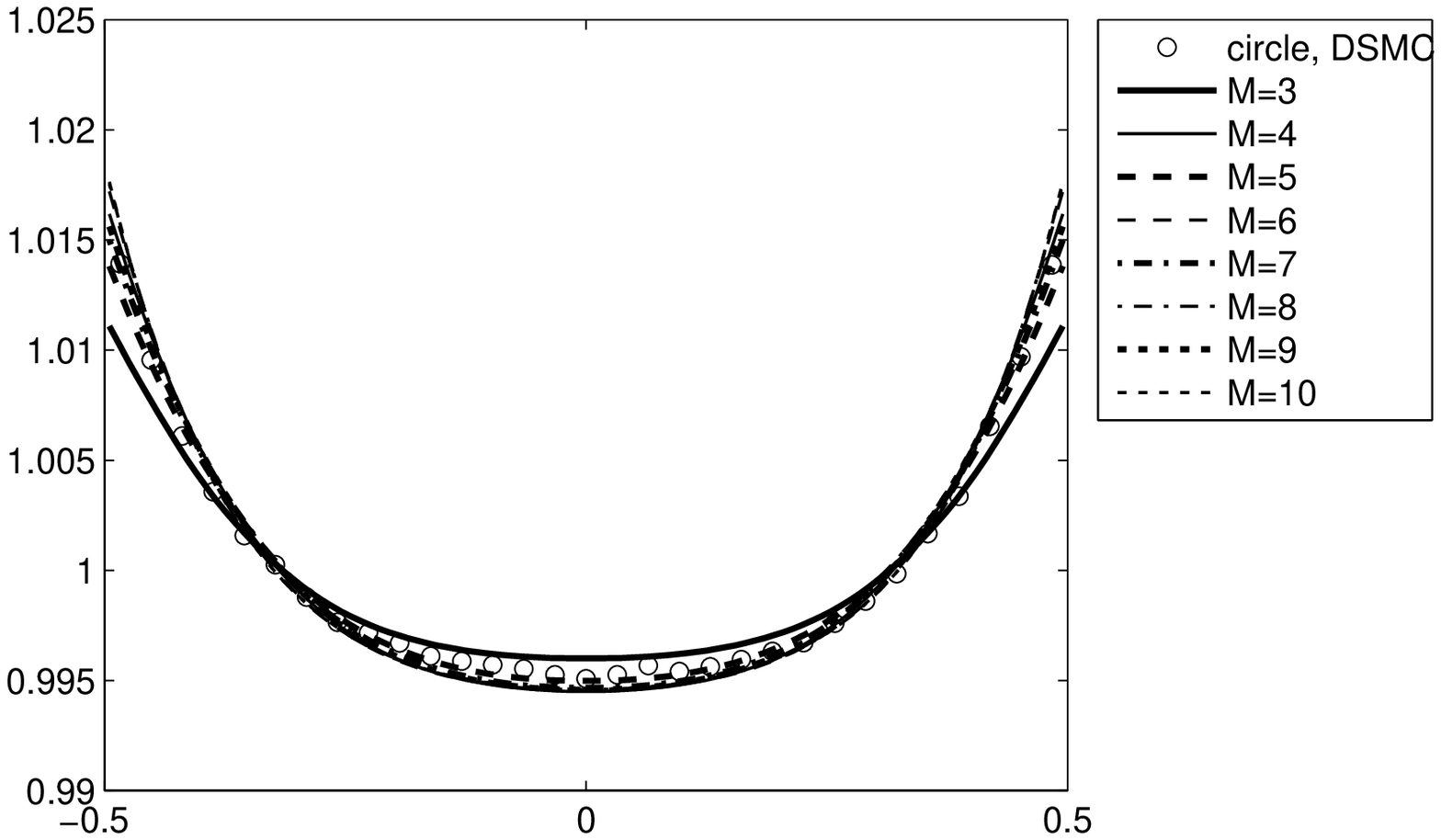}
} \\
\subfigure[Velocity, $u_2$]{
  \includegraphics[scale=.6]{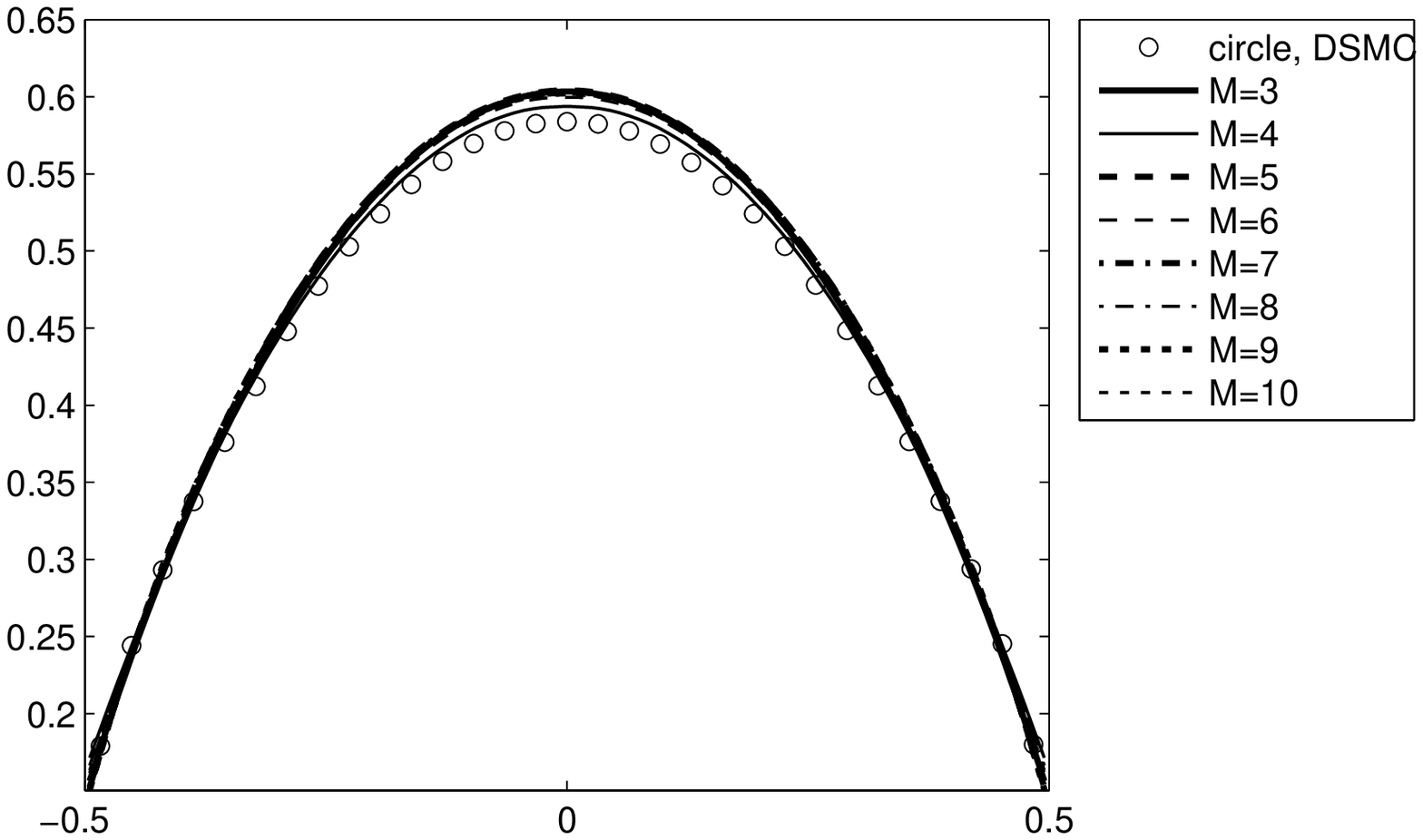}
} \\
\subfigure[Temperature, $\theta$]{
  \label{fig:theta}
  \includegraphics[scale=.6]{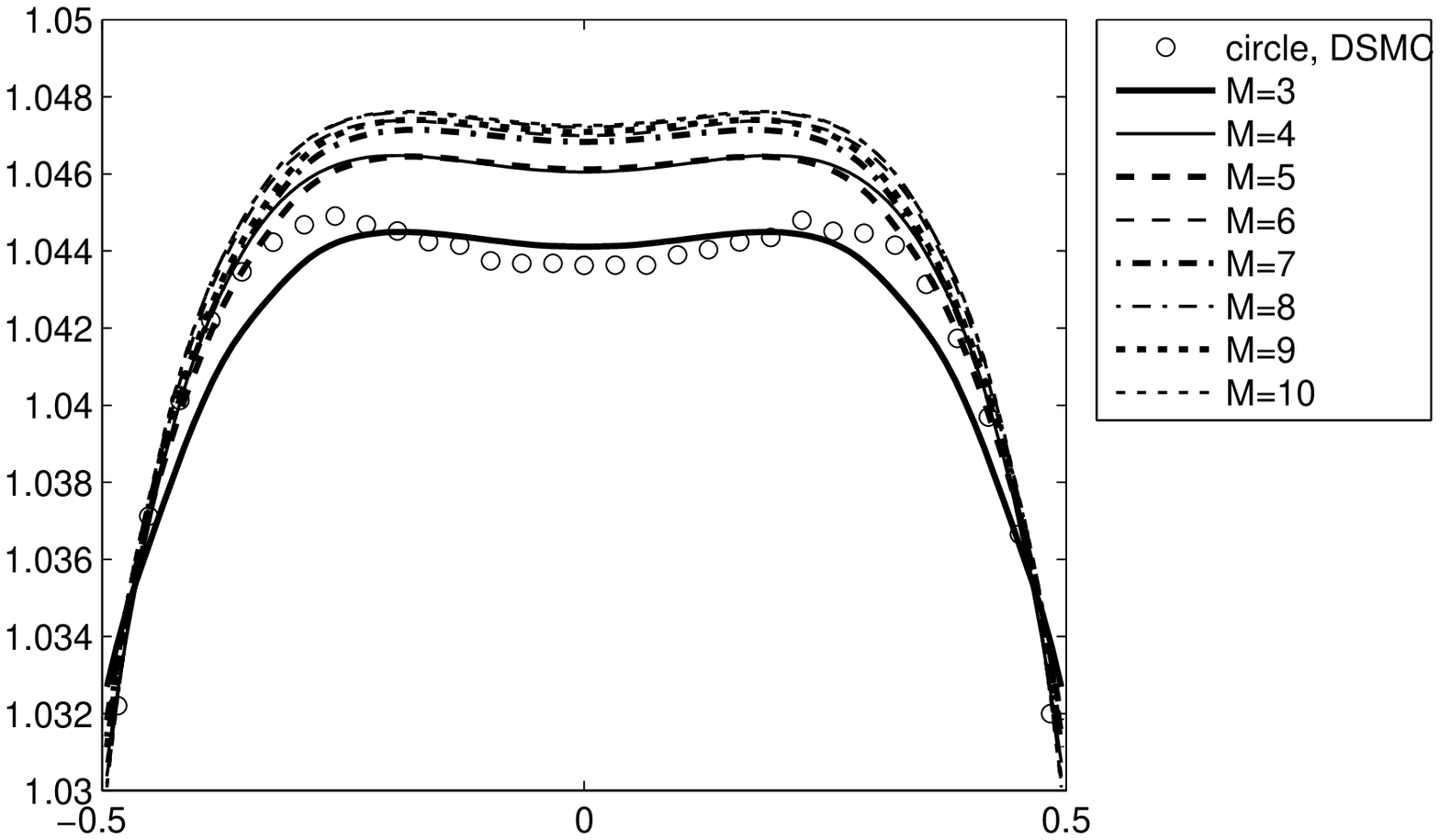}
}
\end{tabular}
\caption{Density, velocity and temperature plots for the planar
Poiseuille flow}
\label{fig:Poiseuille1}
\end{figure}

\begin{figure}[p]
\centering
\setlength\subfigcapskip{10pt}
\begin{tabular}{r}
\subfigure[Shear stress, $\sigma_{12}$]{
  \includegraphics[scale=.6]{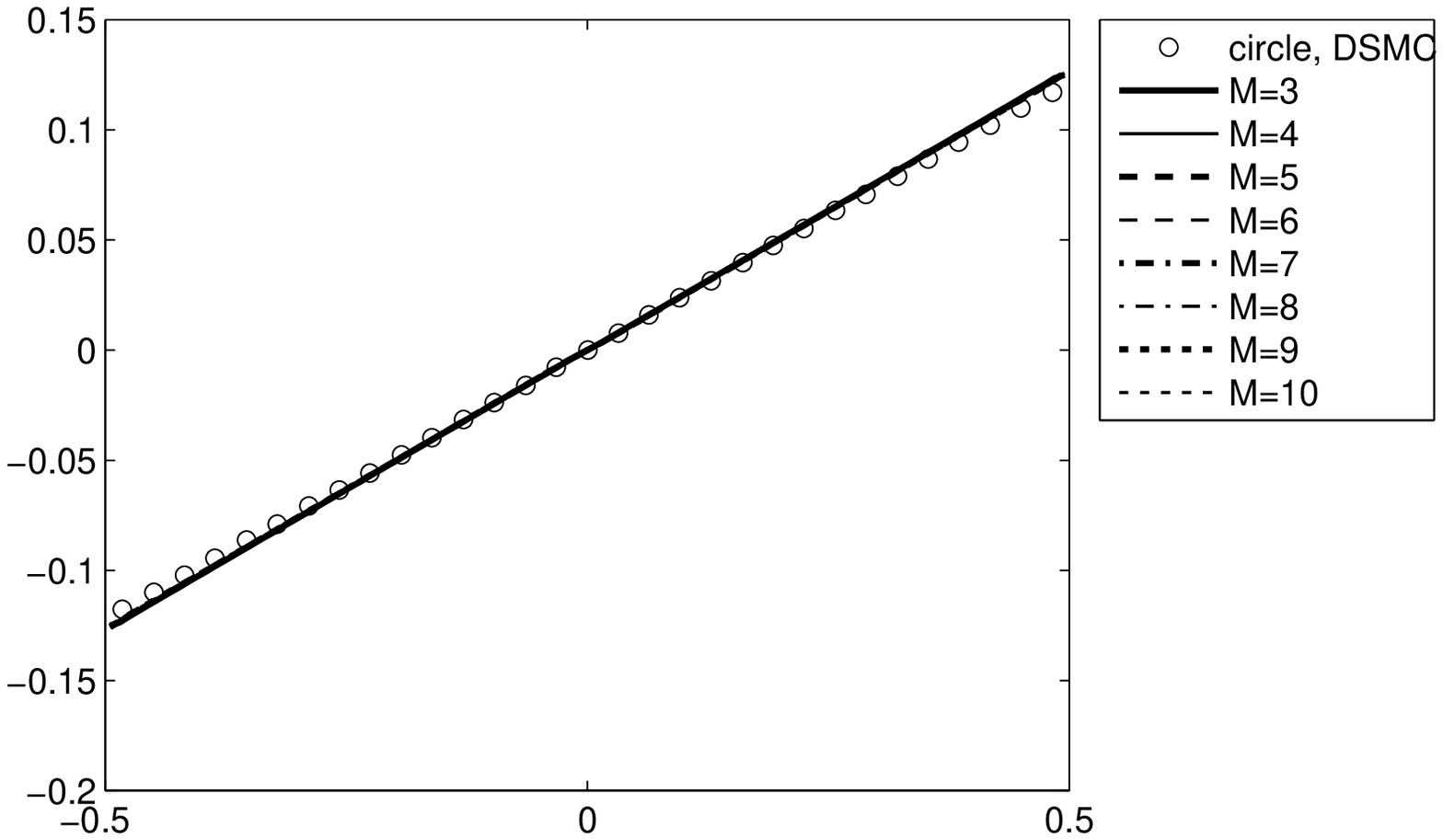}
} \\
\subfigure[Normal stress, $\sigma_{22}$]{
  \includegraphics[scale=.6]{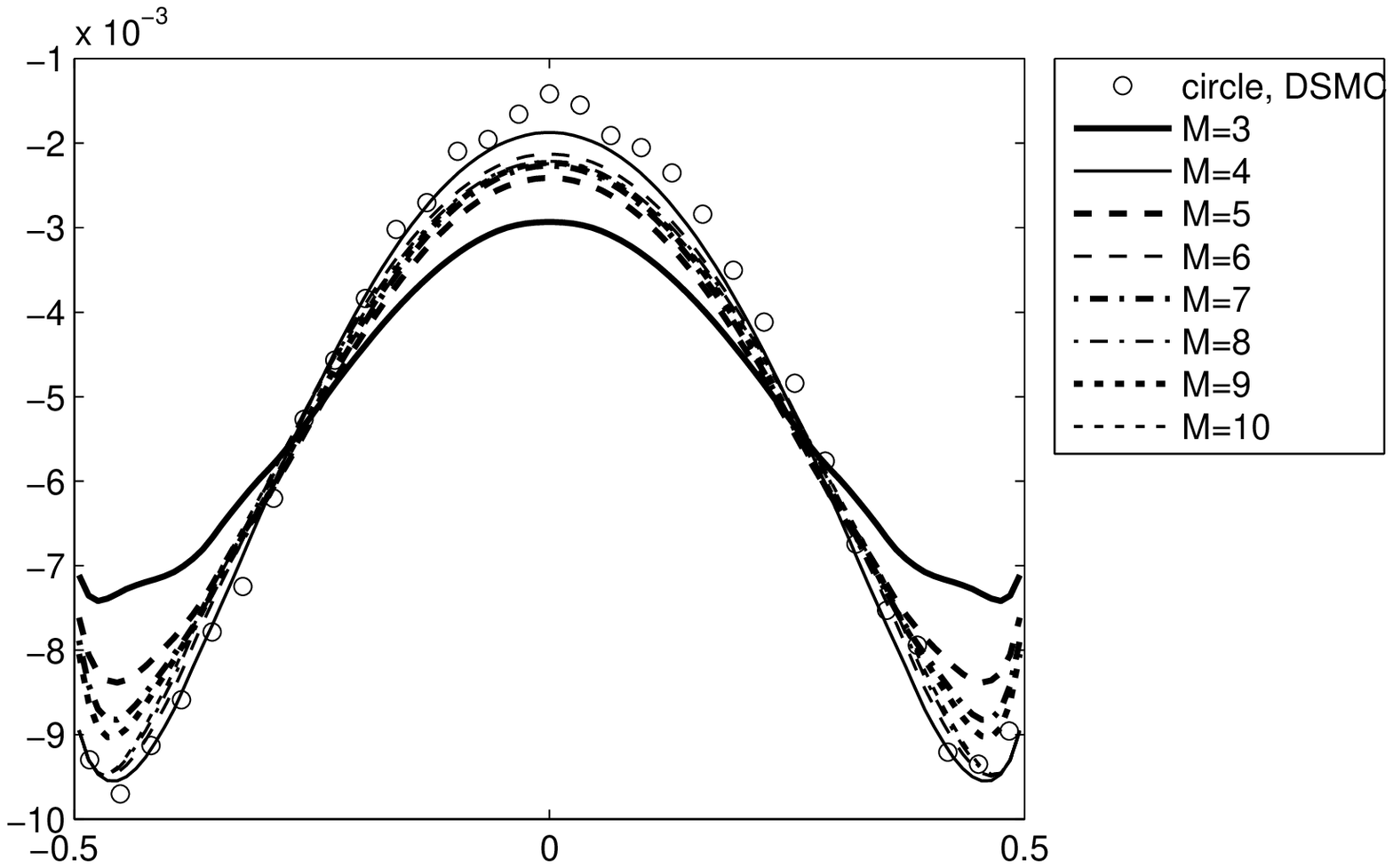}
} \\
\subfigure[Heat flux, $q_1$]{
  \includegraphics[scale=.6]{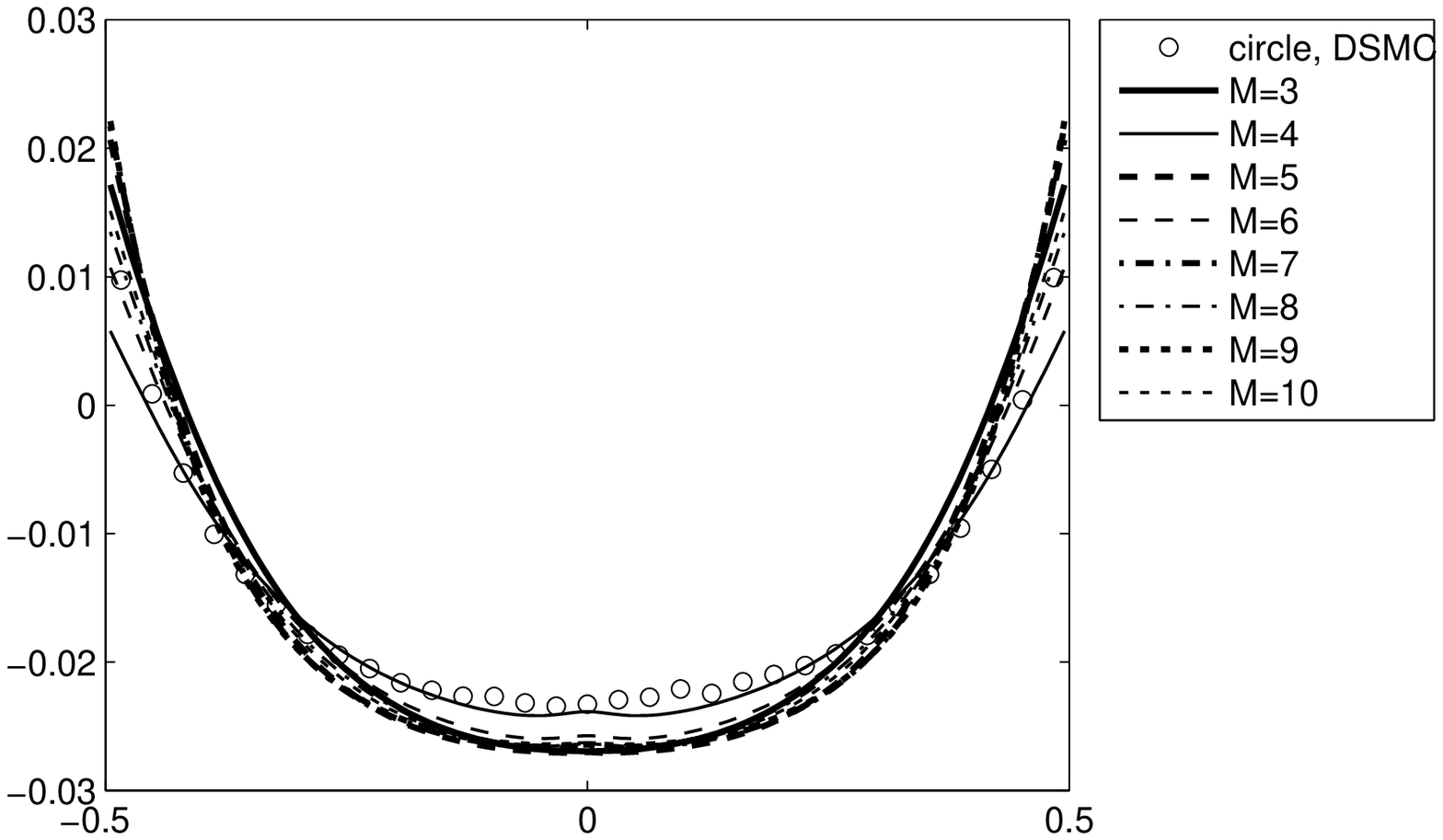}
}
\end{tabular}
\caption{Stress and heat flux plots for the planar Poiseuille flow}
\label{fig:Poiseuille2}
\end{figure}

The numerical results are presented in Figure \ref{fig:Poiseuille1}
and \ref{fig:Poiseuille2}. For all the profiles, the convergence in
the number of moments is legible, while the \NRxx results do not
converge to the results of DSMC. This may due to the difference
between the collision terms of Shakhov model and DSMC. Taking the
temperature plot (Figure \ref {fig:theta}) as an example, the result
of $M=3$ matches DSMC result best, since when $M=3$, the collision
term of Shakhov model is almost the same as that of DSMC. While when
the number of moments increases, the collision term deviates away from
DSMC's gradually. Here the accuracy of collision models is not the
topic of this paper. Even though, two results are very close
quantitatively, which indicates the correctness of the boundary
conditions and the Prandtl number of the \NRxx method.


%%% Local Variables: 
%%% mode: latex
%%% TeX-master: "article"
%%% End: 

% vim: tw=70:spell
\section{Some discussions on the \bNRxx method} \label{sec:discussion}
\subsection{Order of accuracy}
For the macroscopic equations, a basic quantity describing its ability
is the \emph{order of accuracy} with respect the Knudsen number. The
definition of \emph{order of accuracy} can be found in textbook \cite
{Struchtrup}:
\begin{quote} \it%
A set of equations is said to be accurate of order $\lambda_0$, when
the pressure deviator $\sigma_{ij}$ and the heat flux $q_i$ are known
within the order $O(\varepsilon^{\lambda_0})$.
\end{quote}
Here $\varepsilon$ is a small parameter proportional to the relaxation
time $\tau$. As we have discussed in Remark \ref{rem:reg}, in the
view of order of magnitude, the process of Maxwellian iteration for
the Shakhov model is identical to the BGK model. Hence, for an
arbitrary $M \geqslant 3$, the leading order term of $f_{\alpha}$ with
$|\alpha| = M + 1$ is known from the corresponding moment equations
(see \cite{NRxx_new} for details). And it has been deduced in \cite
{NRxx_new} that $f_{\alpha} \sim O(\tau^{\lceil |\alpha| / 3 \rceil})$
for all $|\alpha| \geqslant 4$. Thus, from the analytical form of the
moment equations \eqref{eq:mnt_system}, we immediately have that
$f_{\alpha}$ with $|\alpha| = M$ is known up to $(\lceil (M + 1) / 3
\rceil + 1)$-th order. Subsequently, $f_{\alpha}$ with $|\alpha| = M -
1$ is know up to $(\lceil (M + 1) / 3 \rceil + 2)$-th order, and this
can be done until $|\alpha| = 2$. The general result is
\begin{proposition}
For the moment equations described in Section \ref{sec:gov_eq},
$f_{\alpha}$ has $(\lceil 4(M + 1) / 3 \rceil - |\alpha|)$-th order
accuracy if $2 \leqslant |\alpha| \leqslant M$.
\end{proposition}
\noindent Now, using \eqref{eq:low_order_moments} and the definition
of order of accuracy, we conclude that the \NRxx equations have the
order of accuracy $\lceil (4M - 5) / 3 \rceil$.

For boundary value problems, such discussion is only valid in the
bulk. In the Knudsen layer, which is known to be of width $O(\Kn)$,
we need to use $X = x / \Kn$ as the spatial variable while
investigating the accuracy of moment equations. In this case, if we
consider a steady state problem, the small parameter no longer
appears in the governing equations \eqref{eq:mnt_system}. This means
the order of magnitude for $f_{\alpha}$ does not increase as
$|\alpha|$ increases, as has been found in \cite{Struchtrup2007}.

\subsection{The validity of \bNRxx method for large Knudsen number and
in the Knudsen layer}
\begin{figure}[!ht]
\psfrag{M   =   3}{\footnotesize{$M=3$}}
\psfrag{M   =   6}{\footnotesize{$M=6$}}
\psfrag{M   =   9}{\footnotesize{$M=9$}}
\centering
\subfigure[$\Kn=0.1$]{
  \includegraphics[scale=.6]{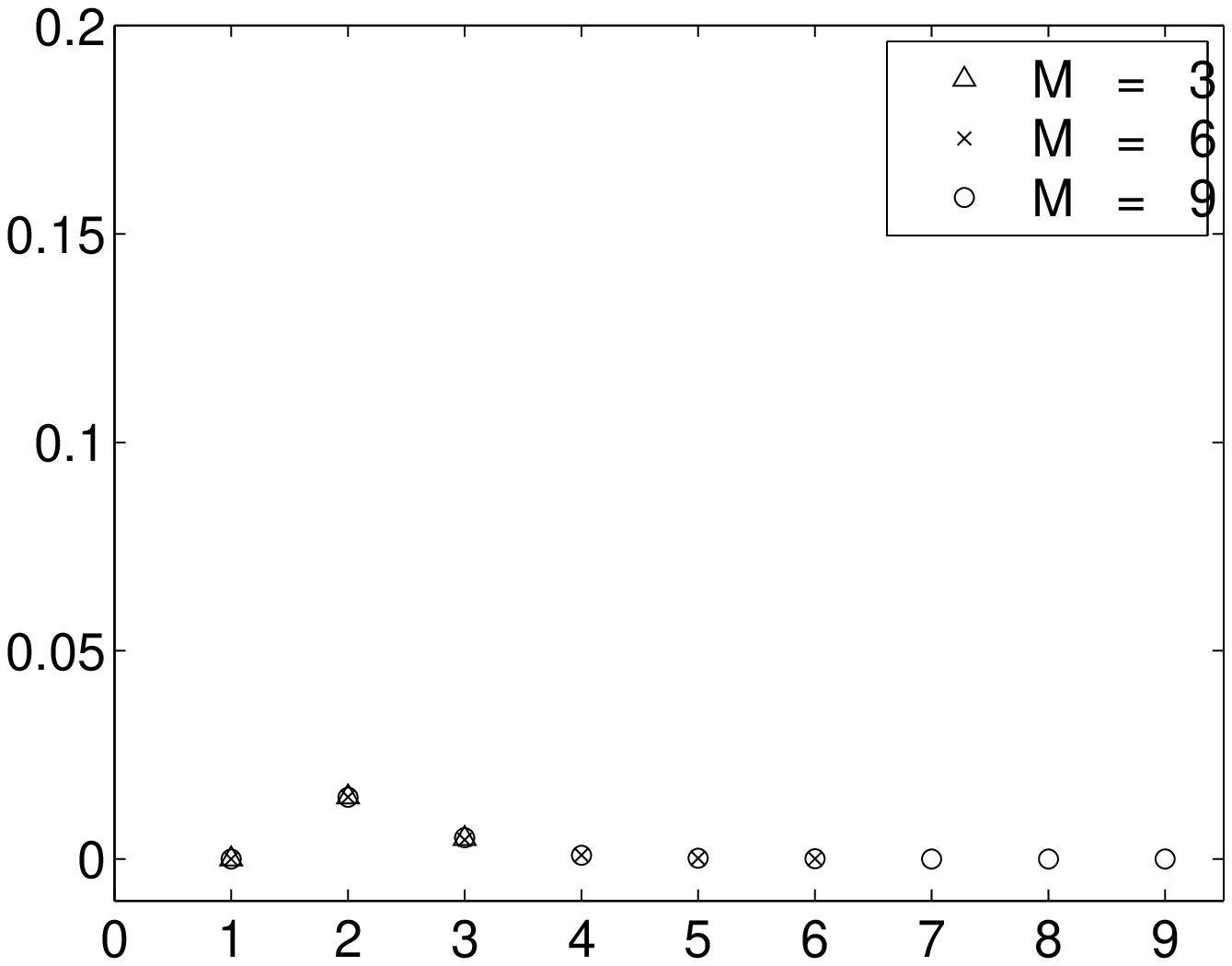}
} \\
\subfigure[$\Kn=1.0$]{
  \includegraphics[scale=.6]{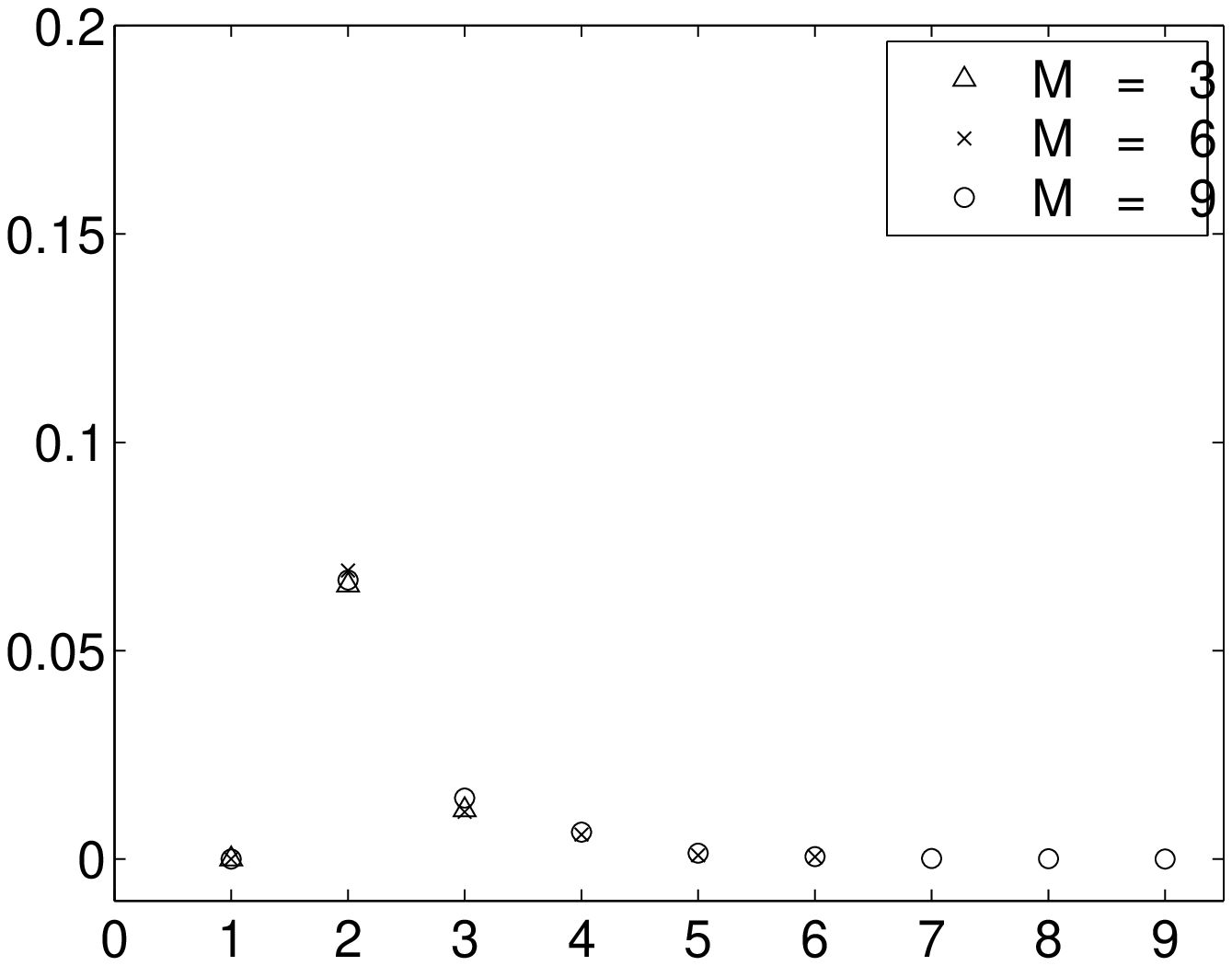}
} \\
\subfigure[$\Kn=10.0$]{
  \includegraphics[scale=.6]{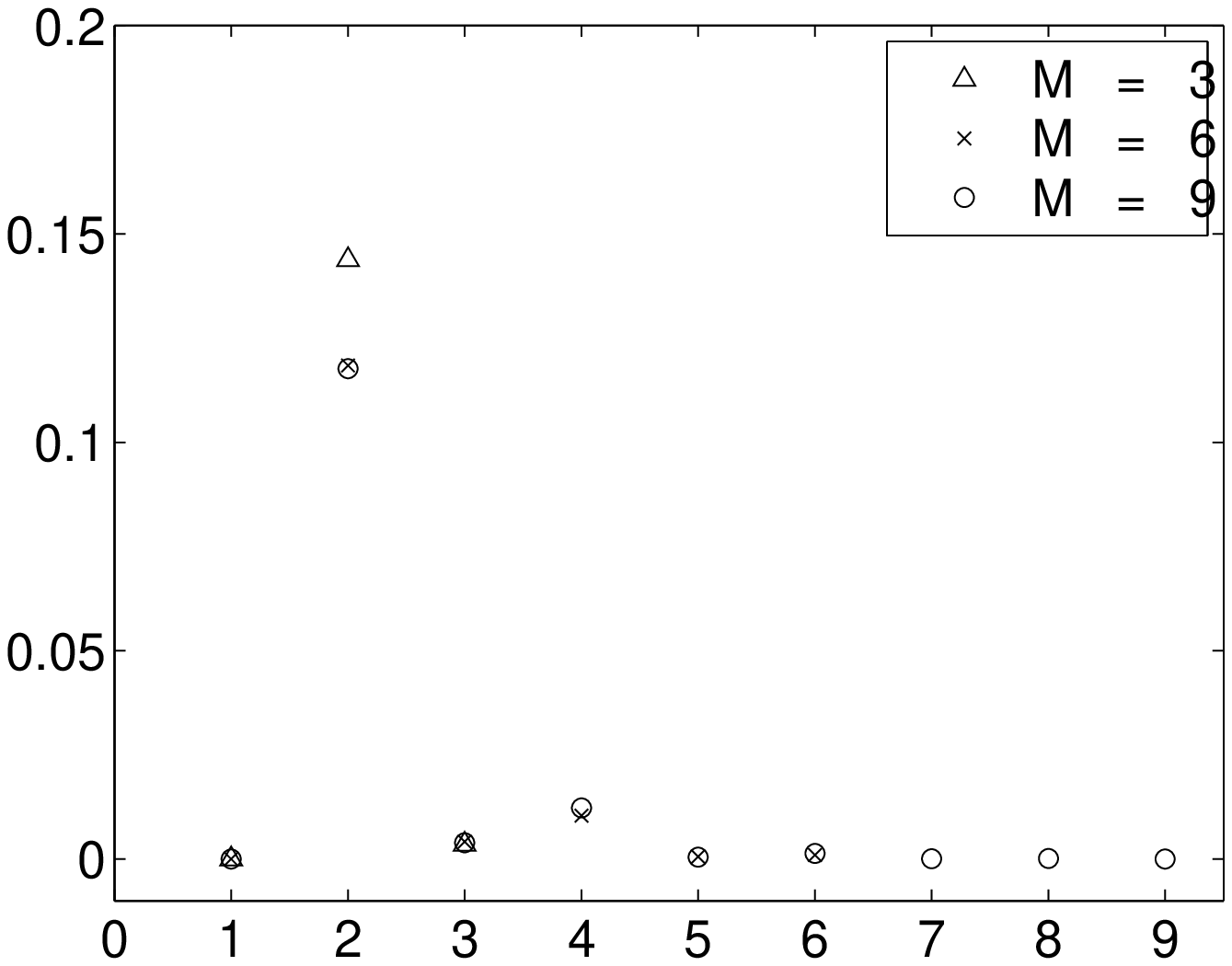}
}
\caption{Average values of $\{|f_{\alpha}|\}_{|\alpha|=k}$ for $k=1$
to $9$. The results are based the \NRxx solutions of the Couette
flow.}
\label{fig:avg_abs_value}
\end{figure}

As we have discussed above, there are two cases when the order of
accuracy is not so meaningful for describing the accuracy of the \NRxx
method:
\begin{enumerate}
\item In the case of $\Kn \sim O(1)$, there are no ``small
  parameters'' in our concept.
\item In the Knudsen layer, the orders of magnitude of moments do not
  increase as they behave in the bulk.
\end{enumerate}
Nevertheless, we can still consider the \NRxx method as a solver for
Boltzmann equation with spectral expansion in the velocity space, and
the method should be valid when $f_{\alpha}$ decays sufficiently fast
as $|\alpha|$ increases. Now we follow \cite{Struchtrup2007} and give
the average absolute values of the moments with the same order for
different $\Kn$ and different $M$ in Figure \ref{fig:avg_abs_value}.
The result is based on the Couette flow problem in Section
\ref{sec:Couette}, and the \NRxx solution at $x = -0.5$ is used in the
these plots.

In these figures, we find that even when the Knudsen number is as
large as $10$, the magnitudes of moments still decay very fast. Thus,
the \NRxx method can still be considered to be valid and efficient.
Although the methodology of regularization which is based on a small
$\tau$ is not valid any more, the regularization term \eqref{eq:reg}
is simply a prediction of higher order moments. Such prediction
differs from Grad equations' guess $f_{\alpha} = 0$, but also has a
uniform expression for all Knudsen numbers. When $M$ goes to infinity,
the regularization term is expected to vanish since $f_{\alpha}$
decays.  On the other hand, this term smooths the profiles of the
macroscopic variables, thus avoids the appearance of some unphysical
phenomena such as subshocks (see \cite{NRxx_new}). This indicates the
meaningfulness of regularization for practical use.

% vim: tw=70:spell
\section{Concluding remarks} \label{sec:conclusion} A uniform
numerical scheme for coupling the \NRxx method and the wall boundary
conditions are developed in this paper, and the \NRxx method is
extended to apply the force term and predict correct Prandtl number by
using the Shakhov collision model. To validate the proposed method,
both steady and unsteady problems are simulated. We are currently
working on applying the \NRxx method to 2D problems.

\section*{Acknowledgements}
We thank Dr. Vladimir Titarev for some useful discussions on the
implementation of conservative discrete velocity method for Shakhov
collision model. The research of the second author was supported in
part by the National Basic Research Program of China (2011CB309704),
the National Science Foundation of China under grant 10731060 and NCET
in China. Z. Qiao is partially supported by the Hong Kong RGC grant
HKBU201710.

%%% Local Variables: 
%%% mode: latex
%%% TeX-master: "article"
%%% End: 

% vim: tw=70:spell
\section*{Appendix}

\appendix

\section{Some properties of Hermite polynomials} \label{sec:Hermite}
The Hermite polynomials defined in \eqref{eq:He} are a set of
orthogonal polynomials over the domain $(-\infty, +\infty)$. Their
properties can be found in many mathematical handbooks such as \cite
{Abramowitz}. Some useful ones are listed below:
\begin{enumerate}
\item Orthogonality: $\displaystyle \int_{\bbR} \He_m(x)
\He_n(x) \exp(-x^2/2) \dd x = m! \sqrt{2\pi} \delta_{m,n}$;
\item Recursion relation: $\He_{n+1}(x) = x \He_n(x) - n \He_{n-1}(x)$;
\item Differential relation: $\He_n'(x) = n \He_{n-1}(x)$.
\end{enumerate}
And the following equality can be derived from the last two relations:
\begin{equation}
[\He_n(x) \exp(-x^2/2)]' = -\He_{n+1}(x) \exp(-x^2/2).
\end{equation}

\section{Calculation of half-space integration} \label{sec:hs_int}
The detailed calculation of $I_{\alpha,\beta}(\theta)$ \eqref
{eq:I_def} will be presented. Using the definition of
$\mathcal{H}_{\theta,\alpha}(\bv)$ \eqref{eq:H}, eq. \eqref {eq:I_def}
can be rewritten as
\begin{equation} \label{eq:I1}
I_{\alpha,\beta}(\theta) = \prod_{k=1}^3 \left[
  \frac{(2\pi)^{-1/2}}{\alpha_k!}
  \theta^{\frac{\alpha_k - \beta_k}{2}}
  \int_{l_k}^{+\infty} \He_{\alpha_k}(v_k) \He_{\beta_k}(v_k)
    \exp \left( -\frac{|v_k|^2}{2} \right) \dd v_k
\right],
\end{equation}
where
\begin{equation}
l_k = \left\{ \begin{array}{ll}
  -\infty, & k=1,3, \\ 0, & k = 2.
\end{array} \right.
\end{equation}
Applying the orthogonality of Hermite polynomials to \eqref{eq:I1}, we
have
\begin{equation}
I_{\alpha,\beta}(\theta) =  \left[
  \frac{(2\pi)^{-1/2}}{\alpha_2!}
  \theta^{\frac{\alpha_2 - \beta_2}{2}}
  \int_0^{+\infty} \He_{\alpha_2}(v_2) \He_{\beta_2}(v_2)
    \exp \left( -\frac{|v_2|^2}{2} \right) \dd v_2
\right] \cdot \delta_{\alpha_1 \beta_1} \delta_{\alpha_3 \beta_3},
\end{equation}
Now it is obvious that \eqref{eq:I} holds if
\begin{equation}
S(m,n) = \frac{1}{\sqrt{2\pi} m!}
  \int_0^{+\infty} \He_m(x) \He_n(x) \exp(-x^2/2) \dd x.
\end{equation}
Some simple cases can be directly worked out as
\begin{equation}
\begin{split}
S(0,0) &=
  \frac{1}{\sqrt{2\pi}} \int_0^{+\infty} \exp(-x^2/2) \dd x = 1/2, \\
S(0,n) &=
  \frac{1}{\sqrt{2\pi}} \int_0^{+\infty} \He_n(x) \exp(-x^2/2) \dd x
  = \frac{1}{\sqrt{2\pi}} \He_{n-1}(0), \quad n \neq 0, \\
S(m,0) &=
  \frac{1}{\sqrt{2\pi}m!} \int_0^{+\infty} \He_m(x) \exp(-x^2/2) \dd x
  = \frac{1}{\sqrt{2\pi}m!} \He_{m-1}(0), \quad m \neq 0.
\end{split}
\end{equation}
This agrees with the first three cases of \eqref{eq:S}. For $m \neq 0$
and $n \neq 0$, we use the differential relation of Hermite
polynomials and get
\begin{equation}
\begin{split}
S(m,n) &= -\frac{1}{\sqrt{2\pi} m!} \int_{x\in [0,+\infty)}
  \He_n(x) \dd [ \He_{m-1}(x) \exp(-x^2/2) ] \\
&= \frac{1}{\sqrt{2\pi} m!} \left[
  \He_{m-1}(0) \He_n(0) +
  n \int_0^{+\infty} \He_{m-1}(x) \He_{n-1}(x) \exp(-x^2/2) \dd x
\right] \\
&= \frac{1}{\sqrt{2\pi} m!} \He_{m-1}(0) \He_n(0)
  + n/m \cdot S(m-1,n-1).
\end{split}
\end{equation}
This is the last case in \eqref{eq:S}.

\section{Expansion of the half-Maxwellian} \label{sec:half_Max}
This section is devoted to the detailed calculation of $p_{\alpha}$
defined in \eqref{eq:p}. Due to \eqref{eq:p_alpha}, only \eqref{eq:J}
and \eqref{eq:tilde_J} need to be evaluated. We first consider $J_s%
(x)$ with $s \geqslant 1$. By applying the recursion relation of
Hermite polynomials, we get
\begin{equation}
\begin{split}
J_s(x) &= \frac{1}{s!} \theta^{\frac{s+1}{2}}
  \int_{-\infty}^{+\infty}
    \frac{1}{\sqrt{2\pi \theta^W}}
    \exp \left( -\frac{|\sqrt{\theta} y - x|^2}{2\theta^W} \right)
    [y \He_{s-1}(y) - (s-1) \He_{s-2}(y)]
  \dd y \\
& = -\frac{\theta}{s} J_{s-2}(x) + \frac{x}{s} J_{s-1}(x) \\
& \qquad + \underline{\frac{\theta^W}{s!} \theta^{\frac{s-1}{2}}
  \int_{-\infty}^{+\infty}
    \frac{1}{\sqrt{2\pi \theta^W}}
    \exp \left( -\frac{|\sqrt{\theta}y - x|^2}{2\theta^W} \right)
    \left(
      \frac{\theta}{\theta^W} y -
      \frac{x}{\theta^W} \sqrt{\theta}
    \right) \He_{s-1}(y)
  \dd y}.
\end{split}
\end{equation}
For the underlined term, we use integration by parts and the
differential relation of Hermite polynomials, and get
\begin{equation}
\begin{split}
J_s(x) &=
  -\frac{\theta}{s} J_{s-2}(x) + \frac{x}{s} J_{s-1}(x) \\
& \qquad + \frac{\theta^W}{s} \cdot \frac{1}{(s-2)!}
  \theta^{\frac{s-1}{2}} \int_{-\infty}^{+\infty}
    \frac{1}{\sqrt{2\pi \theta^W}}
    \exp \left( -\frac{|\sqrt{\theta}y - x|^2}{2\theta^W} \right)
    \He_{s-2}(y)
  \dd y \\
& = \frac{1}{s} \left[
  (\theta^W - \theta) J_{s-2}(x) + x J_{s-1}(x)
\right].
\end{split}
\end{equation}
When $s = 0$ or $s = -1$, the integral \eqref{eq:J_s} can be directly
worked out as \eqref{eq:J_start} since $\He_0(y) \equiv 1$ and
$\He_{-1}(y) \equiv 0$.

The calculation of \eqref{eq:tilde_J} is almost the same as \eqref
{eq:J}. The only difference is that a boundary term will appear when
integrating by parts. So the result becomes
\begin{equation}
\tilde{J}_s(x) = \frac{1}{s} \left[
  (\theta^W - \theta) \tilde{J}_{s-2}(x) + x \tilde{J}_{s-1}(x)
\right]
- \frac{1}{s!} \sqrt{\frac{\theta^W}{2\pi}}
  \theta^{\frac{s-1}{2}} \He_{s-1}(0)
  \exp \left( -\frac{x^2}{2 \theta^W} \right),
  \quad s \geqslant 1
\end{equation}
with initial conditions \eqref{eq:tilde_J_start}. Define
\begin{equation}
H_s(x) = \frac{1}{s!} \sqrt{\frac{\theta^W}{2\pi}}
  \theta^{\frac{s-1}{2}} \He_{s-1}(0)
  \exp \left( -\frac{x^2}{2 \theta^W} \right).
\end{equation}
Then \eqref{eq:tilde_J_s} and \eqref{eq:H_start} are natural. For $s
\geqslant 1$, the recursion relation of $H_s$ can be deduced as
\begin{equation}
\begin{split}
H_s(x) &= \frac{\theta}{s(s-1)} \cdot \frac{1}{(s-2)!}
  \sqrt{\frac{\theta^W}{2\pi}} \theta^{\frac{s-3}{2}}
  [0 \cdot \He_{s-2}(0) - (s - 2) \He_{s-3}(0)]
  \exp \left( -\frac{x^2}{2 \theta^W} \right) \\
&= -\frac{s-2}{s(s-1)} \theta H_{s-2}(x).
\end{split}
\end{equation}

\bibliographystyle{plain}
\bibliography{../article}
\end{document}